\documentclass{emulateapj}
\usepackage{mathrsfs}
\usepackage{IEEEtrantools}
\usepackage{float}
\usepackage{bm}
\usepackage{graphicx}
\usepackage{threeparttable}

\shorttitle{Location and Structure of Dead Zone Inner Boundaries}
\shortauthors{S. Mohanty et al.}

\begin{document}

\def\torb{$t_{\rm orb}$}
\def\etao{$\eta_O$}
\def\etah{$\eta_H$}
\def\etaa{$\eta_A$}
\def\va{$v_{\mathcal A}$}
\def\vaz{$v_{{\mathcal A}z}$}
\def\vasq{$v_{\mathcal A}^2$}
\def\vk{$v_K$}
\def\pgas{$P_{\rm gas}$}
\def\pmag{$P_{\rm B}$}
\def\cs{$c_s$}
\def\cssq{$c_s^2$}
\def\bmin{$\beta_{\rm min}$}
\def\lambdao{$\Lambda_O$}
\def\mstar{$M_{\ast}$}
\def\msun{${\rm M}_{\odot}$}
\def\rsun{${\rm R}_{\odot}$}
\def\mdot{$\dot{M}$}
\def\mdotin{$\dot{M}_{\rm in}$}
\def\mdotout{$\dot{M}_{\rm out}$}
\def\alphain{${\bar{\alpha}}_{\rm in}$}
\def\alphaout{${\bar{\alpha}}_{\rm out}$}
\def\rdzib{$r_{\rm DZIB}$}
\def\amri{\alpha_{\rm MRI}}
\def\adz{\bar\alpha_{\rm DZ}}

\title{Inside-Out Planet Formation. V.\\
Structure of the Inner Disk as Implied by the MRI}

\author{Subhanjoy Mohanty\altaffilmark{1}, Marija R. Jankovic\altaffilmark{1}, Jonathan C. Tan\altaffilmark{2} and James E. Owen\altaffilmark{1}}

\altaffiltext{1}{Dept. of Physics, Imperial College London, 1010 Blackett Lab., Prince Consort Rd., London SW7 2AZ, UK. s.mohanty@imperial.ac.uk, m.jankovic16@imperial.ac.uk.}
\altaffiltext{2}{Departments of Astronomy and Physics, University of Florida, Gainesville, FL 32611}

\begin{abstract}
The ubiquity of Earth to super-Earth sized planets found very close to their host stars has motivated {\it in situ} formation models. In particular, Inside-Out Planet Formation is a scenario in which planets coalesce sequentially in the disk, at the local gas pressure maximum near the inner boundary of the dead zone. The pressure maximum arises from a decline in viscosity, going from the active innermost disk (where thermal ionization yields high viscosities via the magneto-rotational instability (MRI)) to the adjacent dead zone (where the MRI is quenched). Previous studies of the pressure maximum, based on $\alpha$-disk models, have assumed ad hoc values for the viscosity parameter $\alpha$ in the active zone, ignoring the detailed MRI physics. Here we explicitly couple the MRI criteria to the $\alpha$-disk equations, to find steady-state solutions for the disk structure. We consider both Ohmic and ambipolar resistivities, and a range of disk accretion rates ($10^{-10}$--$10^{-8}$\,\msun\,yr$^{-1}$), stellar masses (0.1--1\,\msun), and fiducial values of the {\it non}-MRI $\alpha$-viscosity in the dead zone ($\alpha_{\rm {DZ}} = 10^{-5}$--$10^{-3}$). We find that: {\it (1)} A midplane pressure maximum forms radially {\it outside} the dead zone inner boundary; {\it (2)} Hall resistivity dominates near the inner disk midplane, perhaps explaining why close-in planets do {\it not} form in $\sim$50\% of systems; {\it(3)} X-ray ionization can compete with thermal ionization in the inner disk, because of the low steady-state surface density there; and {\it (4)} our inner disks are viscously unstable to surface density perturbations. 



    
\end{abstract}

\section{Introduction}

The {\it Kepler} mission has discovered more than 4000 exoplanet candidates from observations of their transits \citep[e.g.,][]{mullally15, coughlin16}. One of the great surprises from this dataset is the ubiquity of Earth and super-Earth sized planets in very tight orbits, which have no solar system analogs. Specifically, more than 50\% of sun-like stars appear to harbour one or more planets of size 0.8--4\,\rsun\,\,at orbital periods $P < 85$\,days (i.e., shorter than Mercurys's; Fressin et al.\,2013). Similarly, nearly all M dwarfs seem to host one or more 0.5--4\,\rsun\,\,sized planets at $P < 50$\,days \citep{dressing15}. Note that the single-planet systems included in these statistics may have as yet undetected smaller planets as well. Moreover, a large fraction ($\gtrsim$30\%) of the close-in multi-planet {\it Kepler} systems appear dynamically packed (i.e., cannot admit an additional planet without becoming unstable; Fang \& Margot 2013). Thus, a major, and possibly the dominant, planet formation mechanism in our galaxy produces small planets very close to the central star, with a large fraction of these in tightly-packed multi-planet systems. Two main scenarios have been advanced to explain such planets: {\it (1)} formation in the outer disk followed by inward migration \citep[e.g.,][]{kley12,cossou13, cossou14}; and {\it (2)} formation {\it in situ} \citep[][hereafter CT14]{hansen12, hansen13, chiang13, CT14}.

The inward migration scenario tends to produce planets that are trapped in orbits of low order mean motion resonances, which is not a particular feature of these {\it Kepler} systems \citep{baruteau14, fabrycky14}. Recently discovered trends in the atmospheric photoevaporation of these planets also indicate an Earth-like (rock/iron) core composition, implying formation inwards of the ice-line and thus arguing against significant migration \citep{owen17}. 

The Inside-Out Planet Formation (IOPF) scenario proposed by CT14 is a new type of {\it in situ} formation model. It is based on the fact that the effective viscosity in the disk is expected to decline, moving radially outwards from the innermost disk -- where efficient thermal ionization of alkali metals \citep{umebayashi88} activates the magneto-rotational instability \citep[MRI;][]{balbus91}, leading to high viscosities -- to the adjacent ``dead zone'', where decreasing thermal ionization leads to a suppression of the MRI by Ohmic resistivity, yielding low viscosities \citep{gammie96}. In a steady-state disk, i.e., one with a constant disk accretion rate \mdot, this fall-off in viscosity produces a local maximum in the gas pressure in the vicinity of the dead zone inner boundary (DZIB). The IOPF mechanism proposes that dust grains that have grown to $\sim$cm-sized ``pebbles'' in the outer disk \citep{hu17} and are drifting radially inwards are trapped in this pressure maximum, within which they rapidly coalesce into a protoplanet. The protoplanet itself is also expected to be trapped in this region \citep{hu16}, and thus able to continue growing (especially by pebble accretion), until it becomes massive enough to open a gap a few Hill radii wide in the disk.  

Material interior to the inner rim of this gap will tend to drain rapidly (on a local viscous timescale) onto the star. While some replenishment of this interior region may continue due to gas flowing accross the gap, densities here are expected to decrease, potentially leaving the outer rim of the gap subject to direct stellar X-ray/UV irradiation. This can activate the MRI in disk gas close to the outer rim, over a thickness set by how far stellar ionizing photons penetrate radially into the rim \citep[e.g.,][]{chiang07}. A new DZIB then forms at the outer edge of this MRI-active region, creating a new pressure trap where incoming pebbles can coagulate into another planet. The process continues till the pebble supply from the outer disk is exhausted, leaving behind a system of closely-packed inner planets.  

The formation of gas pressure maxima is thus central to the IOPF model. In particular, the location of the first maximum, controlled by thermal ionization of alkalis in the inner disk, sets the orbital radius of the innermost (so-called ``Vulcan'') planet in the system. The goal of this paper is to investigate the formation of this first pressure maximum. 

There have been several previous works studying pressure traps in the disk created by changes in the viscosity \citep[e.g., CT14;][]{kretke07, kretke09, kretke10}. All of these have been based on a steady-state Shakura-Sunyaev $\alpha$-disk model, wherein the disk accretion rate is constant, viscous heating due to accretion is the main source of energy input, and the disk viscosity is parametrised in terms of the quantity $\alpha$. Crucially, however, these studies have all adopted ad hoc prescriptions of $\alpha$ for computational ease, without accounting for the detailed physics of the MRI. 

Conversely, several groups have investigated the behaviour of active and dead zones in the disk, accounting for the detailed effects of non-ideal MHD and complex gas and dust chemistry on the MRI, either using direct numerical simulations \citep[e.g.,][]{bai11a, bai11b, turner10, bai17} or based on the MRI criteria implied by such simulations \citep[e.g.,][]{perez-becker11a, perez-becker11b, mohanty13}. However, these studies all assume a passive disk (heated and ionized by stellar irradiation), and a pre-determined temperature and surface density profile (usually Minimum Mass Solar Nebula). Consequently, the results are generally neither in steady-state (\mdot\,varies with radius) nor applicable to the inner disk (where viscous heating dominates).   

Our aim here is to marry the two approaches: we wish to solve for the structure of the inner disk assuming a steady-state, viscously heated $\alpha$-disk, but with $\alpha$ determined self-consistently from detailed considerations of the MRI and non-ideal MHD effects. To the best of our knowledge, this is the first such unified disk model (\citet{keith14} present an elegant self-consistent $\alpha$-disk model for circumplanetary disks, but their MRI-$\alpha$ is a more parametrized version than ours, with a saturation value set arbitrarily). As such, the results are germane not only to the IOPF mechanism and the specific purpose of locating a pressure maximum in the inner disk, but also to the broader goal of understanding the structure of viscously heated steady-state disks with MRI-driven accretion and non-ideal MHD. 

In \S2, we provide an overview of our methodology and discuss some critical caveats to our assumption of MRI-driven accretion. In \S3, we summarize the $\alpha$-disk model, and in \S4, we describe our treatment of the MRI. Our technique for calculating $\alpha$ and \mdot\,is detailed in \S\S5 and 6, and our method of determining equilibrium solutions outlined in \S7. We present our results in \S8, and discuss their implications in \S9.   

\section{Overview of Methodology and Caveats}

{\textbf{\textit{Methodology}}}: We wish to investigate the location of the pressure maximum in the inner disk, by solving for the inner disk structure in steady-state (i.e., with constant \mdot) and assuming that the MRI is the dominant magnetically-controlled mechanism for local mass and angular momentum transport. We further wish to do this in the context of the Shakura-Sunyaev $\alpha$-disk model. Consequently, we must solve the {\it coupled} set of equations for the MRI and disk structure: coupled because the effective viscosity parameter $\alpha$ from the MRI, and the attendant \mdot, both depend on the underlying disk structure (as well as on the magnetic field strength $B$), while the disk structure itself, in the Shakura-Sunyaev model, is determined by $\alpha$ and \mdot\,(and stellar parameters).

Briefly, we use a grid-based method of solution. A grid of disk structures is calculated for a desired \mdot\,and a range of input values for $\alpha$ and field strength $B$; the MRI-induced output $\alpha$ and corresponding accretion rate \mdot\,are derived for each of these disk structures; and the chosen solution structure is the one in which the output values of $\alpha$ and \mdot\,match the input ones. We find that a range of such solutions are possible differing in $B$; a unique solution is chosen under the assumption that the MRI is maximally efficient, i.e., generates the largest field it can support (but see `Caveats' below).


{\it Pressure Maximum}: How do our solutions produce a pressure maximum? In the $\alpha$-disk model, the gas pressure is a decreasing function of both $\alpha$ and radius. This leads to a turnover in pressure at the radial location where our derived $\alpha$ falls to its minimum value. What defines this minimum in our methodology? In previous work as well as in this paper, a lower limit (``floor'') on $\alpha$ is set by its value $\alpha_{\rm DZ}$ in the dead zone, where the MRI is quenched but various (non-magnetic) hydrodynamic/gravitational instabilities may still generate viscous stresses. Fiducial values for this floor are chosen based on theory and numerical simulations; we explore the plausible range $\alpha_{\rm DZ}$ = 10$^{-5}$--10$^{-3}$ (discussed in more detail later). The pressure maximum then occurs where the $\alpha$ in the MRI-active zone decreases to this dead zone limiting value. This floor will always be reached if heating due to viscous accretion (a decreasing function of radius for constant \mdot) is the only source of the ionization required to kindle the MRI (as assumed here; but see also X-rays/UV below). 

{\it Simplifications}: In this pilot study, we adopt a number of simplifications: no ionization by stellar photons (X-ray or UV; we only consider thermal ionization due to accretion heating); ionization of a single alkali species (i.e., no complex chemical network); no dust; and a fixed opacity of 10\,cm$^{2}$\,g$^{-1}$. Relaxing these assumptions presents no conceptual difficulties, and we shall do so in a subsequent paper (Jankovic et al.\,in prep.); the inclusion of more physics will certainly change the precise location of the pressure maximum (e.g., dust grains will reduce the MRI efficiency, and X-rays may change the limiting value of $\alpha$; a these effects and others are discussed at appropriate junctures). Nevertheless, as an initial step, the mathematical ease afforded by these simplifications allows us to clearly present our methodology and identify important general trends in the solutions.

{\textbf{\textit{Caveats}}}: Finally, there are crucial caveats, applicable to all work so far on pressure maxima in the inner disk (including this paper), concerning the basic assumption that mass and angular momentum transport are controlled by the MRI. In the innermost disk, where the inductive term in the field evolution equation greatly exceeds the resistive terms, the MRI is indeed likely to be dominant and maximally efficient \citep[e.g.,][]{bai13b}. Further out, however, where the resistivities become non-negligible, the situation is much more complicated. 

Specifically, first, when Ohmic and ambipolar resistivities are both important, vertically stratified 3-D simulations \citep{bai13a, bai13b, gressel15} imply that: {\it (a)} in the absence of any net vertical magnetic flux, the MRI is extremely weak, with an effective viscosity orders of magnitude lower than required to power the observed accretion rates in classical T Tauri stars; and {\it (b)} with even a small net vertical field, MRI turbulence is completely smothered (because, while the MRI is initially present, the field is subsequently amplified to strengths {\it greater} than that at which the MRI can operate under ambipolar diffusion; i.e., the assumption of maximally efficient MRI is no longer valid). The flow over the entire vertical extent of the disk now becomes fully laminar, and a magnetised disk wind develops instead, which efficiently carries angular momentum away from the disk and drives accretion at rates consistent with observations. In other words, where Ohmic and ambipolar effects are both important, mass accretion seems driven primarily by vertical angular momentum transport by magnetised winds, and not radial transport by the MRI. 

Second, introducing the Hall effect into the above situation complicates matters further, depending on whether the net vertical magnetic field is aligned or anti-aligned with the spin-axis of the disk \citep{lesur14, bai14, bai15, simon15, bai17}. When the two are aligned (i.e., $\mathbf{\Omega}\cdot\mathbf{B} > 0$), the Hall shear instability (HSI) generates laminar viscous stresses via the amplification of horizontal components of the field \citep{kunz08}, leading to strong radial angular momentum transport and hence significant mass accretion (in addition to the magnetised-wind-driven accretion at comparable rates).  Conversely, when the field and disk spin-axis point in opposite directions ($\mathbf{\Omega}\cdot\mathbf{B} < 0$), the horizontal field is considerably suppressed, and mass and angular momentum transport are predominantly wind-driven.

At face value, these results suggest that using the Shakura-Sunyaev viscous disk model to search for a pressure maximum, with the expectation that $\alpha$ declines sharply across the interface between the MRI-active innermost disk and the adjacent dead zone-dominated region, might not be a valid exercise for two reasons. First, in the region usually characterised as ``dead zone''-dominated, angular momentum in the aforementioned simulations is mainly transported vertically {\it out} of the disk by wind-related torques, instead of being radially redistributed {\it within} the disk by standard viscous torques (either hydrodynamic/gravitational within the dead zone, or MRI in an overlying active layer). Thus, the Shakura-Sunyaev viscous model is invalid here. Second, when the field and disk spin-axis are aligned, the HSI activates efficient mass and angular momentum transport all the way down to the midplane here (in addition to wind-related transport higher up); i.e., there is no dead zone in any sense.  

Nevertheless, it is premature to write off an inner disk pressure maximum in the standard viscous disk context. All the above simulations are restricted to radii $\gtrsim 1$\,AU, an order of magnitude further out than the presumed location of the pressure maximum at $\lesssim$ few tenths of an AU (Bai 2017's simulation domain formally extends in to 0.6\,AU, but they deem the results at $<$2\,AU to be vitiated by boundary effects). Thus, it remains to be seen whether the above conclusions apply to our region of interest in the inner disk. Concurrently, if the close-in planets we address here are indeed formed {\it in situ} from inward migrating solids, then some sort of pressure trap seems inescapable in this region, in order to corral these solids and prevent their falling into the star. As such, continuing this line of inquiry currently appears justified.  

Finally, even if the wind/Hall results from the simulations extend to much smaller radii, a pressure maximum is still plausible (and, in general, a significant change in disk structure is expected) at the interface between the innermost MRI-active turbulent disk and the adjacent wind-dominated laminar disk, because of the qualitative difference in physical conditions between the two regions. The Shakura-Sunyaev $\alpha$-disk model will not apply across the interface, and the controlling factor for any change in disk structure may be the radial distribution of magnetic flux (since the field ultimately determines the strength of the MRI, the wind and the Hall effect; Bai, X.~N., pvt.\,comm., 2017), rather than the radial behaviour of $\alpha$ as assumed here. Nonetheless, the $\alpha$-disk model will still apply to the MRI region, and insights into the latter gleaned from the present work will remain useful.

\section{Disk model}

A detailed derivation of the steady-state (temporally constant) disk structure within the Shakura-Sunyaev viscous $\alpha$-disk model is given by \citet[hereafter H16]{hu16}. We summarise the main results here. The viscosity parameter $\alpha$ is defined by the relation
\begin{equation}
\nu = \alpha \frac{c_s^2}{\Omega} 
\end{equation}
where $\nu$ is the viscosity, $c_s$ the sound speed and $\Omega$ the Keplerian angular velocity at any given disk radius.  Now, the $\alpha$-disk model is fundamentally derived from vertically-integrated quantities (surface density and accretion rate; see H16); as such, the ``$\alpha$'' that enters into it is more precisely a vertical average. This issue is often elided (e.g., H16 do not discuss it) under the implicit assumption that $\alpha$ is vertically constant or slowly varying. However, in a vertically stratified disk (such as we will find), with MRI-active zones sandwiched between inactive ones, the nature of the viscosity changes with height, and the latter assumption is invalid. In this case, the relevant quantity is the {\it effective} viscosity parameter $\bar\alpha$, defined as the pressure-weighted vertical average of $\alpha$:
\begin{equation}
\bar{\alpha} \equiv \frac{\int_{-\infty}^{+\infty} \alpha P_{\rm gas}\, dz}{\int_{-\infty}^{+\infty} P_{\rm gas}\, dz}  = \frac{\int_{-\infty}^{+\infty} \alpha \rho\, dz}{\int_{-\infty}^{+\infty} \rho\, dz}
\end{equation}
where the second equality (derived using $P_{\rm gas} = \rho c_s^2$ for density $\rho$) holds only for a vertically isothermal disk (so that $c_s^2$ is constant with height; we shall assume such isothermality further below). We show how to calculate $\bar\alpha$ in \S5. We explicitly append the subscript ``gas'' to pressure $P$ to differentiate the gas pressure from the magnetic pressure ($P_{\rm B}$, encountered later); for all other quantities (density, temperature etc.) we drop this subscript, since they always refer to gas alone in this paper. 

With this definition of $\bar\alpha$, the steady-state gas surface density (summing both above and below the midplane) at any orbital radius $r$ in the disk is given by
\begin{IEEEeqnarray}{rCl}
\Sigma(r) &  = & 139.4 \,\, \gamma_{1.4}^{-4/5}\, \kappa_{10}^{-1/5}\, \bar{\alpha}_{-3}^{-4/5}\, M_{\ast,1}^{1/5} \nonumber\\ && \times \left(f_r\dot{M}_{-9}\right)^{3/5}\, r_{\rm AU}^{-3/5} \,\,\,{\rm g \,cm}^{-2}
\end{IEEEeqnarray}
for a normalised stellar mass $M_{\ast,1} \equiv$ \mstar/1\,\msun, accretion rate $\dot{M}_{-9} \equiv$ \mdot/10$^{-9}$\,\msun\,yr$^{-1}$, radial distance $r_{\rm AU} \equiv$ $r$/1\,AU, opacity $\kappa_{10} \equiv$ $\kappa$/10\,cm$^2$g$^{-1}$, adiabatic index $\gamma_{1.4} \equiv$ $\gamma$/1.4 and effective viscosity parameter $\bar{\alpha}_{-3} \equiv$ $\bar{\alpha}/10^{-3}$, with $f_r \equiv \left(1 - \sqrt{R_{in}/r}\right)$ for a disk inner edge located at $R_{in}$ (if the disk extends to the stellar surface, then $R_{in}$ = $R_{\ast}$, the stellar radius). The associated midplane temperature is given by
\begin{IEEEeqnarray}{rCl}
T_0(r) & = & 192.6 \,\, \gamma_{1.4}^{-1/5}\, \kappa_{10}^{1/5}\, \bar{\alpha}_{-3}^{-1/5}\, M_{\ast,1}^{3/10} \nonumber\\ && \times \left(f_r \dot{M}_{-9}\right)^{2/5}\, r_{\rm AU}^{-9/10} \,\,\,{\rm K}
\end{IEEEeqnarray}
if viscous heating in the main source of energy input (i.e., heating by stellar irradiation is ignored). The midplane pressure is then
\begin{IEEEeqnarray}{rCl}
P_{0,{\rm gas}}(r) & = & 0.773 \,\, \gamma_{1.4}^{-7/5}\, \kappa_{10}^{-1/10}\, \bar{\alpha}_{-3}^{-9/10}\, M_{\ast,1}^{17/20} \nonumber\\
&& \times \left(f_r \dot{M}_{-9}\right)^{4/5}\, r_{\rm AU}^{-51/20} \,\,\,{\rm erg \,cm}^{-3}\,\,,
\end{IEEEeqnarray}
and the midplane density (which follows from the ideal gas law $P_{\rm gas} = \rho k_B T/\mu$, for particles with mean molecular mass $\mu \approx 2.34\,m_{\rm H}$, where $m_{\rm H}$ is the atomic mass of hydrogen) is
\begin{IEEEeqnarray}{rCl}
\rho_0(r) & = & \left(1.133\times 10^{-10}\right) \,\, \gamma_{1.4}^{-6/5}\, \kappa_{-10}^{-3/10}\, \bar{\alpha}_{-3}^{-7/10}\, M_{\ast,1}^{11/20} \nonumber\\
&& \times \left(f_r \dot{M}_{-9}\right)^{2/5}\, r_{\rm AU}^{-33/20} \,\,\,{\rm g \,cm}^{-3} \,\,.
\end{IEEEeqnarray}
Crucially, equations (5) and (6) show that the {\it midplane} pressure and density do not depend on the {\it local} $\alpha$, but rather on its vertically averaged value $\bar\alpha$. In other words, the midplane pressure (and thus density) is sensitive to conditions in the entire column pressing down from above (as intuitively expected), not simply local ones. This has the following important consequence. As we will show, the midplane pressure maximum does not form where the dead zone first develops in the midplane (i.e., at the dead zone inner boundary, which is where the midplane $\alpha$ reaches its minimum), as often assumed. Instead, it forms further out radially, where the {\it effective} parameter $\bar\alpha$ reaches its minimum (because the MRI-active zone continues outwards for some distance above the dead zone). Thus, we will find that the midplane pressure maximum is actually located {\it within} the dead zone.

Unlike H16, we assume for simplicity that the disk is vertically isothermal (i.e., $\gamma = 1$). Strictly speaking, this is slightly inconsistent with the derivation of the midplane temperature (equation (4) above) by H16, following the formalism of \citet{hubeny90}, wherein the temperature depends on the vertical optical depth in the disk. However, implementing this dependence couples together the vertical temperature and density profiles in a complicated fashion (H16 avoid this because they are concerned with just midplane values). Moreover, at small optical depths ($\tau \ll 1$), the temperature is also highly sensitive to the details of the appropriate radiative processes \citep[a simplistic treatment of which leads to an infinitely hot disk surface; see discussion by][]{hubeny90}; addressing these is beyond the scope of this paper. On the other hand, at large optical depths ($\tau \gg 1$), $T$ only varies very slowly with depth, as $\tau^{1/4}$ \citep[see][]{hubeny90}. Therefore, since we expect the inner disk to only be active in optically thick regions close to the midplane, we approximate the vertical temperature profile in the region of interest by the midplane values: $T(z, r) \sim T_0 (r)$.       

The (isothermal) sound speed is then $c_s = \sqrt{k_B T_0/\mu}$, and the vertical pressure profile in hydrostatic equilibrium becomes
\begin{equation}
P_{\rm gas} (z, r) = P_{0,{\rm gas}}(r) \,\,{\rm exp}\left(-\frac{z^2}{z_H^2}\right) 
\end{equation}
where the pressure scale height is defined as $z_H \equiv \sqrt{2}\,c_s/ \Omega$. 
Finally, we assume a constant opacity of $\kappa = 10$\,cm$^2$\,g$^{-1}$, approximately the expected value in protoplanetary disks \citep[e.g.,][]{wood02}. H16 use the detailed opacity tables of \citet{zhu12}, where the values depend on the pressure and temperature structure of the disk, and solve for the equilibrium opacities and structure iteratively. In our case, however, the disk structure equations are already coupled to the MRI ones, and the two sets must be solved simultaneously. Introducing a further inter-dependence with opacity adds a level of complexity that we set aside in this exploratory work. We do compare, a posteriori, our constant $\kappa$ to the values implied by \cite{zhu12} for our equilibrium disk structure, to gauge the discrepancy between the two; in general we find our value to be reasonable. 


\section{MRI}
Our treatment of the MRI generally follows that of \cite{mohanty13}, except we consider ionization by thermal collisions instead of by X-rays, and we do not include grains. Here we summarise the major points of our analysis. The physical conditions required for the MRI to operate are set out in \S4.1; our treatment of thermal ionization and recombination is discussed in \S4.2; and the calculation of the various resistivities (Ohmic, ambipolar and Hall), which determine whether or not the MRI criteria are met, is described in \S4.3.  

\subsection{Criteria for Active MRI}
We discuss the necessary conditions for active MRI in Appendix A, and only state the final results here. The Ohmic Elsasser number $\Lambda$ is defined as
\begin{equation}
\Lambda \equiv \frac{v_{{\mathcal A}z}^2}{\eta_O \Omega}
\end{equation}
where $\eta_O$ is the Ohmic resistivity and $v_{{\mathcal A}z}$ the vertical component of the local Alfv\'{e}n velocity ($\equiv B_z/\sqrt{4\pi \rho}$, where $B_z$ is the vertical field strength and $\rho$ the local gas density). Similarly, the ambipolar Elsasser number $Am$ is defined as
\begin{equation}
Am \equiv \frac{v_{\mathcal A}^2}{\eta_A \Omega}
\end{equation}
where $\eta_A$  is the ambipolar resistivity and $v_{\mathcal A}$ the local total Alfv\'{e}n velocity ($\equiv B/\sqrt{4\pi \rho}$, where $B$ is the r.m.s.\,field strength)\footnote{Our reasons for adopting $v_{{\mathcal A}z}$ in equation (8) but $v_{\mathcal A}$ in equation (9) are supplied in the discussion preceding equation (A1) and in footnote [8], in Appendix A.}.  

With these definitions, the conditions for sustaining active MRI are:
\begin{equation}
\Lambda > 1
\end{equation}
and
$$ \beta > \beta_{\rm min}(Am) \,\,. \eqno(11a) $$
Here $\beta \equiv$ \pgas/\pmag\, is the plasma $\beta$-parameter (with magnetic pressure \pmag\, $\equiv B^2/8\pi$), and the minimum allowed value of $\beta$ -- denoted by $\beta_{\rm min}$ -- is a function of the ambipolar Elsasser number $Am$ \citep{bai11a}:
$$ \beta_{\rm min}(Am) = \left[ \left(\frac{50}{Am^{1.2}}\right)^2 + \left(\frac{8}{Am^{0.3}} + 1\right)^2\right]^{1/2} \,\, . \eqno(11b)  $$
Equation (10) encapsulates the reasonable condition that, when Ohmic resistivity dominates, the MRI is sustained when the growth rate of the fastest growing MRI mode exceeds its dissipation rate. When ambipolar diffusion dominates, on the other hand, \cite{bai11a} find that, in the strong-coupling (single-fluid) limit applicable to protoplanetary disks (see discussion preceding equation (A4a) in Appendix A), the MRI can operate at {\it any} value of $Am$, {\it provided} the field is sufficiently weak. Equations (11a,b) then define what ``sufficiently weak'' means: it signifies that the plasma $\beta$-parameter must exceed a minimum threshold $\beta_{\rm min}$. Specifically, it implies that the gas pressure must dominate over the magnetic pressure in the disk for the MRI to function (see discussion following equation (A4b)). An ``active zone'' is where both conditions (10) and (11) are satisfied, allowing efficient MRI; a ``dead zone'' is where condition (10) is not met, so that Ohmic resistivity shuts off the MRI; and a ``zombie zone'' \citep[following the nomenclature of][]{mohanty13} is where condition (11a) is not satisfied, so that ambipolar diffusion quenches the MRI. 

Note that the effects of Hall diffusion are ignored in the above analysis. As discussed in \S2 and Appendix A, in the presence of a net vertical background field, the Hall effect can amplify the MRI or suppress it, depending on whether the field is aligned or anti-aligned with the spin axis of the disk. Quantifying this effect is beyond the scope of this paper. However, we do investigate the Hall effect a posteriori, by calculating the Hall Elsasser number ($\chi \equiv  {v_{\mathcal A}^2} / ({|\eta_H| \Omega})$; see Appendix A and equation (A3)) everywhere in our solutions. In any region where $\chi < 1$, which we call a ``Hall zone'', Hall diffusion has a strong effect on the MRI; we discuss the potentially critical implications of such regions for planet formation.   

\subsubsection{Choice of Magnetic Field Strength}

Both the Ohmic and ambipolar conditions for active MRI, equations (10) and (11), depend on the magnetic field strength: via \vaz\, in \lambdao\, and \pmag\, in $\beta$. Indeed, for a given set of stellar parameters and a fixed accretion rate, we will see that there exist an infinite number of solutions, each corresponding to a different disk structure with a different field strength $B$. 

The question then is how to determine an appropriate $B$. We do so by assuming that: (a) {\it the magnetic field strength is constant with height across the active layer}; and (b) {\it the MRI is maximally efficient, generating the strongest possible field that still allows the MRI to operate (i.e., still satisfies the constraint $\beta > \beta_{\rm min}$)}. 

The same assumptions are made by \citet{mohanty13} and \citet{bai11b}. A roughly constant $B$ across the active layer is expected from MRI-driven turbulent mixing \citep[][and references therein]{bai11b}, justifying (a). Condition (b) encapsulates the notion that (in the absence of any other mechanism) the MRI-turbulence will continue to amplify the field up to some maximum value $B_{\rm max}$ corresponding to $\beta_{\rm min}$, beyond which the MRI is quenched (i.e., the instability is self-regulated). Our implementation of this condition to derive equilibrium disk solutions is described in \S5. 

Finally, we note that numerical simulations of the MRI by \citet{sano04} indicate that the total r.m.s.\,field strength $B$ and its vertical component $B_z$ are related by $B^2 \sim 25\,B_z^2$, a condition we adopt. Thus, though our Ohmic MRI condition is defined in terms of \vaz\, $\propto B_z^2$ while the ambipolar condition is in terms of \pmag\, $\propto B^2$, one need specify only $B$ or $B_z$, not both independently.

\subsection{Thermal Ionization and Recombination}
In the hot inner regions of the disk, ionization is dominated by thermal collisions, with the equilibrium level of thermal ionization of an atomic species $a$ given by the Saha equation:
$$ \frac{n_e \,n_{+,a}}{n_{0,a}} = \frac{1}{\lambda_e^3}\, \frac{g_e \,g_{+,a}}{g_{0,a}} \,\,{\rm exp}\left(\frac{-{\mathcal{I}}_a}{k_B T}\right) \,\, . \eqno(12) $$
Here $n_e$ is the number density of free electrons, and $n_{0,a}$ and $n_{+,a}$ are the number densities of neutral atoms and singly ionized ions respectively of species $a$; $\lambda_e \equiv \sqrt{h^2/(2\pi m_e k_B T)}$ is the thermal de Broglie wavelength of electrons of mass $m_e$; $g_e (=2)$, $g_{0,a}$ and $g_{+,a}$ are the degeneracy of states for free electrons, neutrals and ions; and ${\mathcal{I}}_a$ is the ionization energy.   

We note the following simplifications when only one, singly-ionized species (e.g., an alkali metal; see below) participates in ionization / recombination. In this case, charge conservation requires $n_e = n_{+,a}$ and $n_{0,a} = n_a - n_e$ (where $n_a$ is the total number density of species $a$). Since molecular hydrogen, with number density $n_{\rm H_2}$, forms the vast bulk of the gas, we adopt the standard expressions for fractional ionization, $x_e \equiv n_e / n_{\rm H_2}$, and the abundance of species $a$, $x_a \equiv n_a / n_{\rm H_2}$. Writing the entire R.H.S. of the Saha equation above as ${\mathcal{S}}_a(T)$, a little algebra then yields: $x_e = \left[-1 \pm\sqrt{1 + 4 x_a(n_{\rm H_2}/{\mathcal{S}}_a)}\right] / \left[ 2(n_{\rm H_2} / {\mathcal{S}}_a)\right]$. This leads to two limiting physical solutions: when $n_{\rm H_2} \rightarrow 0$ (more precisely, when $4 x_a(n_{\rm H_2}/{\mathcal{S}}_a) \ll 1$), we get $x_e \approx x_a$; and when $n_{\rm H_2} \rightarrow \infty$ (more precisely, when $4 x_a(n_{\rm H_2}/{\mathcal{S}}_a) \gg 1$), we get $x_e \approx \sqrt{x_a{\mathcal{S}}_a / n_{\rm H_2}}$. Also note that, without any ionization of hydrogen itself, and with hydrogen being the most abundant species by far, we have $n_{\rm H_2}$ $\approx$ $n_n$ (number density of neutrals) $\approx$ $n_{\rm tot}$ (total number density of particles). We use these results later.     

In order of decreasing ionization potential ${\mathcal{I}}_a$, the important elements in the inner disk are He, H, Mg, Na and K \citep{keith14}. The exponential in the Saha equation ensures the on/off behaviour of thermal ionization, wherein most of the atoms of a species $a$ become ionized over a narrow range of temperatures around the ionization temperature $T_a \equiv {\mathcal{I}}_a/k_B$. Thus, since we expect the disk temperature to generally decrease radially outwards and we are concerned with the outer edge of the active zone, we only consider potassium (K) here, which has the smallest ${\mathcal{I}}_a$ and is thus ionized furthest out. Our adopted quantities for K are listed in Table 1; in this pilot study, we neglect its depletion into grains.  

\begin{table}
\centering
\begin{threeparttable}
    \caption{Adopted Parameters for Potassium}
    \begin{tabular}{c c c c}
    \hline
    A$^a$ & $x_{\rm K}$$^{a,b}$ & $\mathcal{I}_{\rm K}$$^a$ & $g_{+,{\rm K}}$/$g_{0,{\rm K}}$$^c$ \\
    (amu) & & (eV) & \\
    \hline
    39.10               & $1.97$\,$10^{-7}$ & 4.34 & 1/2 \\
    \hline
    \end{tabular}
\begin{tablenotes}
\item[a] Atomic mass (A), abundance ($x_{\rm K} \equiv n_{\rm K}/n_{\rm H_2}$) and ionization potential (${\mathcal{I}}_{\rm K}$) from \citet{keith14}. 
\item[b] \citet{keith14} cite the abundance of K relative to H atoms as 9.87$\times$10$^{-8}$; our value is relative to H molecules, and thus double their value.  
\item[c] \citet{rouse61} cites $g_+$/$g_0 = 1/2$ for the alkali metal sodium; we adopt the same value for the alkali potassium. 
\end{tablenotes}
\end{threeparttable}
\end{table}

With a chemical network comprising collisional ionization/recombination of just one singly-ionized element, the recombination rate is simply $dn_e / dt = k_{ei}\, n_e\, n_{+,a} = k_{ei}\, n_e^2$, where $k_{ei} = 3\times 10^{-11}/\sqrt{T}$\,cm$^3$\,s$^{-1}$ \citep{ilgner06} is the rate coefficient for electron-ion collisions, and the second equality follows from charge conservation. The recombination timescale is then \citep[e.g.,][]{bai11b} 
$$ t_{\rm rcb} \sim \frac{n_e}{dn_e/dt} = \frac{1}{k_{ei}\,n_e}\,\, . \eqno(13) $$
We will compare this timescale to the dynamical time $t_{\rm dyn}$ to verify whether our equilibrium solutions are in the strongly-coupled limit described in Appendix A. 

\subsection{Resistivities}
Armed with the equilibrium abundances of electrons, ions and neutrals computed via the Saha equation, we derive the resistivities in the disk, and thus examine where the disk is MRI active by the criteria of \S4.1 (for a field strength $B$ given by the considerations of \S4.1.1). We follow \citet{wardle07} in writing the Ohmic, Hall and Pederson conductivities ($\sigma_O$, $\sigma_H$ and $\sigma_P$ respectively) as
$$ \sigma_O = \frac{ec}{B}\, \sum_j n_j \,|Z_j|\,\beta_j \eqno(14) $$
$$ \sigma_H = \frac{ec}{B}\, \sum_j \frac{n_j \,Z_j}{1+\beta_j^2} \eqno(15) $$
$$ \sigma_P = \frac{ec}{B}\, \sum_j \frac{n_j \,|Z_j| \,\beta_j}{1+\beta_j^2} \eqno(16) $$
where the summation is over all charged species $j$ (in our case, $j = e$ for electrons and $i$ for singly-charged ions of K), with particle mass $m_j$, number density $n_j$ and charge $Z_j e$ (with $Z_j = \pm1$ for us). The Hall parameter $\beta_j$ (not to be confused with the plasma $\beta$ parameter) is the ratio of the gyrofrequency of a charged particle of species $j$ to its collision frequency with neutrals (of mean particle mass $m_n = \mu m_{\rm H}$ and density $\rho_n$):
$$ \beta_j = \frac{|Z_j|e B}{m_j\,c}\frac{1}{\gamma_j\rho_n} \,\,. \eqno(17) $$  
Here $\gamma_j = \langle \sigma v \rangle_j / (m_j + m_n)$ is the drag coefficient and $\langle \sigma v \rangle_j$ the rate coefficient for collisional momentum transfer between charged species $j$ and neutrals, making $\gamma_j\rho_n$ the collision frequency with neutrals. Note that $\beta_i \ll \beta_e$ (since $m_i \gg m_e$). 

The resistivities may then be written as
$$ \eta_O = \frac{c^2}{4\pi \sigma_O} \eqno(18) $$
$$ \eta_H = \frac{c^2}{4\pi \sigma_{\perp}} \frac{\sigma_H}{\sigma_{\perp}} \eqno(19) $$
$$ \eta_A = \frac{c^2}{4\pi \sigma_{\perp}} \frac{\sigma_P}{\sigma_{\perp}} - \eta_O \eqno(20) $$
where $\sigma_{\perp} \equiv \sqrt{\sigma_H^2 + \sigma_P^2}$ is the total conductivity perpendicular to the magnetic field. 

If electrons and ions are the only charged species (which is the case for us, without grains), then the above equations imply: (1) $\eta_H = \beta_e \eta_O$ and $\eta_A = \beta_i\beta_e\eta_O$\,; (2) consequently, while $\eta_O$ is independent of the magnetic field strength $B$, $\eta_H$ and $\eta_A$ scale linearly and quadratically, respectively, with $B$; and (3) the ambipolar Elsasser number in equation (9), $Am \equiv v_{\mathcal A}^2/\eta_A \Omega$, reduces to (using $\beta_i \ll \beta_e$) $Am \approx \gamma_i\rho_i/\Omega$. The three diffusion regimes then correspond to \citep[e.g.,][]{wardle07}: $\beta_i \ll \beta_e \ll 1$ (Ohmic: neither electrons nor ions are tied to the field, being coupled instead to the neutrals through frequent collisions); $\beta_i \ll 1 \ll \beta_e$ (Hall: electrons are tied to the field while ions are not), and $1\ll \beta_i \ll \beta_e$ (ambipolar: both electrons and ions are tied to the field, and drift together through the sea of neutrals).   

To compute the resitivities, we use the rate coefficients from \citet{wardleng99}:
$$ \langle \sigma v\rangle_e = 10^{-15} \left(\frac{128\, k_B T_e}{9\pi m_e} \right)^{1/2}\,\, {\rm cm}^3\, {\rm s}^{-1} \eqno(21) $$
$$ \langle \sigma v\rangle_i = 1.6 \times 10^{-9} \,\,{\rm cm}^3\, {\rm s}^{-1} \eqno(22) $$
where $T_e$ is the electron temperature, assumed here to equal the disk gas temperature given by equation (4).

\section{Calculation of $\bar\alpha$}

Finally, we must connect the MRI formulation of accretion to the $\alpha$-disk model. In particular, we must specify how to calculate the effective viscosity parameter $\bar\alpha$, defined by equations (1) and (2), that goes into the Shakura-Sunyaev disk model. The derivation is supplied in Appendix B; we only state the main results here. At any radius in the disk, we expect a vertically layered structure: in the hot innermost disk close to the star, we expect an MRI-active zone straddling the midplane, bounded by a zombie zone close to the disk upper and lower surfaces; further out, where the disk is cooler, we expect a dead zone straddling the midplane, a zombie zone close to the disk upper and lower surfaces, and an MRI-active zone sandwiched between the two\footnote{Such a layered disk model was first put forward by \citet{gammie96}, and has since been recovered in various semi-analytic studies invoking both Ohmic and ambipolar diffusion and based on local shearing box MHD simulations (e.g., Bai 2011; Mohanty et al.\,2013; Dzyurkevich et al.\,2013), as well as by global stratified 3D simulations invoking only Ohmic dissipation (Dzyurkevich et al.\,2010) (though all these studies concern larger radii in the disk where the ionisation is primarily due to stellar irradiation, e.g., X-rays, instead of being thermally driven as in this paper, the basic physics for active MRI remains the same as outlined in \S4.1.). As noted in \S2, such a model becomes invalid if, in the presence of both Ohmic and ambipolar diffusion and a net vertical field, the MRI is shut off, angular momentum transport is driven by winds, and the entire vertical extent of the disk becomes laminar instead; our models in this paper do not speak to the latter situation.}. For a vertically isothermal disk (as assumed here), $\bar\alpha$ at any radius is then given in general by
$$ \bar\alpha =  \frac{\sum_i \left( N_i\,{\bar\alpha}_i \right)}{N_{\rm tot}} \eqno(23) $$
where the summation is over $i$ = MRI (active zone), DZ (dead zone) and ZZ (zombie zone). Here $N_i$ is the one-sided column-density of the $i$-th zone, $N_{\rm tot} = \sum_i N_i$ is the total one-sided column density of the disk at that radius (i.e., from the surface to the midplane), and ${\bar\alpha}_i$ is the effective viscosity parameter within the $i$-th zone (see below). Thus, for a vertically isothermal disk, $\bar\alpha$ at any radius is the {\it column-weighted mean of the active, dead and zombie effective viscosity parameters}. 

The different ${\bar\alpha}_i$ (${\bar\alpha}_{\rm MRI}$, ${\bar\alpha}_{\rm DZ}$ and ${\bar\alpha}_{\rm ZZ}$) are specified as follow. Within the MRI-active zone, we have (see Appendix B)
$$ \bar\alpha_{\rm MRI} = \frac{2}{3} \left(\frac{1}{2\langle\beta\rangle}\right) \eqno(24) $$
where $\langle\beta\rangle = \langle P_{\rm gas} \rangle/P_{\rm B}$ is the plama beta parameter averaged over the thickness of the active layer (note that we assume $B$ and hence $P_B$ are vertically constant, so the averaging is only over $P_{\rm gas}$).  In the dead and zombie zones, where the MRI is quenched, various hydrodynamical processes can still produce residual ({\it non}-MRI) stresses; numerical simulations of these suggest an associated effective $\alpha$ in the approximate range $\sim$10$^{-5}$--10$^{-3}$ (e.g., Dzyurkevich et al.\,2010; Dzyurkevich et al.\,2013 and references therein; Malygin et al.\,2017 and references therein). Additionally, without carrying out detailed hydrodynamic simulations, we have no concrete way of judging how the effective $\alpha$ in the dead and zombie zones might differ. For simplicity, therefore, we assume that the effective viscosity parameter in the dead and zombie zones is the same (i.e., ${\bar\alpha}_{\rm DZ} = {\bar\alpha}_{\rm ZZ}$), and find equilibrium solutions for the disk structure for three different fiducial values of ${\bar\alpha}_{\rm DZ}$ spanning the range implied by the numerical solutions:
$$ \bar\alpha_{\rm DZ}\, (=  \bar\alpha_{\rm ZZ}) = 10^{-5}\,\,\, {\rm or} \,\,\, 10^{-4} \,\,\, {\rm or} \,\,\, 10^{-3} \,\,. \eqno(25) $$    
Importantly, note that $\bar\alpha_{\rm DZ}$ also sets a minimum value (``floor'') on $\bar\alpha_{\rm MRI}$: when our calculations imply that a region is formally `MRI-active' (i.e., satisfies equations (10) and (11)), but nevertheless has $\bar\alpha_{\rm MRI}$ less than our adopted $\bar\alpha_{\rm DZ}$, we expect that the residual hydrodynamic stresses there will dominate over the MRI stress. We therefore declare such a region to be dead by fiat, and assign it an effective viscosity parameter equal to $\bar\alpha_{\rm DZ}$.

\section{Accretion Rates}

Within a given disk zone (MRI-active, dead or zombie), the local accretion rate (positive inwards) at any radius is $ \dot{M}_i = -2(r\Omega)^{-1} {\partial [2\pi r^2\int_{2h_i} T_{r\phi,i} \,dz]} / {\partial r}$, where $2h_i$ is the thickness of the $i$-th zone (summed over both sides of the midplane) and $T_{r\phi,i}$ the particular shear-stress operating in that zone. For a vertically isothermal disk, this reduces to (see Appendix B)
$$ \dot{M}_i = \frac{12\pi\,\mu\,m_{\rm H}}{r\Omega}\, \frac{\partial}{\partial r} \left(r^2 c_s^2\, N_i {\bar{\alpha}}_i\right) \eqno(26) $$
where $i$ = MRI, DZ or ZZ, $N_i$ is the one-sided column density of the $i$-th zone, and the values of the various ${\bar\alpha}_i$ are specified in the previous section. 

Similarly, the total accretion rate at any radius, i.e., the local sum of the rates through the different vertical layers, is $ \dot{M} \equiv \sum_i \dot{M}_i = -2(r\Omega)^{-1} {\partial [2\pi r^2\int_{-\infty}^{+\infty} T_{r\phi} \,dz]} / {\partial r}$. Now, in a real disk, the chemistry and ionization, and hence the thickness (column) of any zone and the field strength, will generally vary with radius, and there is no physically compelling reason to expect the accretion rate through any given zone (equation (26) above) to be radially or temporally constant. In steady-state, however, the {\it total} accretion rate must by definition be a constant in both time and radius (to prevent temporal changes in the local surface density). Imposing this condition on our solutions, the total accretion rate becomes (see Appendix B)    
$$ \dot{M} = \frac{3\pi\,\bar{\alpha}\,c_s^2\,\Sigma}{f_r\,\Omega} \eqno(27) $$
the standard expression for a constant accretion rate in a vertically isothermal $\alpha$-disk. $\bar\alpha$ here is given by equation (23), $\Sigma$ ($= 2\mu m_{\rm H} N_{\rm tot}$) is the total surface density summed over both sides of the midplane, and $f_r \equiv (1 - \sqrt{R_{in}/r})$. As an aside, note that it is the combination $f_r\dot{M}$ that appears in the disk structure equations (\S3), which is independent of $R_{in}$ by equation (27).

\section{Method for Determining \\The Equilibrium Solution}

At a given disk radius around a fixed stellar mass, specified input values of the accretion rate and mean viscosity parameter (\mdotin\,and \alphain) determine the pressure, temperature and density via the disk structure equations (3--7). The latter quantities, combined with the Saha equation (12), set the fractional ionization. The disk structure and ionization, together with a specified field strength $B$, then determine the resistivities (via equations 14--22) and hence the extent of the active layer via the MRI conditions (10--11). This in turn yields the output mean viscosity parameter and accretion rate (\alphaout\,and \mdotout) implied by the MRI (equations 23--25 and 27). We find self-consistent equilibrium solutions (\mdotout=\mdotin\, and \alphaout=\alphain) through a grid-based technique, as follows. 

For a specified stellar mass \mstar\, and disk radius $r$, and a desired disk accretion rate \mdotin, we determine the disk structure and ionization for a range of input $\bar\alpha$: ${\bar{\alpha}}_{\rm in} = [\adz$,\,1], spanning the gamut of plausible values given the assumed $\adz$ in the dead zone. For each of these disk structures, we then derive the height of the active layer, and thus the MRI-implied \mdotout\, and \alphaout, for a range of field strengths: $B$ = [$10^{-5}$,\,$10^3$]\,G, which covers the plausible range in stellar accretion disks. A self-consistent disk structure solution is then one for which \mdotout\,=\,\mdotin\, and \alphaout\,=\,\alphain\,. 

How exactly such a solution is determined is illustrated in Fig.\,1 for a fiducial case: $M_{\ast} = 1$\,\msun, $\dot{M}_{\rm in} = 10^{-9}$\,\msun\,yr$^{-1}$, $\adz = 10^{-4}$, at radius $r=0.02$\,AU. The $x$- and $y$-axes show \alphain\, and \alphaout\, respectively, while the overplotted greyscale contour map shows the magnetic field strength $B$ (with the white curves marking contours of constant $B$). The overlaid solid black contours are the output accretion rate \mdotout.

We see that, along the locus of equilibrium solutions (solid blue line, along which \alphaout\,=\,\alphain\, {\it and} \mdotout\,=\,\mdotin), increasing $\bar\alpha$ corresponds to increasing field strength $B$ (this is easily seen by noticing that contours of constant $\dot{M}$ are steeper than the contours of constant $B$, so $B$ changes -- increases -- as one marches up the blue solution locus with $\dot{M}$ constant). In other words, for any given $\bar\alpha$, there exists a field strength $B$ which yields an equilibrium solution with the desired \mdot, up to some upper limit in $\bar\alpha$ (corresponding to an upper limit in $B$). How do we choose a unique solution from among these infinite possibilities? We do so by invoking our assumption (see \S4.1.1) that the MRI is maximally efficient, generating the strongest possible field that still allows the MRI to operate. Thus we choose the maximum $B$, and thus the maximum $\bar\alpha$ (marked by a dashed vertical line), for which an equilibrium solution exists.

For a given \mstar\, and \mdot, we repeat the above calculations for a range of radii $r$, to determine $\bar\alpha$ as a function of radius. Our calculations begin at a disk inner edge of $R_{in}$ = $R_{\ast}$. We continue working outwards in radius until our derived equilibrium solution for $\bar\alpha$ falls to the assumed floor value $\adz$. Beyond this radius, there is no active zone any more in our model, and we simply assume a constant $\bar\alpha = \adz$. 

\begin{figure}[H] 
\centering
\includegraphics[width = \columnwidth]{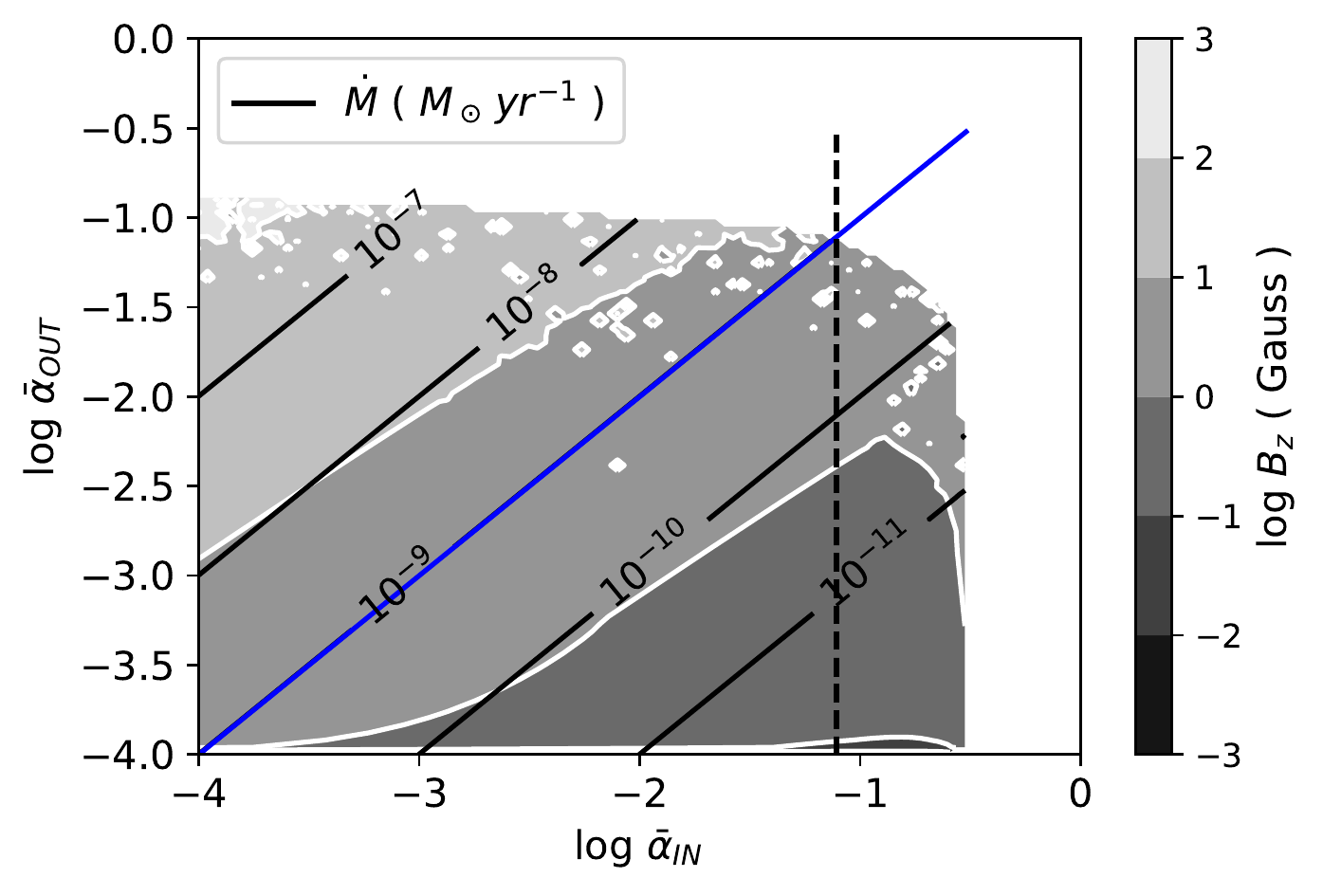} 
\caption{Output values \mdotout\, ({\it solid black lines}) and \alphaout\, ({\it y-axis}) corresponding to input vaues for \alphain\, ({\it x-axis}) and magnetic field strength $B$ ({\it greyscale} with {\it white contours}), for fixed stellar and disk parameters $M_{\ast} = 1$\,\msun, \mdotin\,= $10^{-9}$\,\msun\,yr$^{-1}$, $\adz =10^{-4}$ and radius $r=0.02$\,AU. The {\it solid blue line} indicates the locus of all solutions \alphain\,=\,\alphaout; note that this line also overlies the required accretion rate for a self-consistent solution: \mdotout\,= \mdotin\,= $10^{-9}$\,\msun\,yr$^{-1}$. The intersection of the {\it vertical dashed line} with the blue line marks the position of the final adopted equilibrium solution, corresponding to the largest value of $B$ that still allows the MRI to operate (there are no solutions with active MRI beyond this $B$, which is why this solution lies at the edge of the map). 
}
\label{fig:fig1}
\centering
\end{figure}

\section{Results}
We first present a detailed discussion of our solution for the fiducial case (\mstar\,= 1\,\msun, \mdot\, = 10$^{-9}$\,\msun\,yr$^{-1}$, $\bar\alpha_{\rm DZ} = 10^{-4}$) in \S8.1: the disk structure and location of the pressure maximum (\S\S8.1.1--8.1.4); behaviour of the accretion flow in different layers (\S8.1.5); the appearance of a viscous instability (\S8.1.6); and the validity of various assumptions (\S\S8.1.7--8.1.8). We then briefly discuss the solutions arising from variations in our fiducial parameters ($\alpha_{\rm DZ}$, \mdot\,and \mstar), pointing out any salient differences along the way (\S\S8.2--8.4). Piece-wise polynomial fits to our $\bar\alpha$ and $B$ results as a function of radius are provided in Appendix C for all cases.

\subsection{Fiducial Model: $\boldsymbol M_{\ast} \boldsymbol = \boldsymbol 1\, {\rm {\mathbf M}}_{\boldsymbol \odot}$,\, ${\boldsymbol \dot{\boldsymbol M}} \boldsymbol = \boldsymbol{10}^{\boldsymbol{-9}}\, {\rm {\mathbf M}}_{\boldsymbol \odot}\, {\rm {\mathbf{yr}}}^{\mathbf{-1}}$,\, $ \boldsymbol{\adz} \,\mathbf{ = 10^{-4}}$}
For this \mstar\,= 1\,\msun\,case, the stellar radius and effective temperature are $R_{\ast}$ = 2.33\,R$_{\odot}$ and $T_{\rm eff} = 4350$\,K respectively (using the evolutionary models of Baraffe et al.\,(1998)\footnote{Specifically, the iso.3 models with mixing length = 1.9 $\times$ pressure scale-height, as required to fit the sun.}, for a fiducial age of 1\,Myr). In this and all following solutions, the disk inner radius is situated at the stellar surface (i.e., $R_{in} = R_{\ast}$), and our MRI calculations stop at the radius where the effective viscosity parameter $\bar\alpha$ falls to the floor value $\adz$ (i.e., where the pressure maximum forms). Beyond this radius, the disk structure is calculated assuming that the viscosity parameter remains constant at $\bar\alpha = \adz$.  

\subsubsection{Dominant Resistivities}
Fig.\,2 shows the relative importance of the three resistivities -- $\eta_O$, $|\eta_H|$ and $\eta_A$ -- as a function of location in the inner disk. Ambipolar diffusion dominates over Hall and Ohmic in the surface layers, while Hall resistivity dominates everywhere else at these radii. Ohmic resistivity is not dominant anywhere, though it is larger than ambipolar closer to the midplane at radii $\gtrsim$0.09\,AU. This distribution of resistivities is also depicted more quantitatively in Fig.\,3, where we plot $\eta_O$, $|\eta_H|$ and $\eta_A$ as functions of scale-height at various radii. 

\begin{figure} 
\centering 
\includegraphics[width = \columnwidth, trim={2cm 2cm 2cm 2cm}, clip]{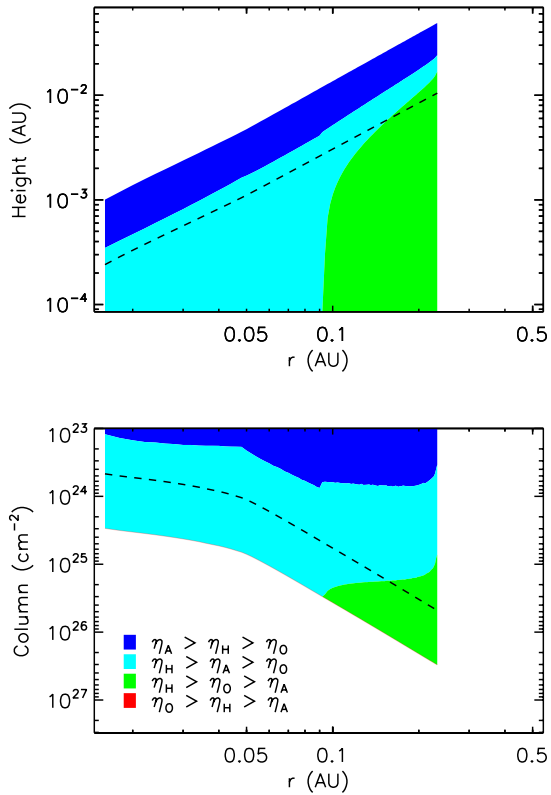} 
\caption{Relative importance of the Ohmic ($\eta_O$), Hall ($|\eta_H|$) and ambipolar ($\eta_A$) resistivities as a function of location in the inner disk, for our fiducial disk model. The {\it top} panel shows vertical location in units of actual height above the midplane; the {\it bottom} panel shows vertical location in units of column density measured from the disk surface. The {\it dashed line} in both panels indicate one pressure scale-height. Note that there is {\it no} region where Ohmic resistivity dominates over both Hall and ambipolar resistivities. See \S8.1.1.
}
\label{fig:fig2}
\centering
\end{figure}

\begin{figure*} 
\centering
\includegraphics[width = 0.8\textwidth, trim={0cm 1cm 0cm 2cm}, clip]{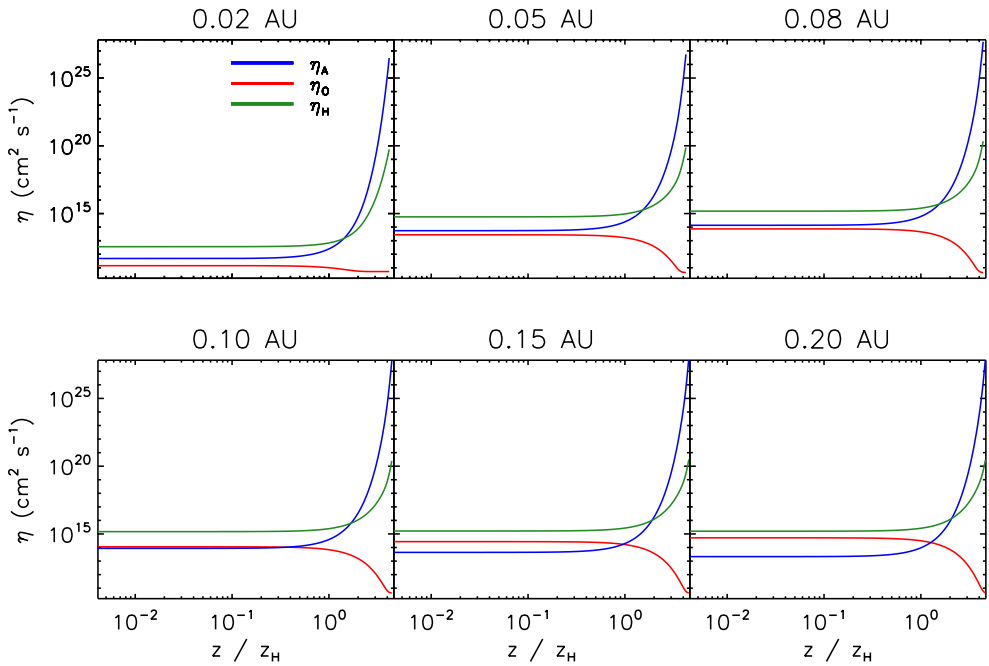} 
\caption{Ambipolar ($\eta_A$), Ohmic ($\eta_O$) and Hall ($|\eta_H|$) resistivities as a function of height above the midplane (in units of the local scale-height $z_H$), at various radii for our fiducial disk model. See \S8.1.1. 
}
\label{fig:fig3}
\centering
\end{figure*}

\begin{figure*}
\centering
\includegraphics[width = 0.8\textwidth, trim={1.7cm 1cm 0cm 2cm}, clip]{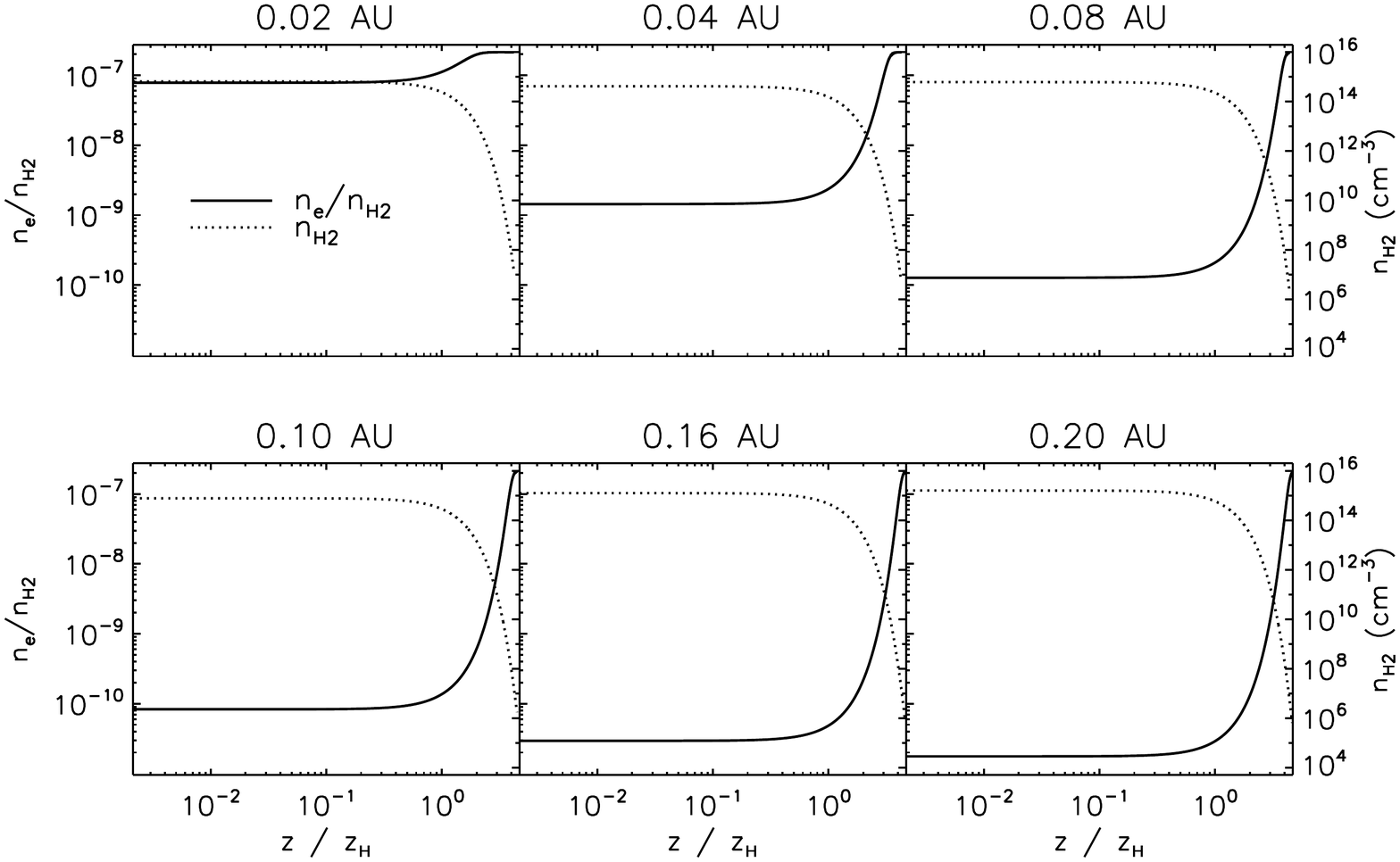} 
\caption{Fractional ionization ($n_e/n_{H_2}$) and molecular hydrogen density ($n_{H_2}$) as a function of height above the midplane (in units of the local scale-height $z_H$), at various radii for our fiducial disk model. See \S8.1.1. 
}
\label{fig:fig4}
\centering
\end{figure*}

The physics underlying the above behaviour can be extracted from Fig.\,4, where we plot the fractional ionization ($x_e \equiv n_e/n_{\rm{H_2}}$) and number density of neutral molecular hydrogen ($n_{\rm{H_2}}$) as functions of scale-height at different radii. Recall that $n_{\rm{H_2}} \approx n_n$ (number density of neutrals) $\approx n_{\rm{tot}}$ (total number density), given the overwhelming relative abundance of hydrogen and the very low ionization fractions in general (since potassium, with total abundance $x_{\rm K} \equiv n_{\rm K} / n_{\rm{H_2}} \sim 2\times10^{-7}$, is the only ionized species here). Combining this with our results from \S4.2 for one singly-ionised species, we get $x_e \approx n_e/n_n \propto \sqrt{{\mathcal{S}}_{\rm K}(T)/n_n}$ when $n_n$ is sufficiently high (with the subscript `K' on $\mathcal{S}$ denoting the specific case of potassium). For the same conditions, and combining the latter relationship with results from \S4.3, we also have: $\eta_O \propto 1/x_e \propto \sqrt{n_n / {\mathcal{S}}_{\rm K}(T)}$; $|\eta_H| \propto B/(x_e n_n) \propto B/\sqrt{{\mathcal{S}}_{\rm K}(T)\, n_n}$; and $\eta_A \propto B^2/(x_e n_n^2) \propto B^2/\sqrt{{\mathcal{S}}_{\rm K}(T)\, n_n^3}$. Thus, at any fixed radius in Fig.\,4 (with ${\mathcal{S}}_{\rm K}(T)$ constant since vertically isothermal), the ionization fraction $x_e$ increases rapidly above a scale-height $z_H$ as hydrostatic equilibrium causes $n_{\rm tot}$$\sim$$n_{\rm{H_2}}$$\sim$$n_n$ to drop, with all the potassium ionized ($x_e \rightarrow x_{\rm K}$ as $n_{\rm H_2} \rightarrow 0$; see \S4.2) by a few$\times$$z_H$. Consequently, at a given radius in Fig.\,3, $\eta_O$ decreases with height above $\sim$$z_H$, while $|\eta_H|$ {\it increases} with height and $\eta_A$ increases even faster (note that the field strength $B$ is vertically constant at fixed radius in our calculations).

In summary, though a large fraction of the alkali atoms are ionized near the disk surface, the total density here is too low to collisionally couple either ions or electrons to the bulk fluid of neutrals, and hence ambipolar diffusion dominates; closer to the midplane, the density increases sufficiently to tie ions (but not electrons) to the neutrals, making Hall resistivity dominant, but the density is still too low for Ohmic resistivity to compete with either Hall or ambipolar diffusion. Beyond $\sim$0.09\,AU, the rising density and falling temperature are sufficient (combined with a declining $B$; see \S8.1.2 below) for Ohmic resistivity to exceed ambipolar diffusion near the midplane, but still not enough to allow Ohmic resistivity to exceed Hall diffusion here.

\subsubsection{Active, Dead and Zombie Zones}
Fig.\,5 shows our derived locations of the MRI-active zone, the dead zone (where Ohmic resistivity shuts of the MRI: $\Lambda < 1$) and the zombie zone (where ambipolar diffusion cuts off the MRI: $\beta < \beta_{\rm min}$). We emphasize that the effects of Hall resistivity on the MRI are not accounted for here: we only consider the effects of Ohmic and ambipolar diffusion, even in regions where $|\eta_H|$ dominates over $\eta_O$ and $\eta_A$. Nevertheless, we also overplot the Hall zone, where the Hall Elsasser number $\chi < 1$: this is where the Hall influence on the MRI is significant (see further below), and {\it should} be accounted for in future work. Fig.\,6 shows the associated field strength $B$ as a function of radius, while Fig.\,7 shows the midplane radial behaviour of the ionization fraction, plasma $\beta$ parameter and Ohmic Elsasser number. 

\begin{figure}
\centering
\includegraphics[width = \columnwidth, trim={2cm 2cm 2cm 2cm}, clip]{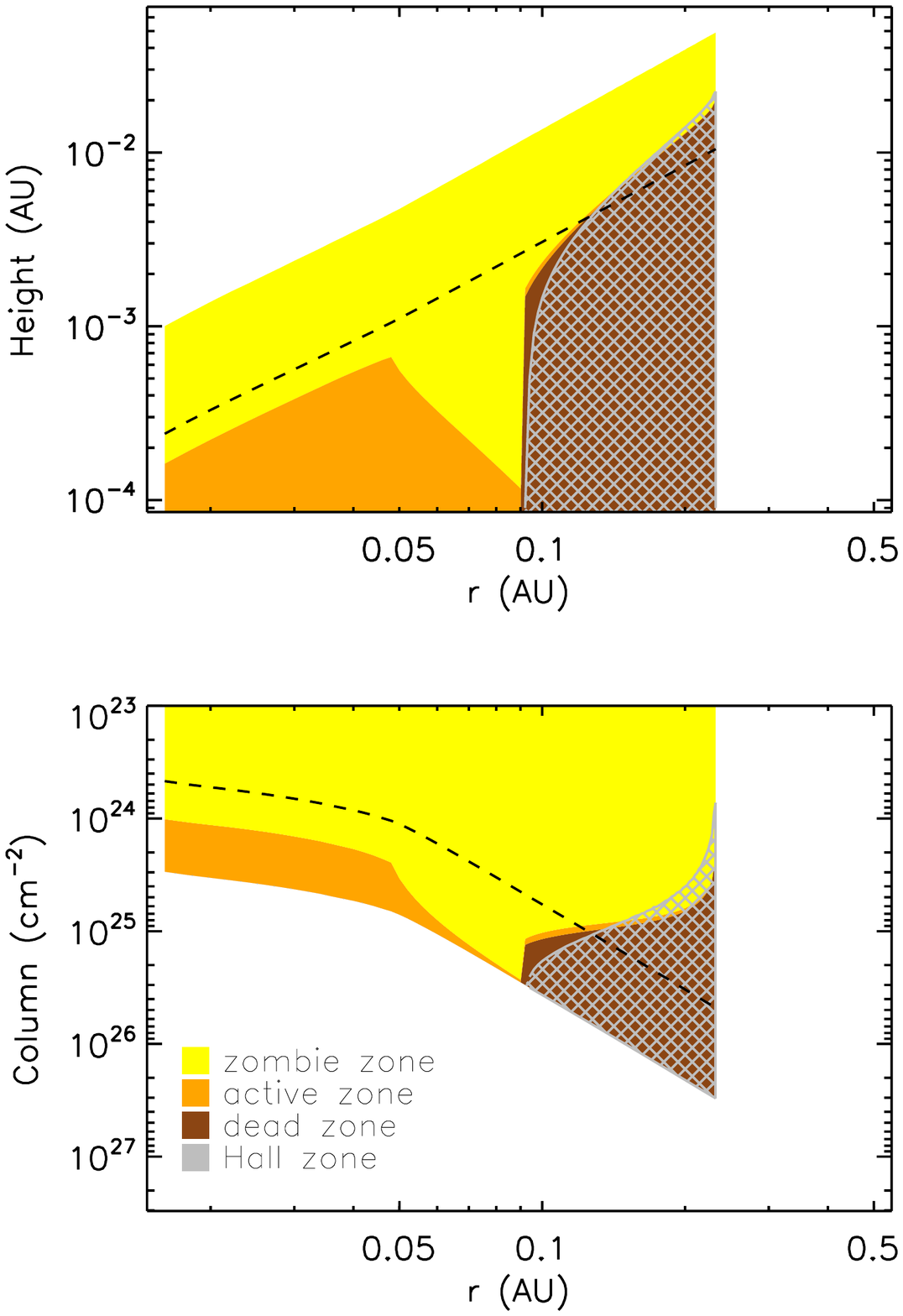} 
\caption{The various MRI zones in our fiducial disk model as a function of disk location. The {\it top} panel shows vertical location in units of height above midplane; the {\it bottom} panel shows it in units of column density. The {\it dashed line} indicate the disk scale-height. {\it Orange} denotes the MRI-active zone (i.e., where $\Lambda_O > 1$ and $\beta > \beta_{\rm min}$); {\it brown} denotes the dead zone (where $\Lambda_O < 1$); {\it yellow} denotes the zombie zone (where $\beta < \beta_{\rm min}$); and the {\it grey hashed} region denotes the Hall zone (where $\chi < 1$). Note that, beyond 0.09\,AU, the active zone rises above the midplane and continues as a thin layer sandwiched between the dead and zombie zones, till it is finally quenched totally at the outer edge of our solution at $\sim$0.25\,AU. See \S8.1.2.
}
\label{fig:fig5}
\centering
\end{figure}

\begin{figure}
\centering
\includegraphics[width = \columnwidth, trim={2cm 1cm 2cm 2cm}, clip]{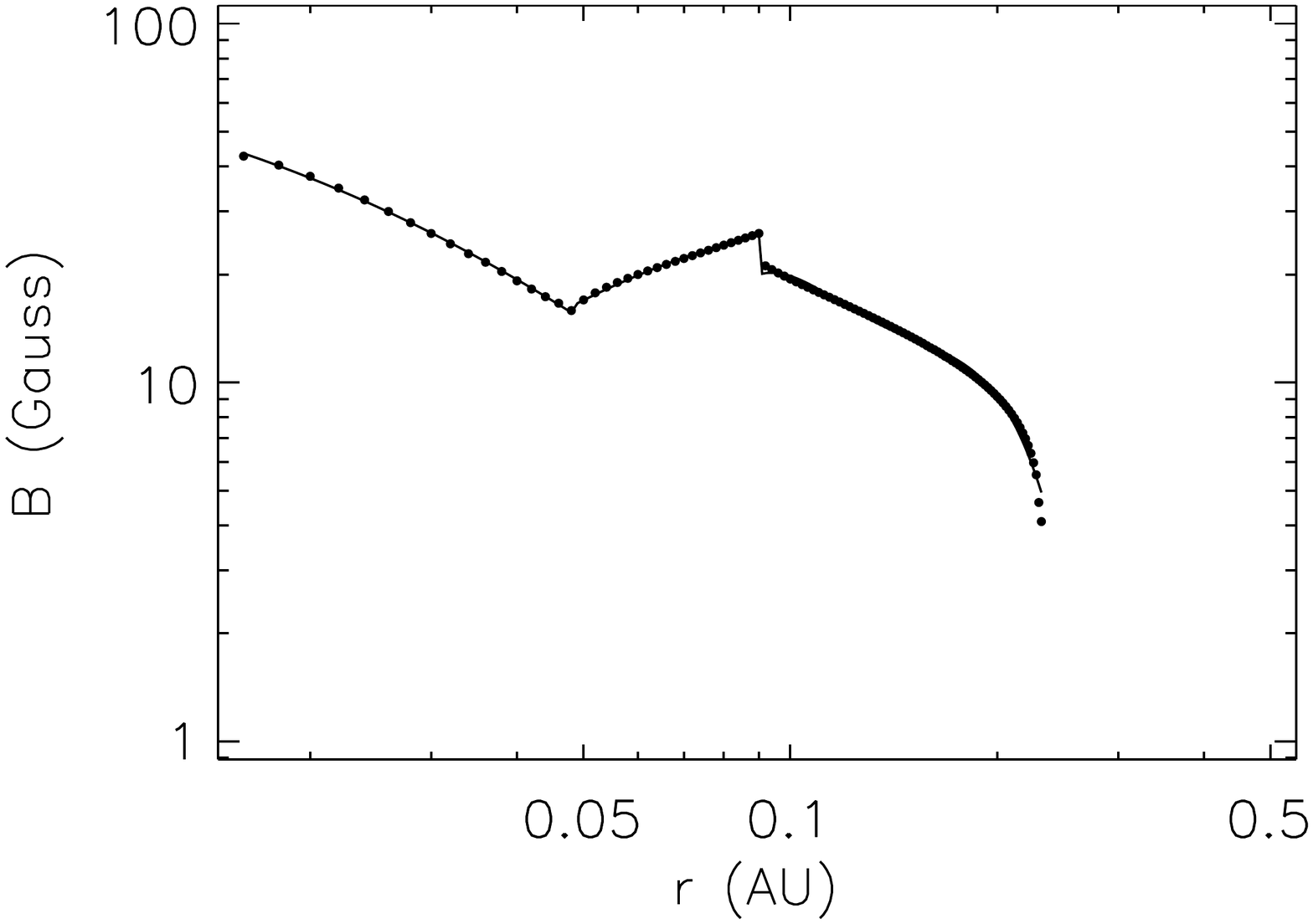} 
\caption{Magnetic field strength $B$ as a function of radius, for our fiducial disk model. {\it Filled circles} are our model results, and the overplotted {\it solid line} is a combined piece-wise polynomial fit to these results. Note that the jump at 0.09\,AU is not a physical discontinuity, but a result of our finite grid radial sampling. See \S8.1.2, and Table 1 in Appendix C for the polynomial fit parameters. 
}
\label{fig:fig6}
\centering
\end{figure}

\begin{figure} 
\centering
\includegraphics[width = \columnwidth, trim={3cm 0cm 0cm 3cm}, clip]{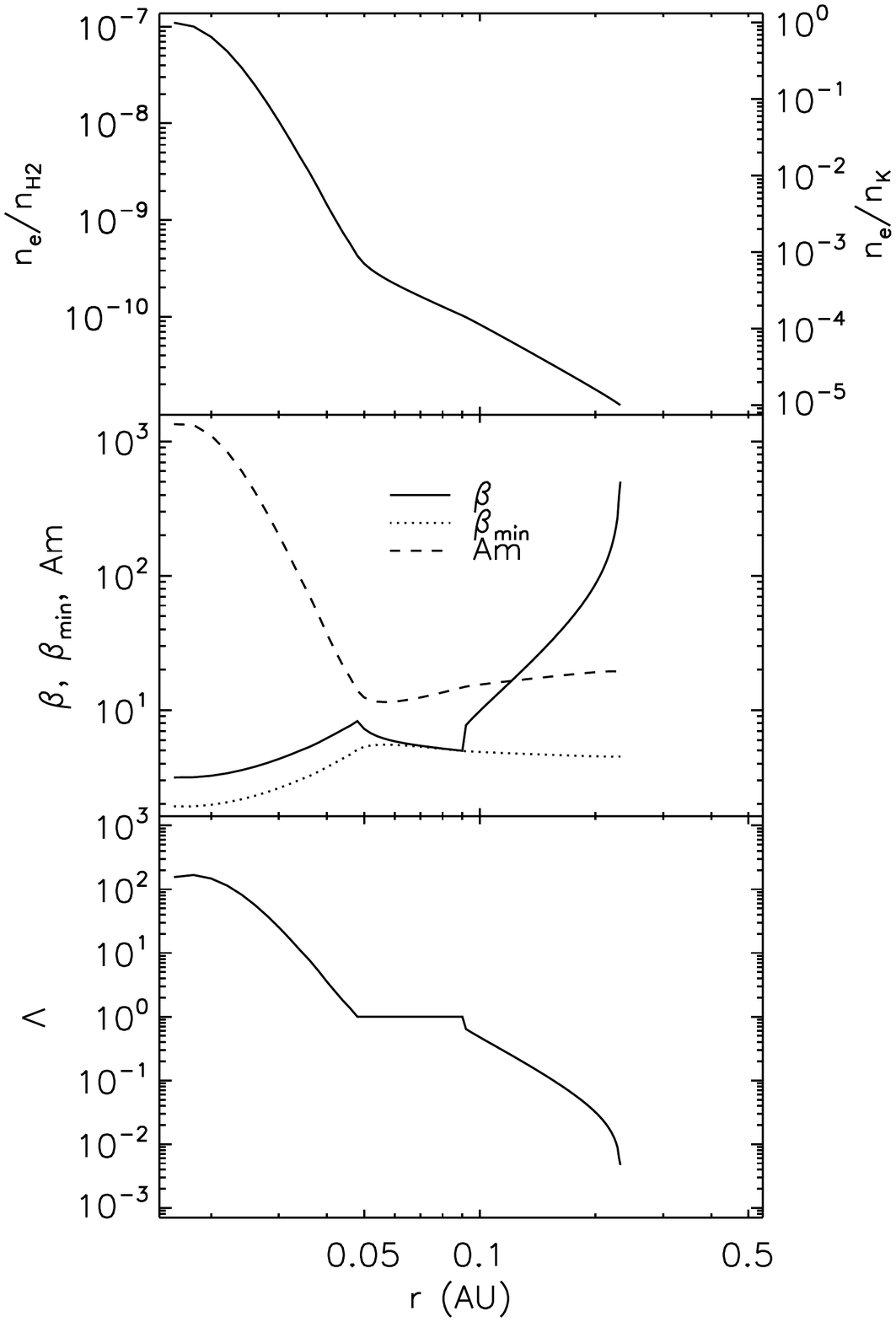} 
\caption{Various MRI related quantities in the midplane, plotted as a function of radius, for our fiducial disk model. {\it Top}: Fractional abundance of electrons, expressed relative to both the number density of hydrogen molecules ($n_e/n_{\rm H_2}$; left axis) and number density of potassium nuclei ($n_e/n_{\rm K}$; right axis). {\it Middle}: Plasma $\beta$ parameter; minimum value $\beta_{\rm min}$ required for active MRI; and ambipolar Elsasser number $Am$. {\it Bottom}: Ohmic Elsasser number $\Lambda$. See \S8.1.2.
}
\label{fig:fig7}
\centering
\end{figure}

We see from Fig.\,5 that, from the inner edge of the disk to $\sim$0.05\,AU, the active zone extends from the midplane up to a roughly constant fraction of the scale-height, bounded above by a zombie zone, while from 0.05 to 0.09\,AU, the active zone narrows considerably, with the zombie zone pushing down increasingly towards the midplane. At 0.09\,AU, a dead zone rises up sharply from the midplane; from here on, the active zone is confined to a very thin and continuously narrowing layer sandwiched between the zombie zone above and dead zone below, until the MRI is completely choked off at $\sim$0.25\,AU, at which point our calculations stop (beyond this radius, we assume a constant $\bar\alpha = \adz$, leading to the formation of a pressure maximum at this radius; see following sections). 

These trends in the active, dead and zombie zones can be understood as follows. At a fixed height (in scale-height units), $\eta_O$ increases while the field strength $B$ declines, going radially outwards from the inner edge to $\sim$0.05\,AU (Figs.\,3 and 6). The combined effect is to decrease the Ohmic Elsasser number $\Lambda$; however, it still remains high enough to allow the active zone to straddle the midplane (e.g., see midplane $\Lambda$ in Fig.\,7, bottom panel). The weakening of the field over this radial span instead serves to keep the plasma $\beta$ sufficiently large, $\beta > \beta_{\rm min}$, so that ambipolar diffusion does not cut off the MRI too close to the midplane and drive \mdot\,below the desired steady-state value (e.g., see midplane $\beta$ and $\beta_{\rm min}$ in Fig.\,7, middle panel). By 0.05\,AU, however, the midplane $\Lambda$ has fallen to unity (Fig.\,7). Now the field $B$ has two choices: either continue to weaken, making $\Lambda < 1$ at the midplane (i.e., creating a dead zone there) and thus driving the active zone upwards; or strengthen instead, thereby keeping the MRI alive around the midplane, but suppressing $\beta$ and thus allowing the zombie zone to descend towards the midplane. Since {\it we assume that the MRI is maximally efficient}, i.e., generates the strongest possible field that still allows the MRI to survive, it is the latter solution that is chosen (Fig.\,6), yielding the observed active and zombie zone shapes in Fig.\,5 over 0.05--0.09\,AU. The quantitative increase in $B$ here (and thus change in $\bar\alpha$ and hence in surface density $\Sigma$; see following sections) is such that the \mdot\,(by equation (27)) remains at the required value.    

By 0.09\,AU, however, the zombie zone has descended all the way to the midplane (i.e., $\beta = \beta_{\rm min}$ at the midplane; Fig.\,7). Now the field has no choice but to weaken again (Fig.\,6), in order to maintain any active zone at all. As $B$ decreases, a dead zone develops at the midplane, the zombie zone lower boundary is impelled upwards, and a thin active layer forms between the dead and zombie regions (Fig.\,5). This situation cannot continue indefinitely, though, since the dead zone upper boundary keeps rising with radius (as $\eta_O$ continues to grow; Fig.\,3). Finally, at $\sim$0.25\,AU, the MRI-active zone is squeezed shut completely, as the upper edge of the dead zone meets the lower edge of the zombie zone. No further changes in $B$ can alter this, since the dead region would expand upwards for smaller $B$, and the zombie region would expand downwards for larger $B$. Thus, this is the radius where the effective viscosity parameter $\bar\alpha$ falls to its minimum value $\adz$ (since the disk is now fully MRI-dead vertically), and hence where the midplane pressure maximum forms.  

The above result raises an important point missed in most earlier work: the midplane gas pressure does not reach its maximum {\it at} the inner edge of the dead zone (i.e., at $\sim$0.09\,AU in this example), but rather somewhat radially {\it beyond} this edge (at $\sim$0.25\,AU here). In other words, the midplane gas pressure achieves its maximum value {\it within} the dead zone. This is a straightforward consequence of two facts: {\it (1)} the midplane pressure in the Shakura-Sunyaev model is not a function of simply the local midplane value of $\alpha$, but rather its {\it vertically averaged} value $\bar\alpha$ (see equation (5) and discussion in \S3); and {\it (2)} the active zone does not abruptly come to an end when a dead zone appears in the midplane, but instead climbs above the dead zone and continues outwards for some distance, thereby pushing the location of minimum $\bar\alpha$ (and so maximum midplane pressure; see Figs.\,8 and 9 further below) beyond the dead zone inner boundary. As such, pebbles drifting inwards along the midplane will become trapped within the dead zone itself, where conditions are less turbulent than at the active/dead zone interface further in, with potentially important implications for planet formation.       

In the context of the location of the pressure maximum, we now discuss the potential importance of some effects ignored in our simplified treatment here.

{\it Other Ionized Elements}: We have only treated potassium here, with the justification that -- as the element with the lowest ionization potential (4.34\,eV) among the important species in the inner disk (\S4.2) -- it remains ionized furthest out, and is thus most relevant to the location of the pressure maximum. Nevertheless, other elements with slightly higher ionization potentials may plausibly matter because their abundances are much higher. To check this, we carried out calculations for our fiducial model with sodium instead, which has an ionization potential (5.14\,eV) only slightly larger than potassium's but is $\sim$16 times more abundant. We find (not plotted) that, while the greater abundance of Na yields a significantly higher ionization fraction in regions where our original simulations showed K to already be highly ionized (in surface layers, and near the midplane close to the disk inner edge), the pressure maximum occurs slightly inwards of its position with K; i.e., the latter is still set by the difference in ionization potentials. As such, while the precise shape of the active, dead and zombie zones will vary somewhat when other atomic species are included with K, we do not expect the position of the pressure maximum to shift substantially. Implementing more complex chemical networks (with additional atomic and molecular species and grains) {\it will} be important for increasing the recombination rate and ensuring that we are in the strongly-coupled limit (see \S8.1.8 further below); we shall tackle this in an upcoming paper.    

{\it Importance of Dust}: Dust grains affect both the opacity of the disk and the efficiency of the MRI. While our calculations are dust-free -- in the sense that grain effects on the MRI are ignored -- we have nonetheless assumed a constant opacity of 10\,cm$^2$\,g$^{-1}$, which is a reasonable value for the warm inner regions of dusty protoplanetary disks \citep[see][]{hu17}. Concurrently, an a posteriori calculation of the opacities in our disk solution, using detailed opacity tables including grains, yields values very close to our assumed contant in all regions of interest except very close to the disk inner edge (see \S8.1.7 further below). As such, grains are effectively included in our opacities, and treating them more precisely via opacity tables should not alter our results appreciably. 

Inclusion of dust is very likely to be important for the MRI, however. Grains can drastically suppress the MRI, by soaking up electrons and thereby reducing the amount of negative charge tied to the magnetic field (since all but the very smallest grains (see below) are collisionally decoupled from the field themselves; e.g., Perez-Becker \& Chiang 2011a; Bai 2013; Mohanty et al.\,2013). Enhanced recombination on the charged grain surfaces also removes positive charge from the gas, further hampering the MRI. Lastly, MRI damping is exacerbated by the incorporation of the alkali atoms (which are the primary charge suppliers) into grains, and their adsorption onto grain surfaces; we have currently ignored this effect, which can deplete metal abundances by an order of magnitude or more (e.g., Keith \& Wardle 2014; Jenkins 2009). Concurrently, as Fig.\,9 shows, the disk temperatures in our solution are well below the dust sublimation temperature of $\sim$1500\,K (at the extant densities) at radii $\gtrsim$0.05\,AU; as such, the pressure maximum as well as the dead zone inner boundary in our current solution sit squarely within the radial range where dust is thermodynamically allowed. Moreover, while the pressure maximum traps relatively large grains (``pebbles'') -- the whole reason for invoking it for planet formation -- smaller ones are increasingly well-coupled to the gas and can thus flow through the trap; it is moreover these small grains that have the greatest impact on the MRI (because of their large collective surface area for electron adsorption). Therefore we expect small grains to exist in our solution space, damping the MRI to some extent and moving both the dead zone inner boundary and the pressure maximum radially inwards of our currently predicted locations. 

The {\it magnitude} of this effect depends, on the one hand, on the relative abundance of grains versus electrons. For dust grains with number density $n_d$ and a fixed radius $a$, the grain abundance $x_d \equiv n_d / n_{\rm H_2}$ may be expressed as $x_d = (3 R \,\mu m_{\rm H}) / (4\pi \rho_{\rm gr}\, a^3) $, where $R$ is the dust-to-gas ratio by mass and $\rho_{\rm gr} \approx 3$\,g\,cm$^{-3}$ is the density of a single grain. For a standard ISM value of $R = 10^{-2}$, very small grains of size $a = 0.1$\,$\mu$m then imply $x_d \approx$ 3$\times$10$^{-12}$: $\sim$10--30 times smaller than the ionization fraction $x_e \sim$ few$\times$10$^{-11}$--10$^{-10}$ that we infer over most of the active zone (both close to the midplane at radii $\gtrsim$0.05\,AU, and higher up, at $\sim$1--2\,$z_H$, once a dead zone forms in the midplane; see Fig.\,4). Such grains will therefore put a significant dent in the number density of free electrons, and thus affect the MRI activity, if the adsorbed negative charge per grain is of order $-$10. Slightly smaller grains, $a = 0.03$\,$\mu$m, imply $x_d = 10^{-10} \gtrsim x_e$, and so will have a severe impact on the MRI even with $\lesssim$1 electron adsorbed per grain on average. Such grain sizes and charging are not unrealistic in disks (e.g., Perez-Becker \& Chiang 2011a). We note that this calculation assumes that {\it all} the dust is sequestered in grains of a single size; a more realistic grain size distribution will reduce the effective dust-to-gas ratio in small grains, and thus decrease $x_d$. This is plausibly a small correction though, since the grain number density is likely to be dominated by the smallest particles (e.g., standard MRN distribution: $n_a \propto a^{-3.5}$; but see Birnstiel et al.\,2011). 

Furthermore, we have compared grain abundances here to the electron abundance derived assuming no depletion of potassium in the gas phase. If a sizeable fraction of K is sequestered in dust instead (both by inclusion in molecules that make up dust grains, and by the adsorption of neutral K atoms onto grains), then the $x_e$ due to thermal ionization will be much smaller than we have inferred to start with, further reducing the MRI (though this effect will be tempered somewhat by ion and thermionic emissions, whereby neutral K collisions with grains {\it produce} free K$^+$ ions and/or electrons; see Desch \& Turner 2015).     

On the other hand, MRI-damping by grains is mitigated to the extent that they are tied to the field (and thus act like ions), instead of being knocked off by collisions with neutrals. The Hall parameter $\beta_j$ (equation (17)) is a measure of the strength of the field-coupling for any species $j$; noting that grains are much more massive than neutral gas particles, the relative coupling strength for grains versus ions is thus: $\beta_{\rm gr} / \beta_i = (|Z_{\rm gr}| \langle\sigma v\rangle_{i}) / (|Z_i| \langle\sigma v\rangle_{\rm gr})$. The rate coefficient for ion-neutral collisions $\langle\sigma v\rangle_i$ is given in equation (22), while that for grain-neutral collisions is (Wardle \& Ng 1999) $\langle\sigma v\rangle_{\rm gr} = \pi\,a^2 \sqrt{(128 k_B T_n)/(9\pi m_n)}$ cm$^3$\,s$^{-1}$, where $T_n$ is the neutral temperature, which we assume equals the gas temperature $T$, and $|Z_i| = 1$ in our case. We thus get: $\beta_{\rm gr} / \beta_i \approx 4\times 10^{-4}\, |Z_{\rm gr}|\,(T/10^3\,{\rm K})^{-1/2}\,(a / 0.1\,\mu{\rm m})^{-2}$. Hence, at the $T$ $\sim$ 1000--2000\,K in our disk solution (Fig.\,9), the 0.03--0.1\,$\mu$m grains considered above will be far more decoupled from the field than the ions, even for grain charges $|Z_{\rm gr}| \sim 10$. We conclude that the net effect of abundant very small grains will be to significantly suppress the MRI, and thus shift the pressure maximum inwards of where we currently find it to be. 

{\it Relevance of X-rays}: Here we have only considered thermal ionization, and ignored photoinization by stellar X-rays. We estimate the effect of the latter as follows. Igea \& Glassgold (1999; hereafter IG99) have calculated the ionization rate $\zeta_X$, due to X-rays with photon energies of a few keV and ignoring grain effects, as a function of column density. They find that, while $\zeta_X \propto L_X/r^2$ (where $L_X$ is the stellar X-ray luminosity and $r$ the radial distance from the star), as expected, it is also ``universal'', in the sense that $\zeta_X$ plotted as a function of (vertical) column density is independent of the precise density structure of the disk. Moreover, in the absence of grains, the ionization fraction is given simply by $x_e = \sqrt{\zeta_X / (n_{\rm H_2} k_{ie})}$, where $k_{ie}$ is the recombination rate coefficient for ion-electron recombinations for the relevant dominant ions (e.g., see expressions for $x_e$ in various limiting cases derived by Perez-Becker \& Chiang 2011a). We use these facts to scale directly from IG99's results (correcting for the fact that they supply column densities in terms of hydrogen nuclei while we use hydrogen molecules instead). 

The column density in our active region close to the midplane, at a mean radial distance $r \sim 0.05$\,AU, is $N_{\rm H_2} \sim$ 3$\times$10$^{24}$\,cm$^{-2}$, while in the active region above the dead zone, at a mean $r \sim 0.1$\,AU, it is $N_{\rm H_2} \sim 10^{25}$\,cm$^{-2}$ (see Fig.\,5). At the same active region locations, we also have $x_e \sim$ 3$\times$10$^{-10}$ and 10$^{-10}$ respectively due to thermal ionization, and  $n_{\rm H_2} \sim 10^{14}$\,cm$^{-3}$ (Fig.\,4). Concurrently, at 1\,AU, for $L_X = 10^{29}$\,erg\,s$^{-1}$ and photon energies of 5\,keV, IG99's Fig.\,5 implies $\zeta_X \sim$ 3$\times$10$^{-17}$\,s$^{-1}$ and $3\times$10$^{-18}$\,s$^{-1}$ at $N_{\rm H_2} \sim$ 3$\times$10$^{24}$\,cm$^{-2}$ and 10$^{25}$\,cm$^{-2}$ respectively (results for 3\,keV and 8\,keV photons are only marginally different). Assuming as IG99 do that molecular ions, specifically HCO$^+$, are dominant, and thus using a dissociative recombination rate coefficient of $k_{{\rm HCO}^+,e} =$ 2.4$\times$$10^{-7} / (T / 300\,{\rm K})^{0.69}$\,cm$^3$\,s$^{-1}$ (Woodall et al.\,2007; Perez-Becker \& Chiang 2011a), and scaling to our radii of interest, where $T \sim 10^3$\,K, we then find X-ray ionization implies: $x_e \sim$ 3$\times$10$^{-11}$ in our active region at 0.05\,AU, and $x_e \sim$ 5$\times$10$^{-12}$ in the active region at 0.1\,AU; these are roughly an order of magnitude smaller than $x_e$ from thermal ionization cited above. We note that Bai \& Goodman (2009) provide an analytic fit to IG99's $\zeta_X$ curves; we get the same results using their fitting formula. 

However, while X-rays first produce molecular ions, charge transfer to metals is so rapid that it is metal ions that comprise the dominant ionic species, {\it if} the metal abundance is high (as it is in our non-depleted grainless case)\footnote{Note that it is the metal {\it abundance}, and not the ionization potential, that is the controlling factor here (because the keV X-ray energies greatly exceed the electron binding energies in the metals). As such, the relevant metal here is magnesium (with attendant ions Mg$^+$), and not potassium as in our thermal ionization calculations, since Mg is far more abundant than K: $x_{\rm Mg}/x_{\rm K} \approx 4\times10^2$ (e.g., Keith \& Wardle 2014).} (e.g., Fujii et al.\,2011; Keith \& Wardle 2014). In that case, in the absence of grains, it is the metal ion (M$^+$)-electron recombination rate coefficient, $k_{{\rm M}^{+},e} = 2.8$$\times$10$^{-12}/(T/300\,{\rm K})^{0.86}$\,cm$^3$\,s$^{-1}$ (see \S4.2), that must be used to calculate the X-ray-driven $x_e$. At the relevant temperatures $T \sim 10^3$\,K, we see that $k_{{\rm M}^{+},e} \approx 10^{-5}$$\times$$k_{{\rm HCO}^+,e}$ (i.e., metal ions recombine vastly slower than molecular ones); consequently, the $x_e$ due to X-rays in our metal-abundant active regions will be more than 2 orders of magnitude higher than inferred above using HCO$+$, completely swamping the $x_e$ from thermal ionization. Of course, metals may be severely depleted when grains are present; however, this will decrease the thermal ionization fraction too, so we expect X-rays to remain highly competitive with thermal ionization in activating the MRI in the inner disk. 

Note however that, once a dead zone forms in the midplane, the midplane column density quickly exceeds that in the overlying active zone by more than an order of magnitude (Fig.\,5). IG99's results then imply an X-ray induced miplane $x_e$ at least 3 orders of magnitude smaller than that deduced from X-rays in the active zone, and much smaller than the midplane $x_e$ from thermal ionization. As such, X-ray ionization will not change our result that a dead zone eventually forms in the midplane and the active zone climbs up above it. However, by enhancing the ionization in the overlying active zone, and thus increasing the effective viscosity parameter $\bar\alpha$, X-rays will alter the location of the pressure maximum. These effects will be quantified in our upcoming work including X-rays (Jankovic et al., in prep.). 
 
Finally, we point out that, in past work, X-ray ionization has widely been stated to be unimportant in the inner disk, with thermal ionization of alkali metals being the dominant process instead. Why then do we find X-rays to be at least as important as thermal collisions? The reason is that previous studies have drawn their conclusions based on the assumption of a surface density distribution that monotonically increases radially inwards (e.g., the Minimum Mass Solar Nebula (MMSN); Igea \& Glassgold 1999). In that case, the surface density in the innermost regions is indeed too high for X-rays to penetrate to any significant depth in the disk. Here, however, we examine a posteriori the degree of X-ray ionization in our steady-state disk solution\footnote{Where the solution has been derived using the standard assumption of thermal ionization alone.}, in which the surface density $\Sigma$ is considerably {\it lower} inwards of the pressure maximum (see Fig.\,9 further below). Such a turnover in the radial $\Sigma$ profile is in fact a generic feature of steady-state models that invoke a radially changing $\alpha$-viscosity to produce a pressure maximum in the disk (because the higher viscosity inwards of the pressure maximum requires a lower $\Sigma$ to drive a given \mdot, by equation (27); e.g., see solutions by Kretke \& Lin 2007, 2010). The severely depressed surface density in the inner disk then allows much greater X-ray penetration and ionization. Therefore, if protoplanetary disks {\it start} with a standard monotonic $\Sigma(r)$ profile, we conjecture they will evolve as follows: initially, thermal ionization will dominate in the inner disk, driving it towards the steady-state solution we find, and thereby reducing the surface density in these regions; once the $\Sigma$ here falls sufficiently (i.e., column densities drop to $\sim$10$^{25}$--10$^{24}$\,cm$^{-2}$), X-ray ionization will begin to complement, and perhaps overtake, the ionization due to thermal collisions, enhancing the MRI and thus the effective $\alpha$. As argued above, we do not expect this to alter the qualitative features of our disk solution, but do expect the precise locations of the dead zone inner boundary and the pressure maximum to change from our current results.

{\it Hall Effect}: Here we have neglected the effects of Hall resistivity on the MRI. This does not prevent us though from calculating the Hall Elsasser number $\chi$ everywhere within our solution disk. The results are shown in Fig.\,5, where the cross-hatched region denotes the Hall zone, i.e., where $\chi < 1$, and hence where the Hall effect {\it is} important. We see that the Hall zone essentially overlaps with the Ohmic dead zone, and also extends into the overlying active zone at radii $\gtrsim$0.15\,AU. Thus, if the net vertical field is anti-aligned with the disk spin-axis, we do not expect our solution to change very much: in this field configuration, the Hall effect damps magnetically-driven radial angular momentum transport, so the active zone will end at (and the pressure maximum will thus be located at) $\sim$0.15\,AU instead of $\sim$0.25\,AU, while the dead zone (where Ohmic resistivity already quenches the MRI) will remain dead. If the net vertical field is aligned with the spin-axis, on the other hand, the HSI can activate magnetically-driven radial transport within the entire dead zone. 

This suggests an explanation for the fact that close-in Earths / super-Earths are {\it not} seen around $\sim$50\% of stars. In general, one expects a net vertical background magnetic field threading the disk, due to either the stellar field or an external interstellar field. Morever, one expects the alignment / anti-alignment of this field to be random relative to the disk angular momentum vector, with a roughly equal distribution of either geometry. Thus, in roughly half the systems, alignment between the field and disk spin axis should lead to the HSI activating the dead zone, which will remove the pressure barrier and thus suppress the formation of close-in small planets; in the other half of systems, anti-alignment will damp the HSI, allow the pressure barier to form, and thus promote the formation of such planets. 

We shall address this mechanism quantitatively in future work; we only note here that our result -- that $\chi < 1$ within the Ohmic dead zone -- is in qualitative agreement with that of Bai 2017, who finds that the Hall effect is critical within the classical Ohmic dead zone (albeit at much larger radii than in our solutions).

\subsubsection{$\bar{\alpha}(r)$}

Fig.\,8 shows our solution for the vertically averaged viscosity parameter $\bar\alpha$ as a function of radius. In the innermost disk, $\bar\alpha$ saturates at $\sim$0.08 as the potassium becomes almost entirely ionized (see top panel of Fig.\,7). It them falls smoothly by nearly 3 orders of magnitude, reaching our adopted floor value of $\adz = 10^{-4}$ at $\sim$0.25\,AU. Beyond this point, there is no MRI-active zone any more, and we assume a constant $\bar\alpha = \adz$ (depicted by the dashed horizontal line in Fig.\,8).

\begin{figure} 
\centering
\includegraphics[width = \columnwidth, trim={2cm 1cm 2cm 2cm}, clip]{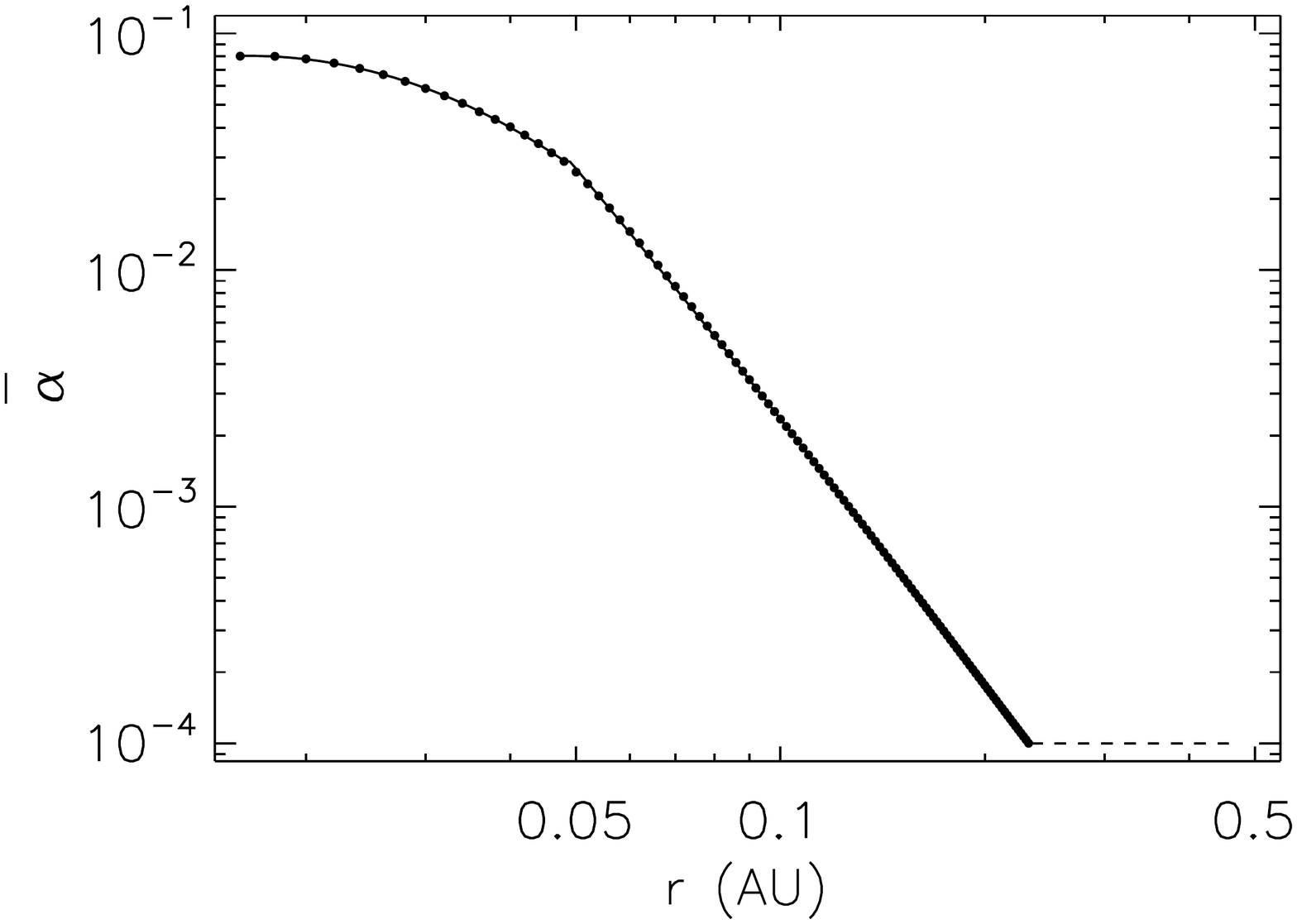} 
\caption{Vertically-averaged viscosity parameter $\bar\alpha$, plotted as a function of radius for our fiducial disk model. {\it Filled circles} represent our model results; the overplotted {\it solid line} is a combined piece-wise polynomial fit to these results. Our MRI calculations end at $\sim$0.25\,AU, the radius at which $\bar\alpha$ falls to our adopted floor value $\adz$ = $10^{-4}$. Beyond this radius we assume a constant $\bar\alpha=\bar\alpha_{\rm DZ}$, as indicated by the {\it dashed horizontal line}. See Table 1 in Appendix C for the polynomial fit parameters. 
}
\label{fig:fig8}
\centering
\end{figure}

\subsubsection{Disk Structure and Pressure Maximum}

Fig.\,9 shows the (vertically isothermal) temperature, midplane density, midplane pressure and surface density as functions of radius for our fiducial disk model. Beyond $\sim$0.25\,AU, where $\bar\alpha$ falls to $\adz$, we calculate these quantities assuming a constant $\bar\alpha = \adz$ (as depicted by the dashed lines in Fig.\,9). 

The salient results are: {\it (a)} There is a clear maximum in the midplane gas pressure (and midplane gas density) at $\sim$0.25\,AU, where $\bar\alpha$ reaches its floor value of $\adz$. Note that this location is radially well beyond the dead zone inner boundary (DZIB), which is located at $\sim$0.09\,AU; thus the midplane pressure maximum is situated {\it within} the dead zone, for the reasons discussed earlier. {\it (b)} The surface density declines sharply inwards of the pressure maximum, falling by 2 orders of magnitude towards the disk inner edge. This is a straightforward consequence of $\bar\alpha$ increasing inwards in this region coupled with a constant \mdot, as discussed previously. {\it (c)} The temperature varies quite slowly in the inner disk in this fiducial model, by less than a factor of 2, and in particular remains lower than the dust sublimation temperature of $\sim$1500\,K except near the disk inner edge. As such, small dust grains (which will be coupled to the gas rather than being trapped in the pressure maximum) are expected to have a significant effect on the MRI in this region, which we examine in a subsequent paper.    

\begin{figure} 
\centering
\includegraphics[width = \columnwidth, trim={4cm 0cm 1cm 1cm}, clip]{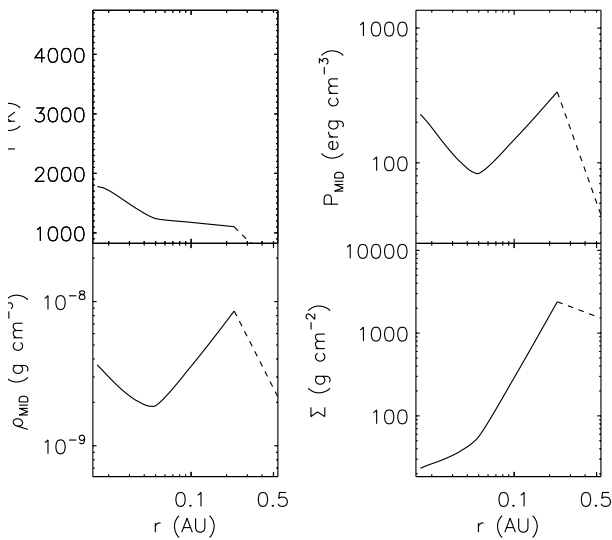} 
\caption{Various disk structure quantities plotted as a function of radius, for our fiducial disk model. {\it Top left}: (Vertically constant) temperature. {\it Top right} and {\it bottom left}: Midplane pressure and midplane density. {\it Bottom right}: Surface density. {\it Solid lines} represent our model calculations, which end at the radius where $\bar\alpha$ falls to $\adz$. Beyond this radius we assume a constant $\bar\alpha=\bar\alpha_{\rm DZ}$, obtaining the results shown here by the {\it dashed lines}.
}
\label{fig:fig9}
\centering
\end{figure}

\subsubsection{Accretion Rates in Active, Dead and Zombie Zones}

The total inward accretion rate (which by definition is radially constant in our steady-state solutions) is, at every radius, the sum of the accretion rates within the individual vertical layers of the disk (active, dead and zombie). We calculate these individual \mdot\,using equation (26); the results are plotted in Fig.\,10. We see that the inward \mdot\,through the active layer is practially the sole contributor to the total from the innermost radii out to $\sim$0.09\,AU, where the dead zone in the midplane first develops; the \mdot\,through the overlying zombie zone (due to non-MRI torques) steadily increases over this radial span, but is negligible compared to the active zone value. Once the Ohmic dead zone forms, the inward accretion through it (again, due to non-MRI torques) rapidly increases (as the thickness of this layer grows), while the \mdot\,in the active and zombie zones correspondingly decrease. Indeed, beyond $\sim$0.15\,AU, the inward \mdot\,in the dead zone {\it exceeds} the total value; this is compensated for by {\it decretion} (outward flow of mass) in the active and zombie zones, which ensures that the total inward accretion rate remains constant at the desired value (10$^{-9}$\,\msun\,yr$^{-1}$ here). 

A little reflection shows that in a non-trivial and non-pathological disk, i.e., one in which the disk properties vary radially in a physically plausible manner, such inconstancy of the accretion rates within the individual layers is unavoidable if the total \mdot\, is to remain fixed: If we demand that the total value be invariant, then we do not have any separate justifiable knobs to turn to ensure that the individual contributing rates remain constant as well.  

Does this phenomenon represent a growing instability? Certainly the buildup of mass at some locations, and excavation at others, that the radially varying accretion rates will generate in the individual layers will tend to drive the disk away from our equilibrium solution. However, these changes in the vertical density profile will occur over a local viscous timescale, given by $t_{\rm visc} \sim r^2/{\bar\nu}$ (where $\bar\nu$ is the vertically averaged local viscosity). By equations (1) and (2), ${\bar\nu} = {\bar\alpha}c_s^2/\Omega$ in our vertically isothermal disk, so $t_{\rm visc} \sim {\bar{\alpha}}^{-1}(z_H/r)^{-2}\,\Omega^{-1} \sim {\bar{\alpha}}^{-1}(z_H/r)^{-2}\,t_{\rm dyn}$, where $t_{\rm dyn} \sim 1/\Omega$ is the dynamical timescale. Simultaneously, the disk will tend to relax back to a hydrostatic equilibrium vertical profile (which is assumed in our solution) on a timescale given by $t_H \sim z_H/c_s \sim 1/\Omega \sim t_{\rm dyn}$. Note that the instantaneous perturbations in the vertical density profile here do {\it not} represent a change in the total surface density $\Sigma$ at any location: the latter remains constant (by equation (27), since the total \mdot\, is fixed at our steady-state value); i.e., the density perturbations sum to zero vertically. Thus, the disk will tend to relax to the {\it same} hydrostatic equilibrium vertical profile as in our solution. Now, in a normal thin disk, the disk aspect ratio $z_H/r \ll 1$, so for a standard ${\bar{\alpha}} < 1$, we have $t_{\rm visc} \gg t_{\rm dyn}$. Fig.\,11 demonstrates this explicitly for our disk: we see that $t_{\rm visc}$ is orders of magnitude larger than $t_{\rm dyn}$ over our radii of interest. Consequently, we expect the density perturbations introduced by the variable accretion rates to be vertically smoothed out, and hydrostatic equilium re-established, much more rapidly than these perturbations can grow; our steady-state solution will then remain valid in a (dynamical) time-averaged sense.    

\begin{figure} 
\centering
\includegraphics[width = \columnwidth, trim={1cm 2cm 1cm 5cm}, clip]{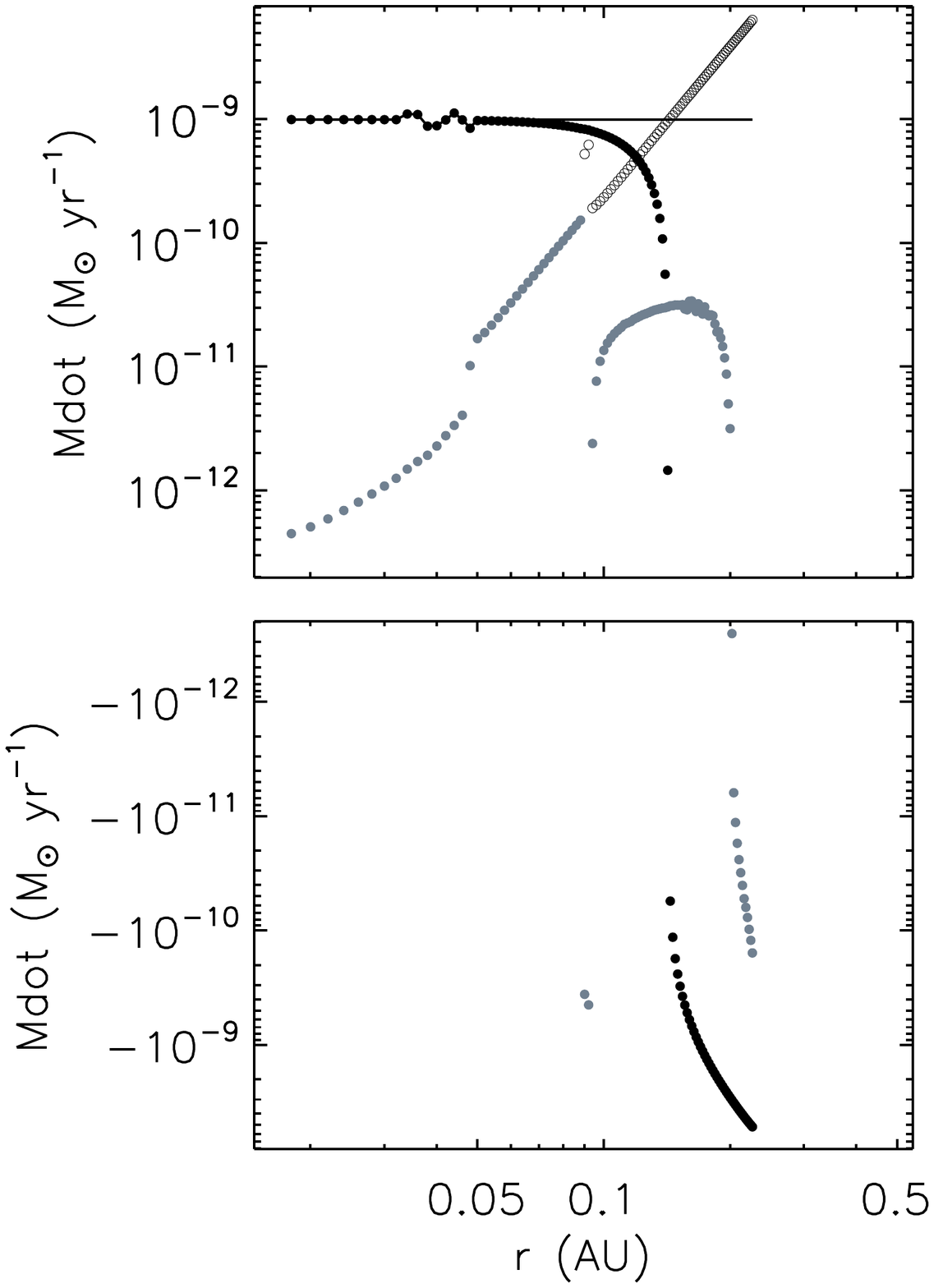} 
\caption{Accretion rates through the active zone ({\it filled black circles}), zombie zone ({\it filled gray circles}) and dead zone ({\it empty circles}) for our fiducial model. The {\it solid black line} represents the sum of the three rates (i.e., the total accretion rate through the disk, held fixed at \mdot\, = 10$^{-9}$\,\msun\,yr$^{-1}$ in this model). The {\it top} panel shows inward (positive) accretion rates, and the bottom shows outward (negative) rates. A few small anomalies -- the minor jitter in the active zone (and thus total) rate around 0.04\,AU, and the anomalously large first two points in the dead zone accretion rate, at 0.09\,AU -- result from our finite grid resolution at locations where the disk resistivities undergo sharp changes (we have left them in to show the limits of our precision).   See \S8.1.5. 
}
\label{fig:fig10}
\centering
\end{figure}

\begin{figure} 
\centering
\includegraphics[width = \columnwidth, trim={2cm 1.5cm 2cm 2cm}, clip]{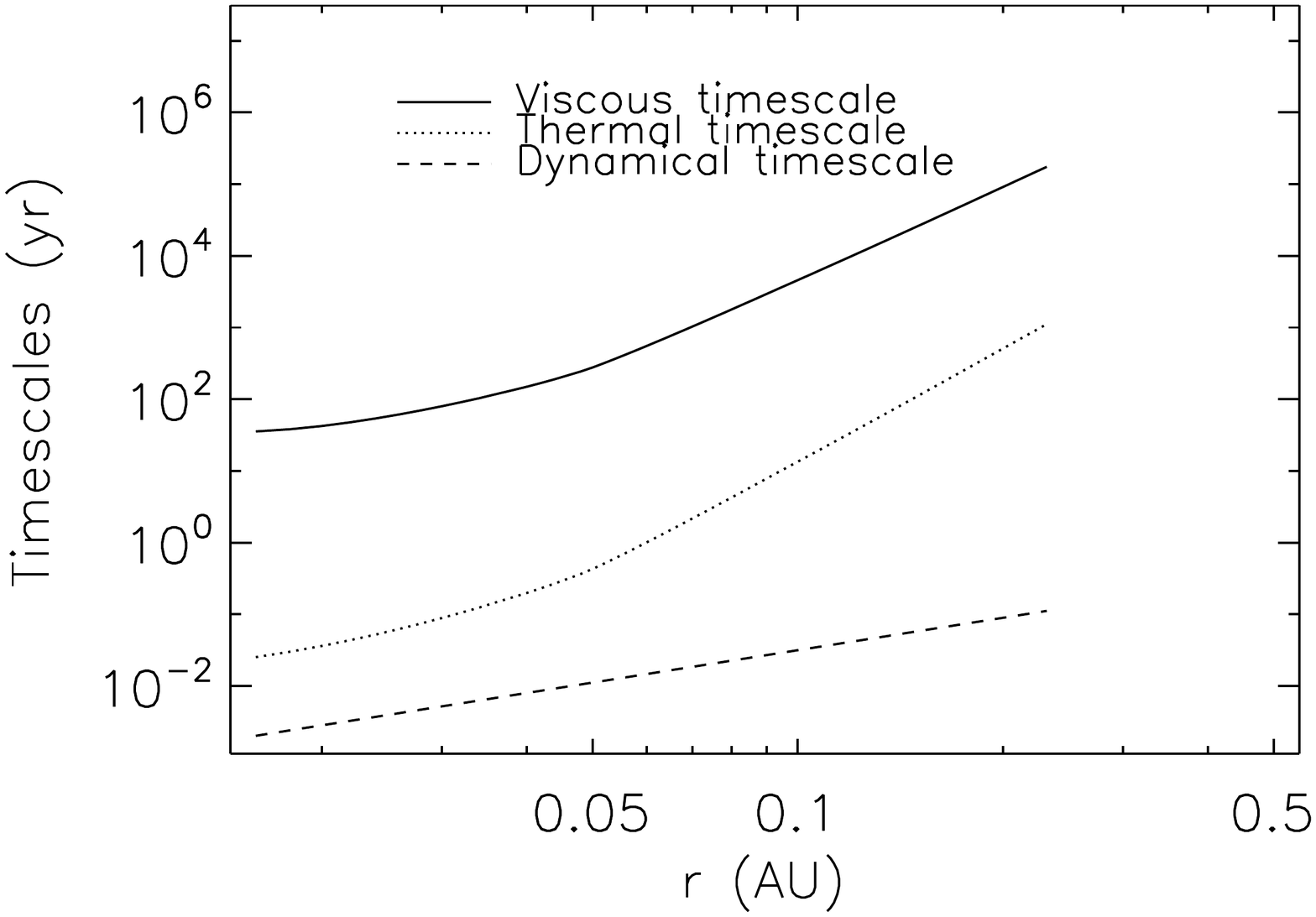} 
\caption{Viscous, thermal and dynamical (orbital) timescales as a function of radius for our fiducial disk model. See \S\S 8.1.5 and 8.1.6.
}
\label{fig:fig11}
\centering
\end{figure}

\subsubsection{Viscous Instability}

In our steady-state solutions, the surface density $\Sigma$, and hence the accretion rate, are temporally constant. Perturbations in $\Sigma$, however, may lead to a viscous instability as follows (see Pringle 1981). The general evolution equation for the disk surface density is
\setcounter{equation}{27}
\begin{equation}
\frac{\partial{\Sigma} }{\partial{t}} = - \frac{1}{r}\frac{\partial}{\partial{r}}\left[\left(\frac{dj}{dr}\right)^{-1} \frac{\partial}{\partial{r}}\left(r^3\,{\bar{\nu}}\,\Sigma\,\frac{d\Omega}{dr}\right)\right]
\end{equation}
where $j$ is the specific angular momentum at any disk location, and $\bar{\nu}$ is again the vertically averaged viscosity. For the specific case of a Keplerian disk, we have $\Omega = \sqrt{GM_{\ast}/r^3}$ and $j = r^2\Omega$, and the above reduces to
\begin{equation}
\frac{\partial{\Sigma} }{\partial{t}} =  \frac{1}{r}\frac{\partial}{\partial{r}}\left[3\, r^{1/2} \frac{\partial}{\partial{r}}\left(r^{1/2}\,{\bar{\nu}}\,\Sigma\right)\right] \,\,. 
\end{equation}
Changes in $\Sigma$ will occur on a viscous timescale. We have already noted that vertical hydrostatic equilibrium is established over a timescale $t_{\rm dyn} \ll t_{\rm visc}$. Similarly, the disk will relax to thermal equilibrium over a time given by the ratio of the thermal energy content per unit area to the rate of viscous heating (= rate of cooling in equilibrium) per unit area: $t_{\rm th} \sim (P_{\rm gas}\, z_H) / ({\bar\nu}\Sigma\,\Omega^2) \sim c_s^2/({\bar\nu}\,\Omega^2) \sim {\bar\alpha}^{-1}\,t_{\rm dyn}$. Thus, for ${\bar\alpha} < 1$, we have $t_{\rm dyn} < t_{\rm th} \ll t_{\rm visc}$ (as Fig.\,11 explicitly shows for our disk), and we expect the disk to be in both thermal and hydrostatic equilibrium over the timescales on which $\Sigma$ varies. In this situation, the mean viscosity at a fixed radius will depend only on the local surface density, i.e., ${\bar{\nu}} = {\bar{\nu}}(\Sigma,r)$, and equation (29) is a non-linear diffusion equation for $\Sigma$. For steady-state solutions, the L.H.S. of equation (29) is zero; we wish to investigate the effect of a small perturbation about any such equilibrium solution $\Sigma_0$. Define $x \equiv {\bar{\nu}}\,\Sigma$. Then any small variation in the surface density, $\Sigma_0 \rightarrow \Sigma_0 + \delta\Sigma$, implies a variation $x_0 \rightarrow x_0 + \delta x$, with $\delta x = \left(\partial{x}/\partial{\Sigma}\right)\delta\Sigma$. Inserting the perturbed value of $\Sigma$ into equation (29) then gives the time evolution equation for the perturbation $\delta x$:
\begin{equation}
\frac{\partial{(\delta x)} }{\partial{t}} = \left(\frac{\partial x}{\partial\Sigma}\right) \frac{1}{r}\frac{\partial}{\partial{r}}\left[3\, r^{1/2} \frac{\partial}{\partial{r}}\left(r^{1/2}\,\delta x\right)\right] \,\,.
\end{equation}
This linear diffusion equation for $\delta x$ is well-behaved {\it if and only if} the diffusion constant $\partial{x}/\partial{\Sigma}$ is positive; instability results otherwise. Hence, using $\bar{\nu} = {\bar{\alpha}}c_s^2/\Omega$ in our disk to evaluate the diffusion constant, we arrive at the viscous instability condition:
\begin{IEEEeqnarray}{rCl}
Instability & \iff & \frac{\partial{x}}{\partial{\Sigma}} < 0 \nonumber\\
& \iff & \frac{\partial{({\rm ln}\,{\bar{\alpha}})}}{\partial {({\rm ln}\,\Sigma)}} + 2\,\frac{\partial{({\rm ln}\,{c_s})}}{\partial {({\rm ln}\,\Sigma)}} < -1\,\,.
\end{IEEEeqnarray}
A negative diffusion constant implies that surface density inhomogeneities will be amplified: overdense regions will grow denser while underdense ones will become even more rarefied. In other words, an axisymmetric disk will tend to break up into rings. 

To investigate whether our inner disk is viscously unstable, we proceed as follows. We {\it assume} that, given a local perturbation in surface density $\Sigma$, the local disk parameters ($\bar\alpha$, $c_s$, $x$, \mdot) tend towards their {\it steady-state} values corresponding to the {\it perturbed} value of $\Sigma$. This allows us to evaluate the instability criterion by comparing the different equilibrium solutions we have calculated. We also find it useful to change variables from $x$ to \mdot, in order to connect to our steady-state solutions for different values of \mdot. 

In general, $\partial{\Sigma}/\partial{t} = (2\pi r)^{-1} \partial{\dot{M}}/\partial{r}$. For steady-state, \mdot\,must be radially constant; in this case, combining the latter expression with equation (29) yields the equilibrium solution $\dot{M}_0 = 3\pi\,{\bar{\nu}}\,\Sigma/f_r$ (equivalent to equation (27) with our definition of $\bar{\nu}$). Thus $\dot{M}_0 \propto x_0$ (with the constant of proportionality independent of $\Sigma$), and the instability condition $\partial{x}/\partial{\Sigma} < 0$ may be expressed as $\partial{\dot{M}}/\partial{\Sigma} < 0$, or equivalently as $\partial{\Sigma}/\partial{\dot{M}} < 0$. Evaluating the latter expression, we can write the instability criterion as:
\begin{IEEEeqnarray}{rCl}
Instability & \iff & \frac{\partial{\Sigma}}{\partial{\dot{M}}} < 0 \nonumber\\
& \iff & \frac{\partial{({\rm ln}\,{\bar{\alpha}})}}{\partial {({\rm ln}\,{\dot{M}})}} + 2\,\frac{\partial{({\rm ln}\,{c_s})}}{\partial {({\rm ln}\,{\dot{M}})}} > 1\,\,.
\end{IEEEeqnarray}

In Fig.\,12, we plot the steady-state $\Sigma$ solution as a function of radius, for various \mdot\,spanning $\pm$0.3\,dex around our fiducial value of 10$^{-9}$\,\msun\,yr$^{-1}$. We immediately see that, at any fixed radius beyond $\sim$0.035\,AU, $\Sigma$ increases as \mdot\,decreases; i.e., $\partial{\Sigma}/\partial{\dot{M}} < 0$. Thus, {\it most of the disk is viscously unstable}. This is shown more explicitly in Fig.\,13, where we plot $\partial{\Sigma}/\partial{\dot{M}}$ (calculated by deriving the steady-state $\Sigma$ for \mdot\,=10$^{-9}$\,\msun\,yr$^{-1}$\,$\pm$1\%) against radius; the quantity is negative over all but the innermost disk regions. By equation (32), the instability criterion may also be expressed as a condition on the summed change in $\bar{\alpha}$ and $c_s^2$ as a function of the change in \mdot. In Fig.\,14, we plot each of these two terms separately. It is apparent that the instability is caused primarily by the large change in $\bar\alpha$ with \mdot, with the change in sound speed making only a minor contribution. We shall see explictly how $\bar\alpha$ changes with accretion rate in \S8.3.

\begin{figure}
\centering
\includegraphics[width = \columnwidth, trim={2cm 1.5cm 2cm 4cm}, clip]{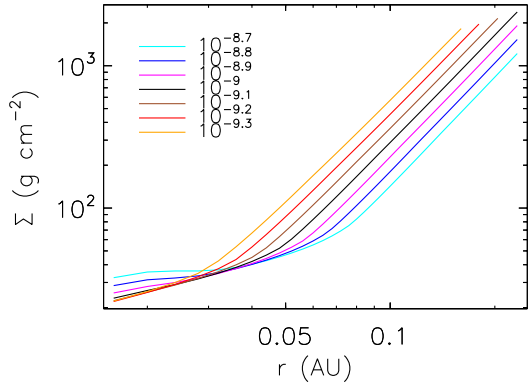} 
\caption{Steady-state solutions for the surface density $\Sigma$ as a function of radius, for model parameters \mstar\,= 1\,\msun, $\adz = 10^{-4}$ and varying accretion rates \mdot\, = 10$^{-9.3}$--10$^{-8.7}$\,\msun\,yr$^{-1}$ in steps of $0.1$\,dex. Over most of the disk (except the innermost regions), the surface density increases with decreasing accretion rate. See \S8.1.6.
}
\label{fig:fig12}
\centering
\end{figure}

\begin{figure} 
\centering
\includegraphics[width = \columnwidth, trim={2cm 0cm 2cm 2cm}, clip]{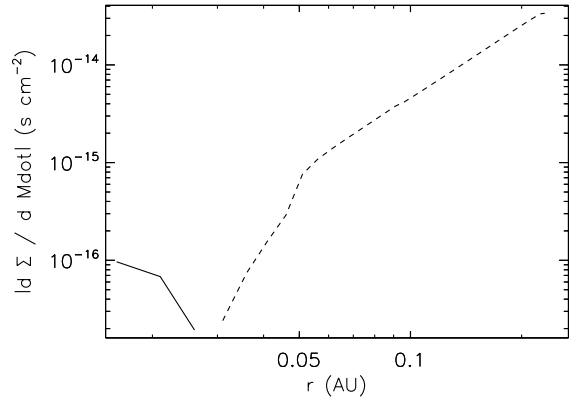} 
\caption{$|\partial{\Sigma}/\partial{\dot{M}}|$ as a function of radius for our fiducial disk model. The {\it solid line} denotes $\partial{\Sigma}/\partial{\dot{M}} > 0$ (viscously stable) while the {\it dashed line} denotes $\partial{\Sigma}/\partial{\dot{M}} < 0$ (viscously unstable). The disk is thus unstable at radii $r \gtrsim 0.03$\,AU. See \S8.1.6.
}
\label{fig:fig13}
\centering
\end{figure}

\begin{figure}
\centering
\includegraphics[width = \columnwidth, trim={2cm 1.5cm 2cm 2cm}, clip]{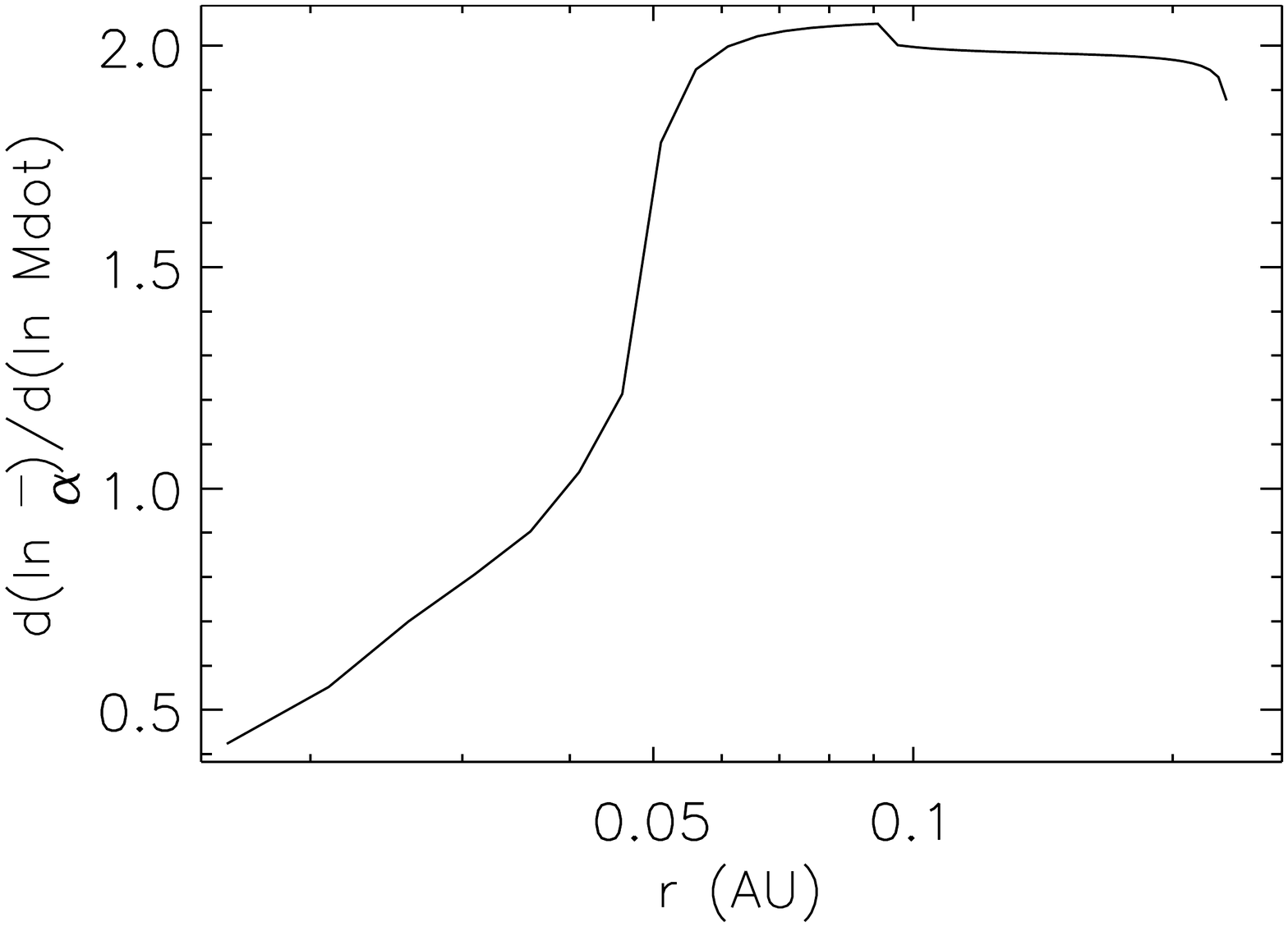}
\includegraphics[width = \columnwidth, trim={2cm 1.5cm 2cm 2cm}, clip]{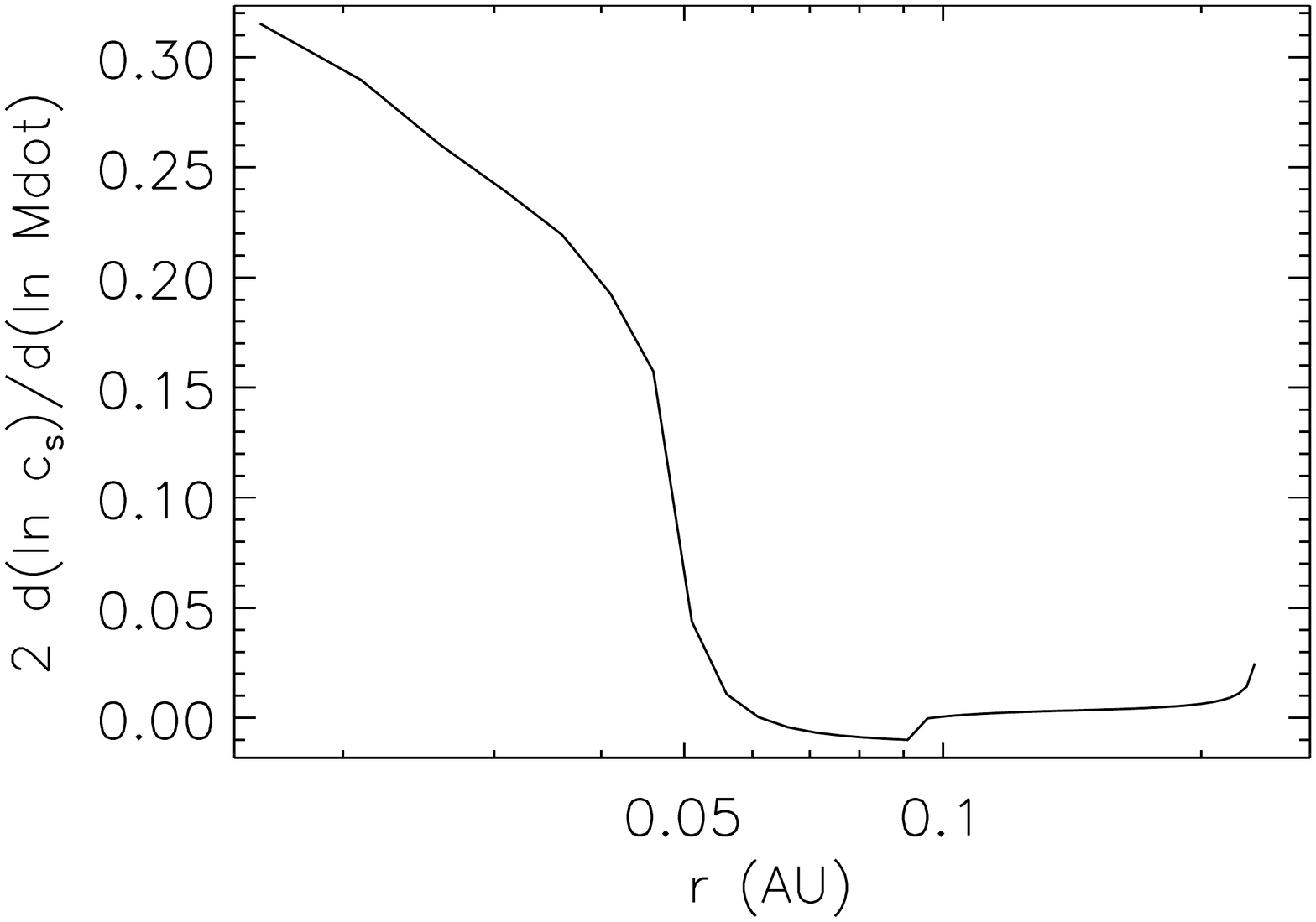}  
\caption{$\partial({\rm ln}{\bar{\alpha}})/\partial({\rm ln}{\dot{M}})$ ({\it top}) and $2\partial({\rm ln}{c_s})/\partial({\rm ln}{\dot{M}})$ ({\it bottom}) as a function of radius, for our fiducial disk model. See \S8.1.6.
}
\label{fig:fig14}
\centering
\end{figure}

\subsubsection{Opacity}
In this work, we have {\it assumed} a constant opacity of 10\,cm$^2$\,g$^{-1}$ throughout our calculation domain. 	Given the pressure and temperature structure derived thereby for our solution disk, we check the validity of this assumption a posteriori, by using the detailed tables of \citet{zhu12} to compute the opacities {\it predicted} as a function of pressure and temperature.       

The results are plotted in Fig.\,15. We see that the predicted opacity over the bulk of our disk solution is 5--10\,cm$^2$\,g$^{-1}$ (primarily due to grains; see below), very close to our assumed value. The only exception is the innermost disk, at $\lesssim$0.03\,AU, where the expected opacities are 1-2 orders of magnitude lower (as grains disappear). However, this small inner region is not consequential to our results at larger radii, e.g., regarding the dead zone inner boundary and the pressure maximum. In summary, therefore, our disk solution is overall self-consistent vis-\`{a}-vis the adopted opacity.  

Note that we have not explicitly included grains in our calculations. Nevertheless, our assumed opacity of 10\,cm$^2$\,g$^{-1}$ is the fiducial value adopted widely for dusty accretion disks, and is validated over most of the disk by the opacity calculations above that do account for grains. In other words, grains are {\it implicitly} included in our opacities. On the other hand, dust will also markedly influence the chemistry and the MRI (see \S8.1.2); these grain effects are ignored in this work (we treat them in a subsequent paper; Jankovic et al., in prep.). 

\begin{figure}
\centering
\includegraphics[width = \columnwidth, clip]{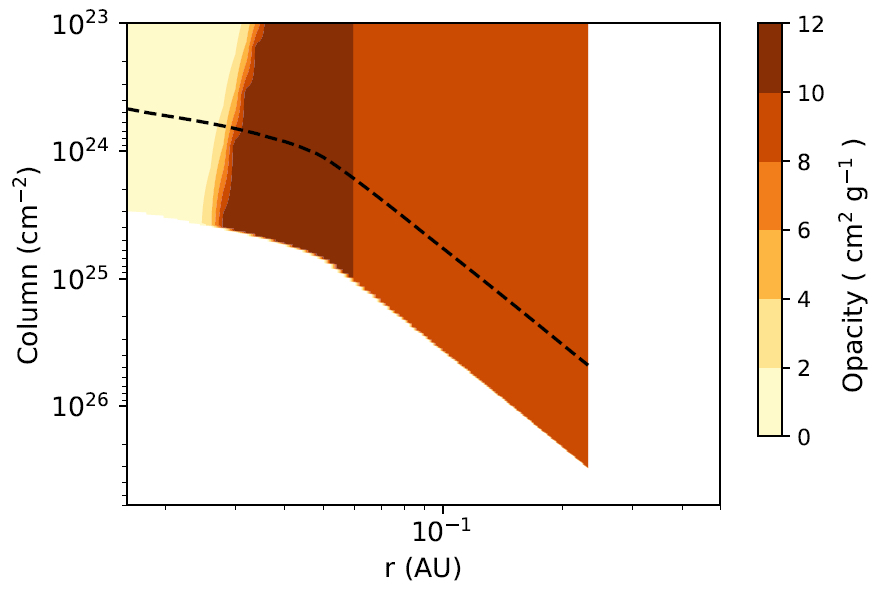} 
\caption{Rosseland mean opacity (in cm$^2$\,g$^{-1}$) calculated {\it a posteriori} for our fiducial disk model, plotted as a function of disk location (with height in units of vertical column). The dashed curve denotes one pressure scale-height. Over most of our region of interest in the disk, the derived opacity is within a factor of two of 10 cm$^2$\,g$^{-1}$, consistent with our {\it a priori} adoption of this value everywhere. See \S8.1.7.
}
\label{fig:fig15}
\centering
\end{figure}


\subsubsection{Validity of the Strong-Coupling Limit}
The criteria we use for active MRI in the presence of ambipolar diffusion (equations [11a,b]), derived from the MRI simulations of \citet{bai11a}, require that we be in the strong-coupling limit, i.e., in the single-fluid regime. The conditions for the latter are (see Appendix A): {\it (1)} $\rho_n \gg \rho_i$ (which is always satisfied in our case wherein potassium is the only ionised species, since the abundance of K puts a hard upper limit of $\sim$$10^{-7} \times m_K/m_{\rm{H_2}} \ll 1$ on $\rho_i/\rho_n$); and {\it (2)} $t_{\rm{rcb}} \ll t_{\rm{dyn}}$, where $t_{\rm{rcb}}$ is the recombination timescale. The latter condition expresses the requirement that ionization-recombination equilibrium be established on timescales shorter than the dynamical time on which other relevant disk physics (such as field amplification by Keplerian shear) occurs. Since ionization is generally very fast, it is the recombination time that forms the bottleneck in establishing ionization equilibrium; hence the criterion $t_{\rm rcb} \ll t_{\rm dyn}$. If this is not satisfied, then the MRI simulation results do not represent a steady-state.   

We use equation (13) to calculate $t_{\rm rcb}$ everywhere in our solution disk, and compare it to the local $t_{\rm dyn}$; the results are shown in Fig.\,16. We find that in fact the required condition on $t_{\rm rcb}$ is met {\it only} in the innermost disk close to the midplane, and nowhere else. The reason is clear: with effectively only a single chemical species (K), there is only one, relatively slow, recombination channel; thus, $t_{\rm rcb}$ ($\propto \sqrt{T}/n_e$) only becomes small enough to fall below $t_{\rm dyn}$ at the smallest radii, where $n_e$ is highest (see Fig.\,4; the weaker dependence on $T$, combined with the relatively small variation in $T$ in our solution -- see Fig.\,9 -- means that the temperature does not alter $t_{\rm rcb}$ very much). As such, our disk solution is to be interpreted only as an idealised case that holds {\it if} ionisation equilibrium is established with a single alkali species. Whether such an equilbrium can indeed be reached, or maintained, when the disk and field are otherwise evolving on much shorter dynamical timescales, is unclear\footnote{Answering this question rigorously requires a general 2-fluid simulation (of which the 1-fluid regime is a special case), including source and sink terms for the ions in order to account for an evolving ionisation fraction (we thank X. Bai for useful discussions on this issue). Note that the idealised 2-fluid simulations of Hawley \& Stone (1998; hereafter HS98; see Appendix A) assume a {\it fixed} ion fraction, so do not address this issue directly. Nonetheless, if $t_{\rm rcb} \gg t_{\rm dyn}$, then the ionisation fraction may be assumed to be approximately constant over $t_{\rm dyn}$, with all the relevant species being completely ionised (since the ionisation timescale alone is very short). In this sense, the HS98 results may be applied to a disk like ours, with only a single alkali species, with the specification that all the alkali atoms be ionised. We cannot, however, apply the HS98 results to the ionisation fractions {\it we} have derived assuming Saha equilibrium, because $t_{\rm rcb} \gg t_{\rm dyn}$ means that Saha equilibrium is simply not established over the dynamical timescales relevant to the HS98 simulations. At any rate, as discussed in the main text above, we do not expect a chemical network comprising only one alkali to be generally representative of real disks, so we do not pursue this line of inquiry further here.}.  

Nevertheless, our disk solutions are useful for two reasons. First, actual disks should support far more complex chemical networks, including both molecular ions and grains in addition to metal ions. With the much larger number of recombination channels available in such physically realistic circumstances, we do expect the time to attain ionization equilibrium to usually be shorter than the dynamical one \citep[e.g.,][]{bai11b}.  In that case, as long as $\bar\alpha$ follows the general form in our solutions (high value at very small radii, and tapering off with increasing distance), our results, regarding the behaviour of the various zones and the trends in the MRI and accretion rates, should remain qualitativaly applicable (though the quantitive locations of the pressure maximum and so forth will certainly change). Second, our analysis provides a general {\it method} for self-consistently solving the problem of an $\alpha$-disk coupled to the MRI (and for checking the validity of the solution a posteriori, as done here). This methodology will remain applicable, whatever the specifics of the chemical network.  

\begin{figure}[H] 
\centering
\includegraphics[width = \columnwidth, trim={2cm 1.5cm 2cm 2cm}, clip]{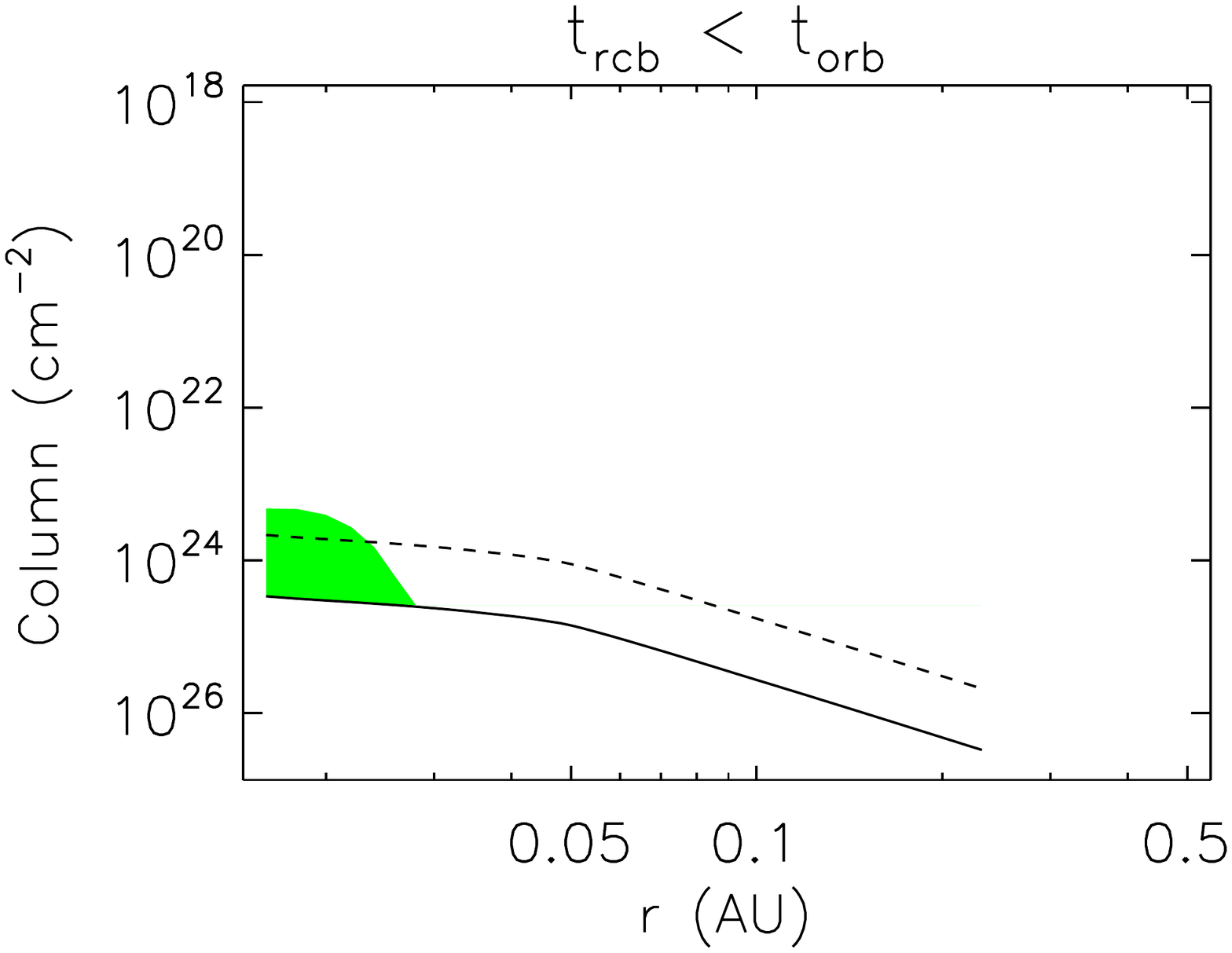} 
\caption{Recombination timescale ($t_{rcb}$) versus dynamical timescale ($t_{dyn}$) as a function of location in our fiducial disk model. The {\it solid black curve} denotes the disk midplane, and the {\it dashed curve} denotes one pressure scale-height. The {\it green} region is where $t_{\rm rcb} < t_{\rm dyn}$ (and thus where the single-fluid approximation is valid); in the rest of our fiducial disk, $t_{\rm rcb} > t_{\rm dyn}$. See \S8.1.8.
}
\label{fig:fig16}
\centering
\end{figure}

\subsection{Variations in ${\boldsymbol \adz}$}
Figs.\,17--19 show our disk solutions for the same \mstar\,and \mdot\,as the fiducial case, but with $\adz = 10^{-3}$ and $10^{-5}$ (instead of $10^{-4}$). These results closely resemble the fiducial solution, but with a couple of important quantitative differences. 

\begin{figure*} 
\centering
\includegraphics[width = \columnwidth, trim={2cm 2cm 2cm 2cm}, clip]{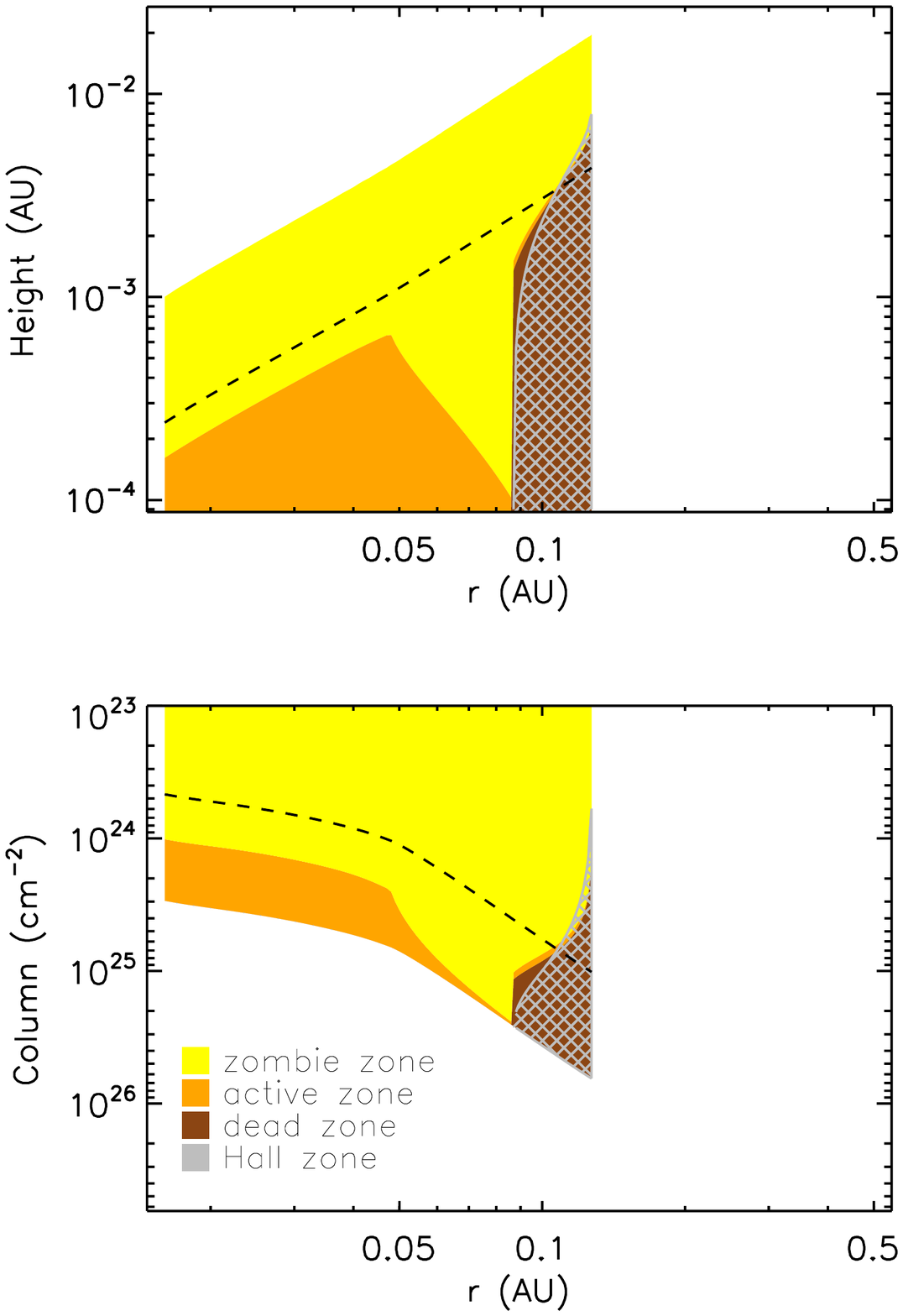}
\includegraphics[width = \columnwidth, trim={2cm 2cm 2cm 2cm}, clip]{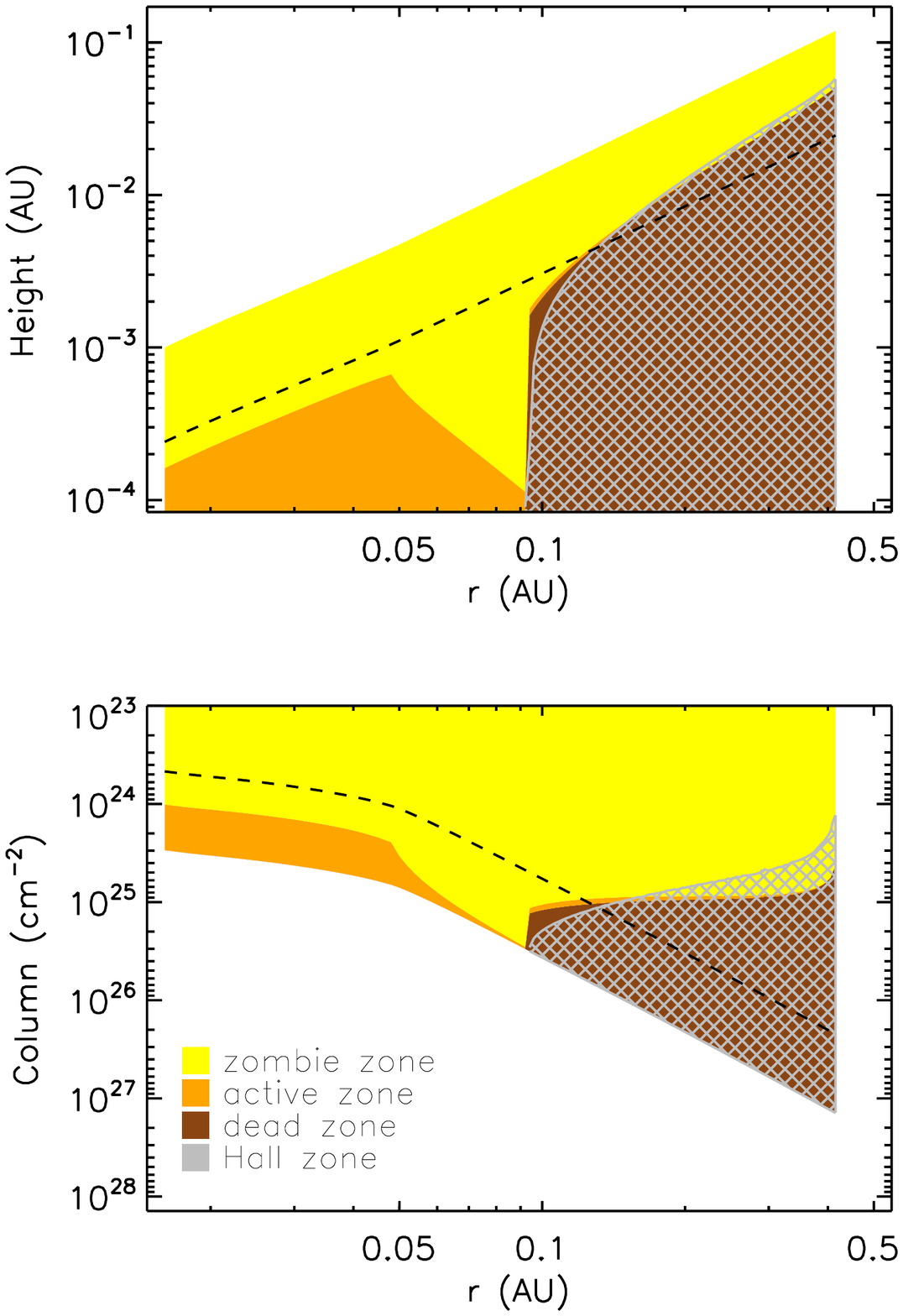}
\caption{Same as Fig. \ref{fig:fig5}, but now for $\adz = 10^{-3}$ ({\it left}) and $10^{-5}$ ({\it right}).
}
\label{fig:fig17}
\centering
\end{figure*}

\begin{figure*}
\centering
\includegraphics[width = \columnwidth, trim={2cm 1cm 2cm 2cm}, clip]{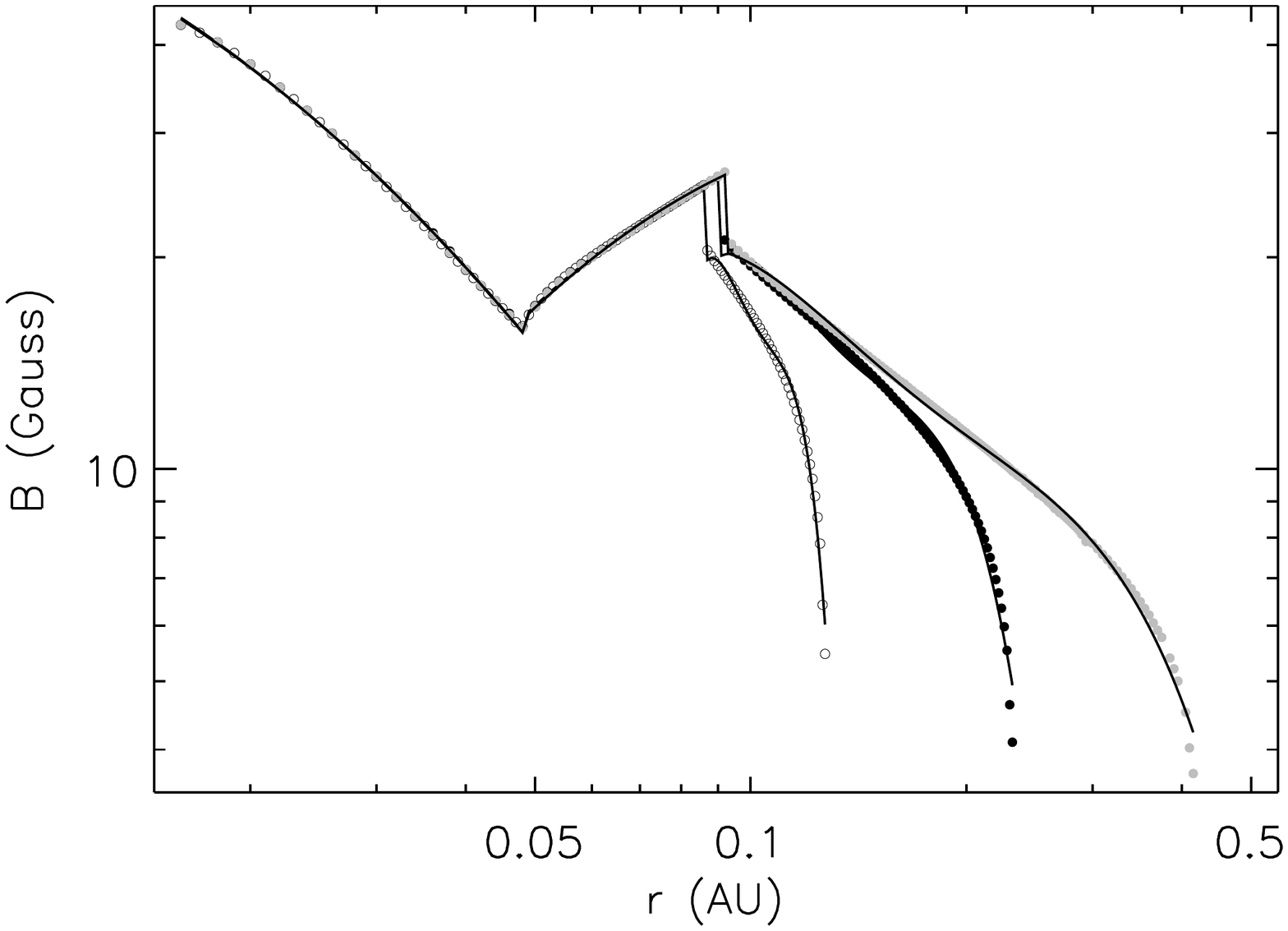}
\includegraphics[width = \columnwidth, trim={2cm 1cm 2cm 2cm}, clip]{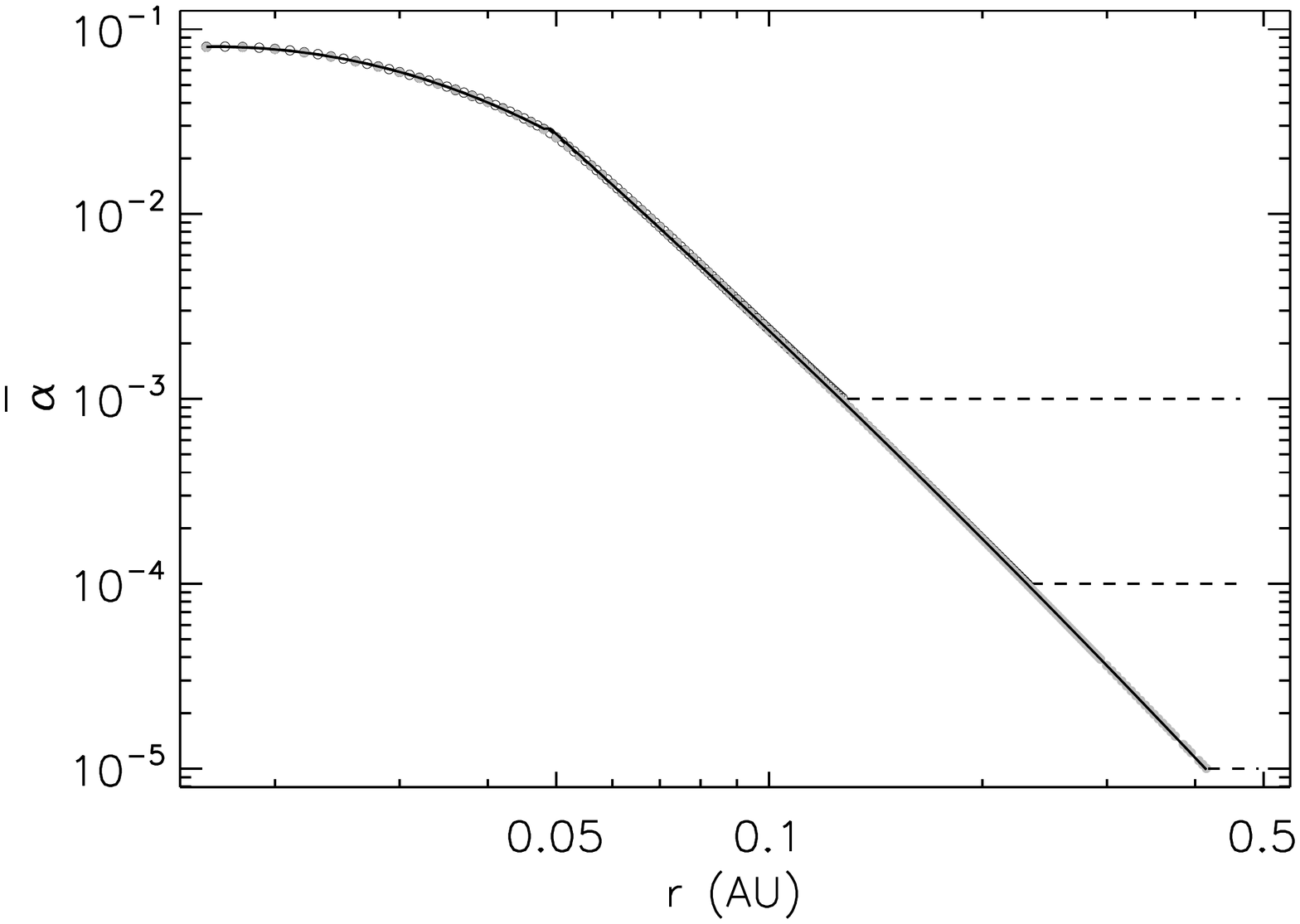}
\caption{{\it Left}: Field strength $B$ as a function of radius. {\it Right}: $\bar\alpha$ as a function of radius. In both plots, results for $\adz = 10^{-3}$ ({\it empty circles}) and $\adz = 10^{-5}$ ({\it filled grey circles}) are overplotted on the results for our fiducial model with $\adz = 10^{-4}$ ({\it filled black circles}; the fiducial results are the same ones shown in Figs.\,6 and 8 respectively).
}
\label{fig:fig18}
\centering
\end{figure*}
 
\begin{figure*}
\centering
\includegraphics[width = \columnwidth, trim={4cm 0cm 1cm 1cm}, clip]{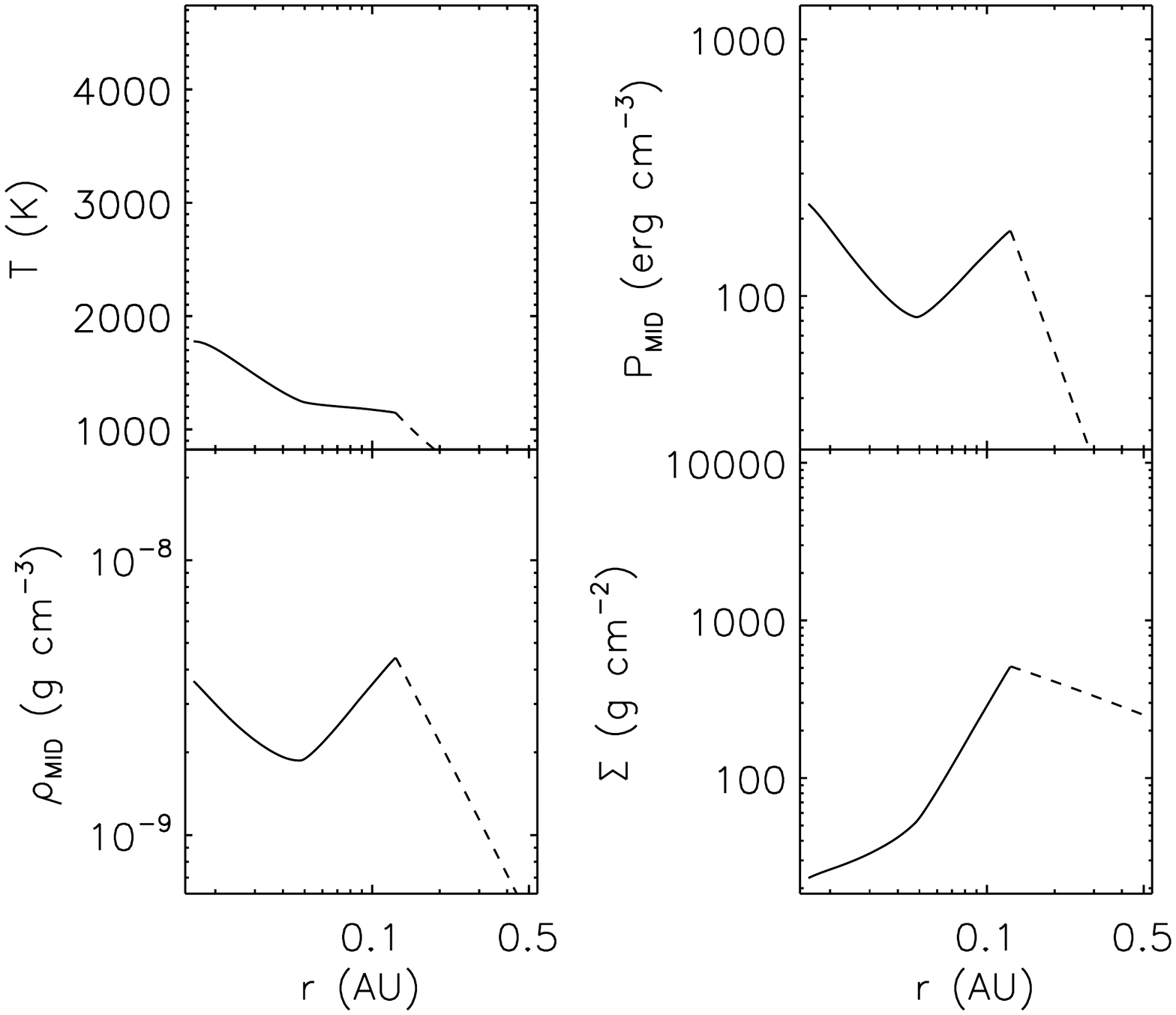}
\includegraphics[width = \columnwidth, trim={4cm 0cm 1cm 1cm}, clip]{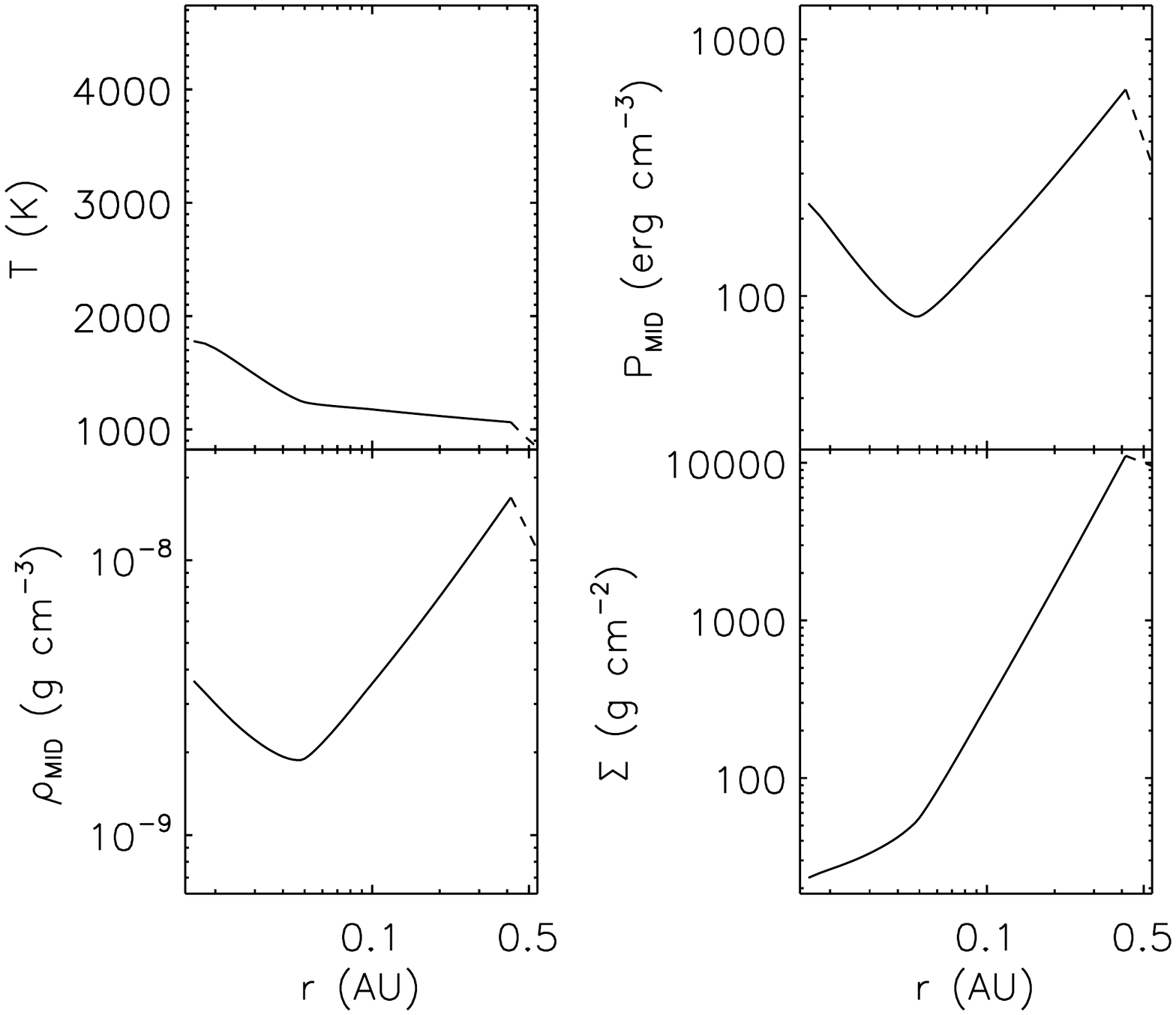} 
\caption{Various disk paramaters as a function of radius: same as Fig.\,9, but now for $\adz = 10^{-3}$ ({\it left}) and $10^{-5}$ ({\it right}).
}
\label{fig:fig19}
\centering
\end{figure*}

\begin{figure} 
\centering
\includegraphics[width = \columnwidth, trim={2cm 1cm 2cm 2cm}, clip]{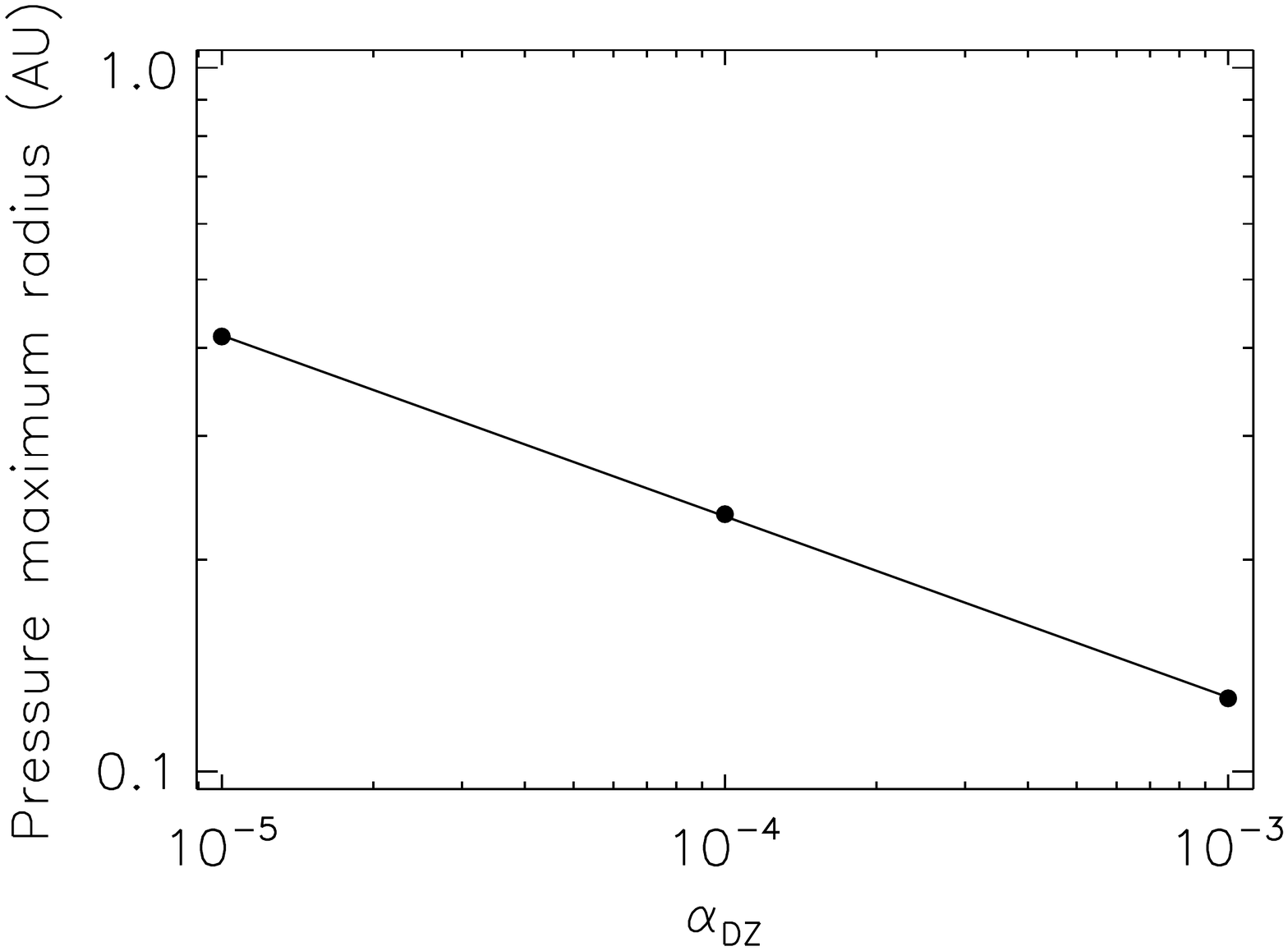}
\caption{Radial location of the pressure maximum as a function of $\adz$, showing the approximate power law dependence $r_{P_{\rm max}} \propto \adz{^{-1/4}}$.
}
\label{fig:fig20}
\centering
\end{figure}

First, because $\bar\alpha$ declines with increasing radius, the pressure maximum (located where $\bar\alpha$ hits the floor value $\adz$) occurs at a smaller radius ($\sim$0.12\,AU) for $\adz = 10^{-3}$ compared to the fiducial case ($\sim$0.25\,AU); conversely, it is at a larger radius ($\sim$0.4\,AU) for $\adz = 10^{-5}$. in fact, as Fig.\,20 shows, the radial location of the pressure maximum as a function of the floor value $\adz$ is approximately a power law: $r_{P_{\rm max}} \propto \alpha_{\rm DZ}^{-1/4}$. This follows from the fact that our solutions $\bar\alpha$ as a function of radius approximatelly decrease as power laws beyond 0.1\,AU, and the location of the pressure maximum corresponds to the radius at which this power law falls the floor value $\adz$ (right panel of Fig.\,18).

Second, while all three $\adz$ solutions are quite similar at radii inwards of the pressure maxima, they are not exactly the same: specifically, the field strengths in the three cases diverge beyond $\sim$0.09\,AU (left panel of Fig.\,18), which is where the dead zone first arises (compare Figs.\,5 and 17). This stems from the fact that the total \mdot\,at any radius is the sum of the individual accretion rates through the active, dead and zombie layers at that location. As discussed in \S8.1.5 for the fiducial case, the accretion rate in the low density zombie zone makes a negligible contribution to the total; hence, at radii where the active zone dominates in the midplane, the \mdot\,through it (controlled by the $B$-field) is essentially constant at the fixed total rate. Thus, the field strength at these radii remains the same for the three $\adz$ cases considered (which all have the same total \mdot). Once a dead zone forms in the high-density midplane, however, the accretion rate through it makes a significant and radially increasing contribution to the total rate; the rate through the active zone then compensates (in order to keep the total \mdot\,fixed) by declining rapidly with radius, facilitated by a steep decrease in the field strength (see Fig.\,10). Since the accretion rate through the dead zone increases with $\adz$, a higher (lower) $\adz$ leads to a steeper (shallower) fall-off in field strength (and thus in the active zone accretion rate) with radius, as depicted in Fig.\,18. 

\subsection{Variations in ${\boldsymbol \dot{\boldsymbol M}}$}
Figs.\,21--24 show our disk solutions for the same \mstar\,and $\adz$ as the fiducial case, but with \mdot\, = $10^{-8}$ and $10^{-10}$\,\msun\,yr$^{-1}$ (instead of $10^{-9}$\,\msun\,yr$^{-1}$). The salient deviations here from the fiducial case are all rooted in the fact that a higher \mdot\,elevates the viscous heating rate, leading to a larger ionisation fraction at a given location.


\begin{figure*}
\centering
\includegraphics[width = \columnwidth, trim={2cm 2cm 2cm 2cm}, clip]{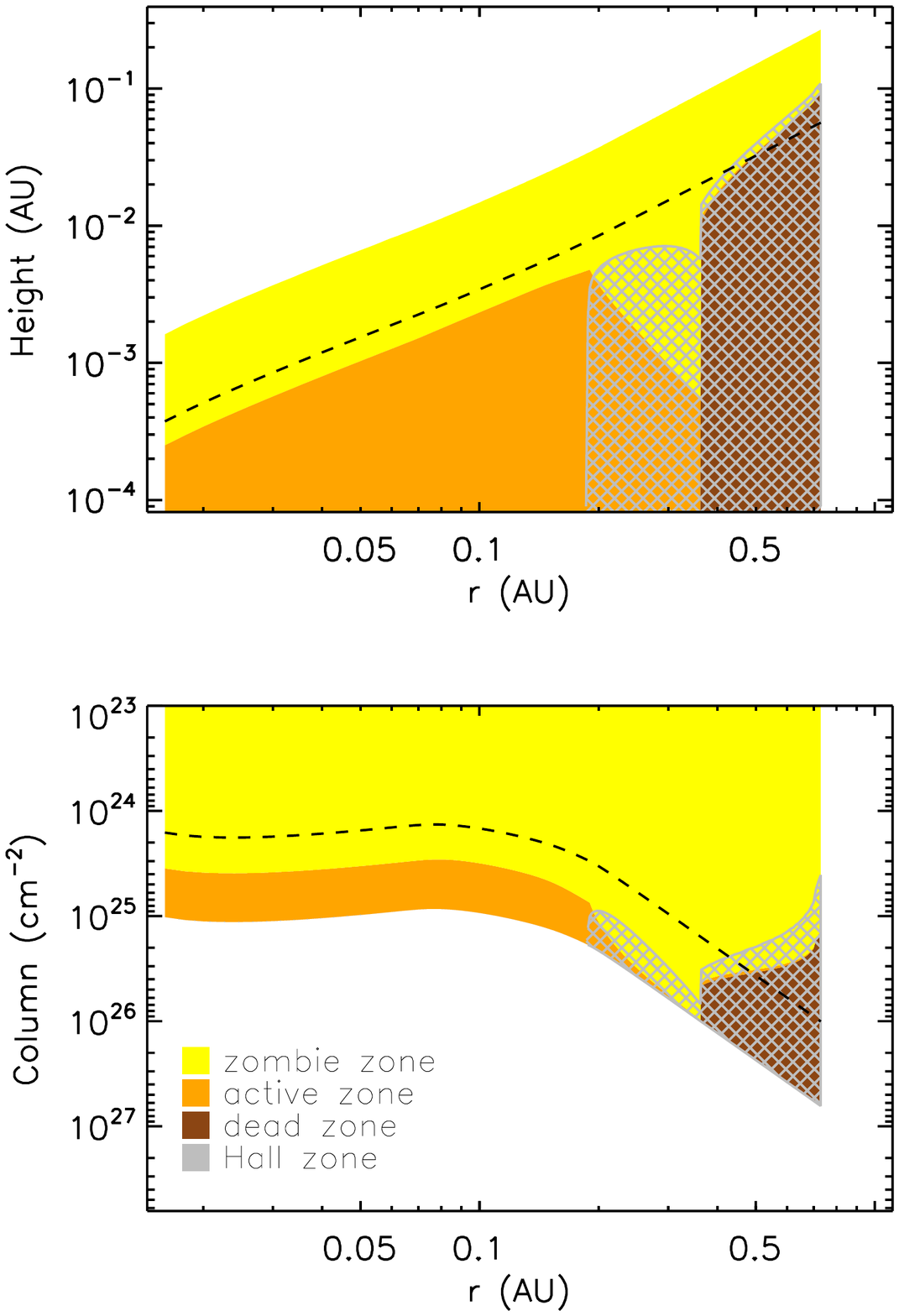}
\includegraphics[width = \columnwidth, trim={2cm 2cm 2cm 2cm}, clip]{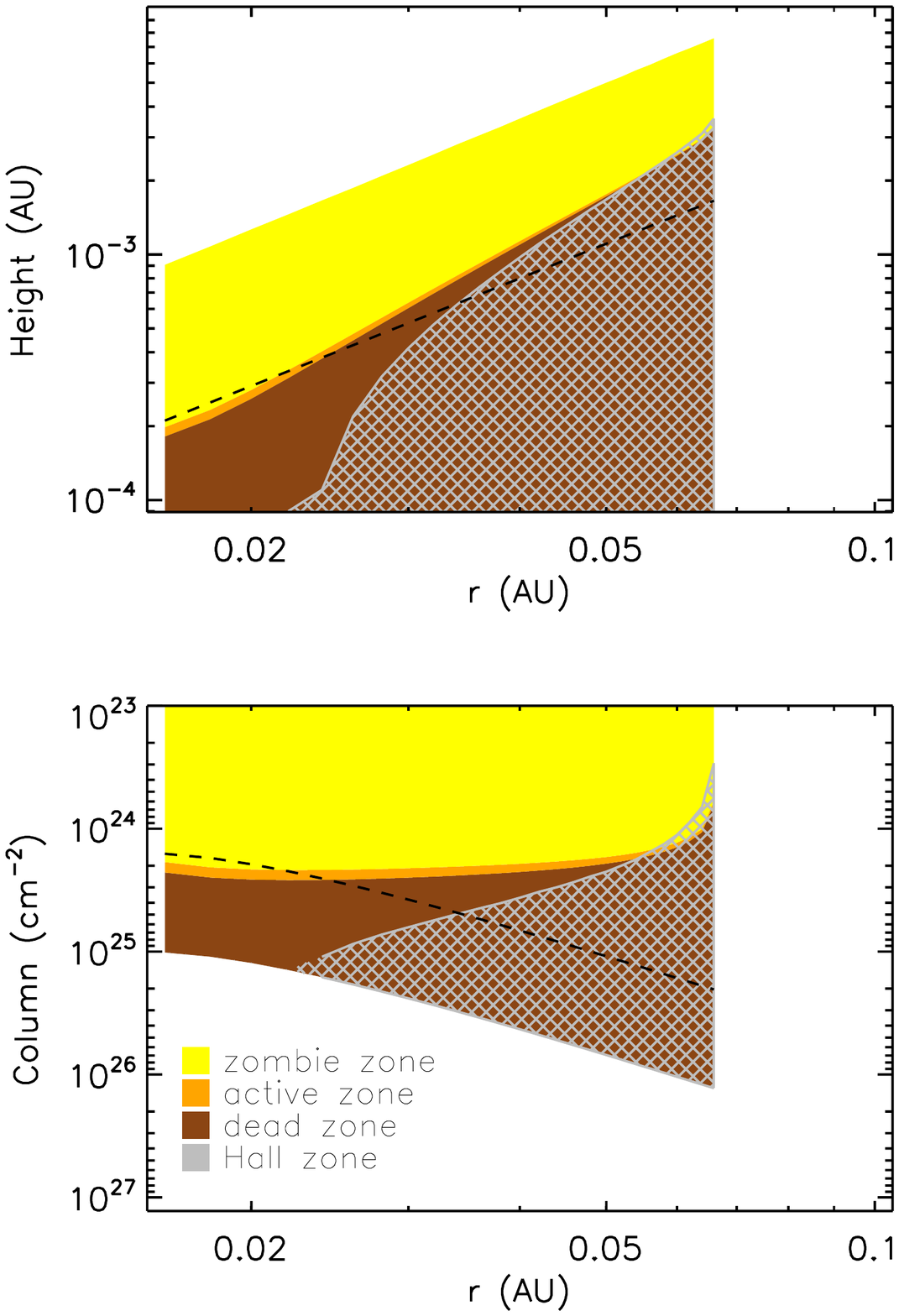} 
\caption{Same as Fig.\,5, but now for an accretion rate \mdot\, = 10$^{-8}$\,\msun\,yr$^{-1}$ ({\it left}) and 10$^{-10}$\,\msun\,yr$^{-1}$ ({\it right}).
}
\label{fig:fig21}
\centering
\end{figure*}

\begin{figure}
\centering
\includegraphics[width = \columnwidth, trim={2cm 2cm 2cm 2cm}, clip]{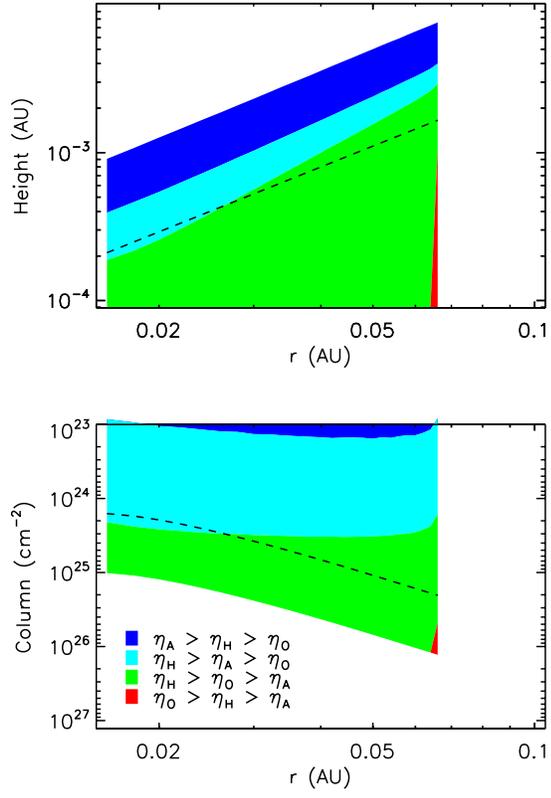} 
\caption{Same as Fig.\,2, but for an accretion rate \mdot\, = 10$^{-10}$\,\msun\,yr$^{-1}$. This is the only one of our various disk models in which an Ohmic-dominated region arises ({\it red} sliver in bottom right corner of both panels). 
}
\label{fig:fig22}
\centering
\end{figure}

\begin{figure*}
\centering
\includegraphics[width = \columnwidth, trim={2cm 1cm 2cm 2cm}, clip]{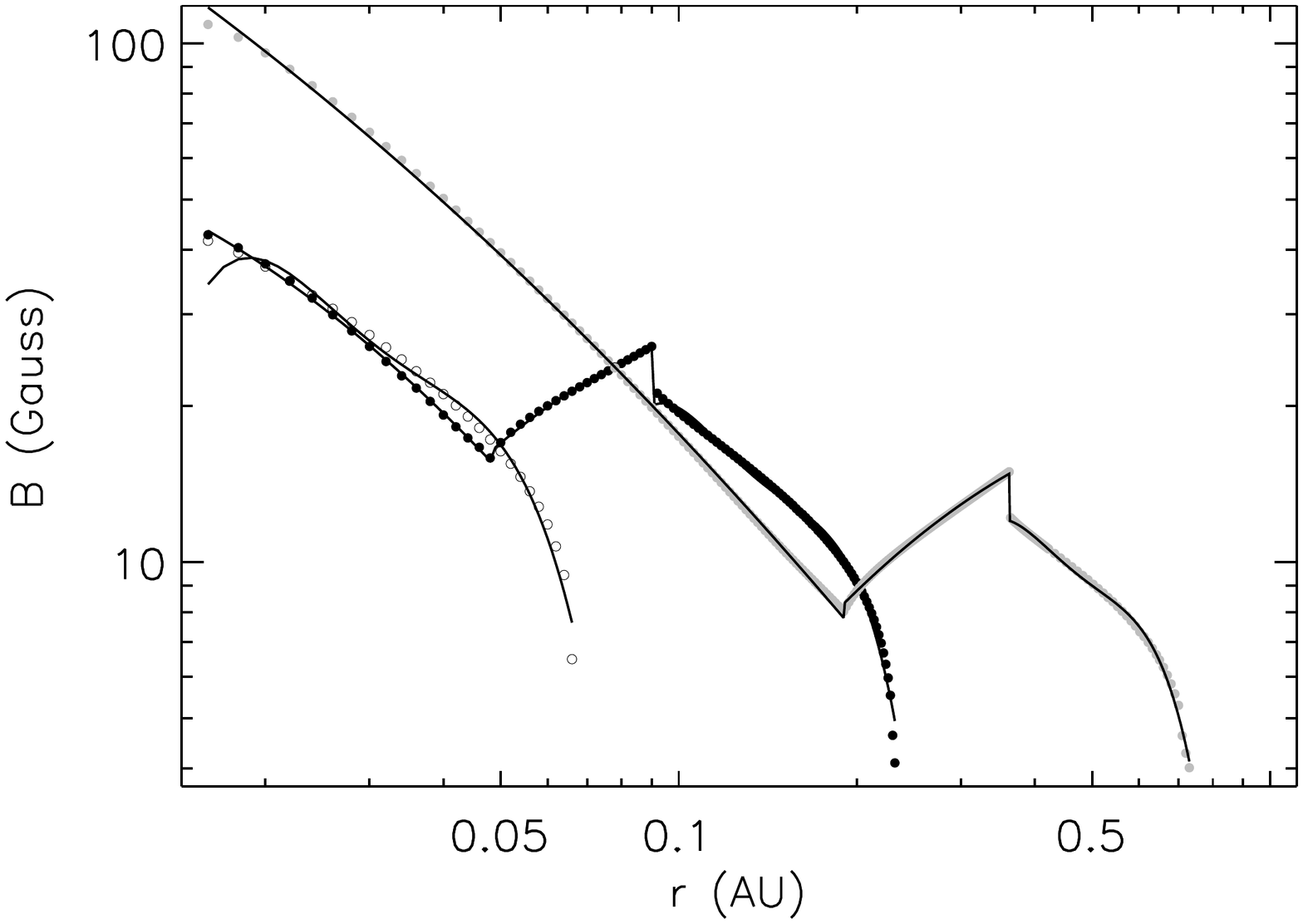}
\includegraphics[width = \columnwidth, trim={2cm 1cm 2cm 2cm}, clip]{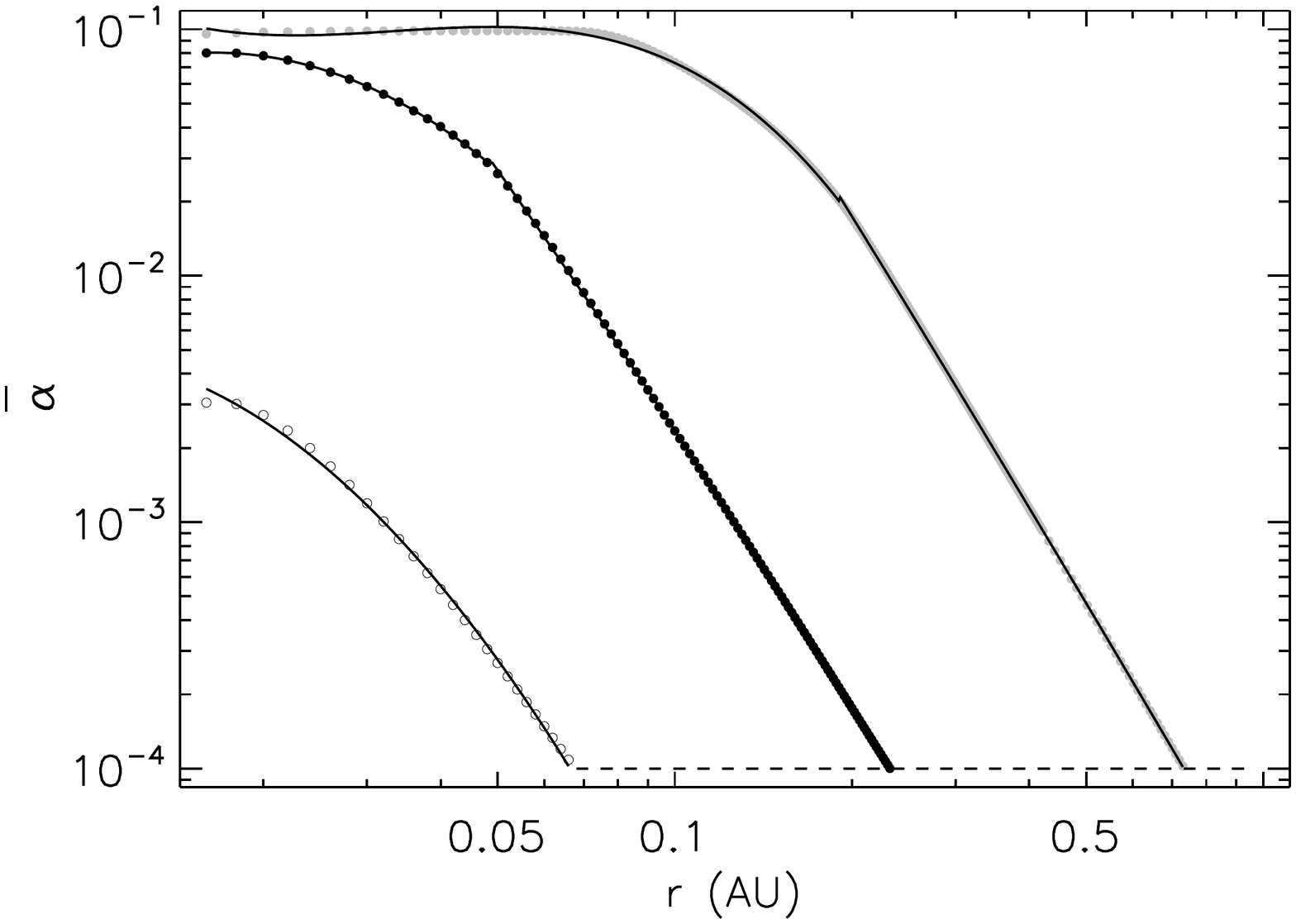} 
\caption{{\it Left}: Field strength $B$ as a function of radius. {\it Right}: $\bar\alpha$ as a function of radius. In both plots, results for \mdot\, = 10$^{-10}$\,\msun\,yr$^{-1}$ ({\it empty circles}) and \mdot\, = 10$^{-8}$\,\msun\,yr$^{-1}$ ({\it filled grey circles}) are overplotted on the results for our fiducial model with \mdot\, = 10$^{-9}$\,\msun\,yr$^{-1}$ ({\it filled black circles}; the fiducial results are the same ones shown in Figs.\,6 and 8 respectively).
}
\label{fig:fig23}
\centering
\end{figure*}

\begin{figure} 
\centering
\includegraphics[width = \columnwidth, trim={4cm 0cm 1cm 1cm}, clip]{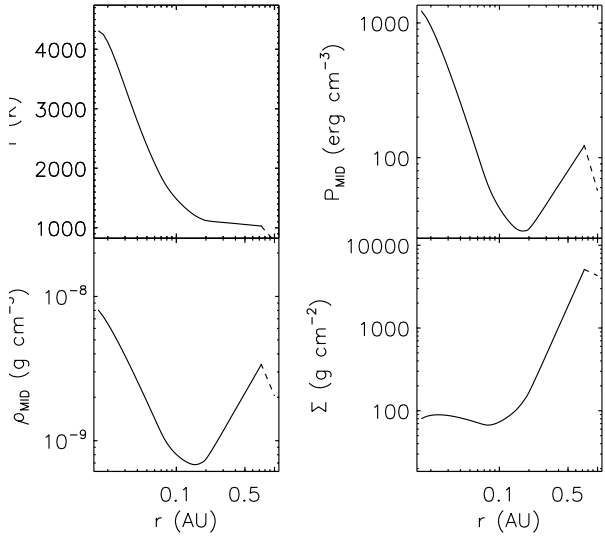}
\includegraphics[width = \columnwidth, trim={4cm 0cm 1cm 1cm}, clip]{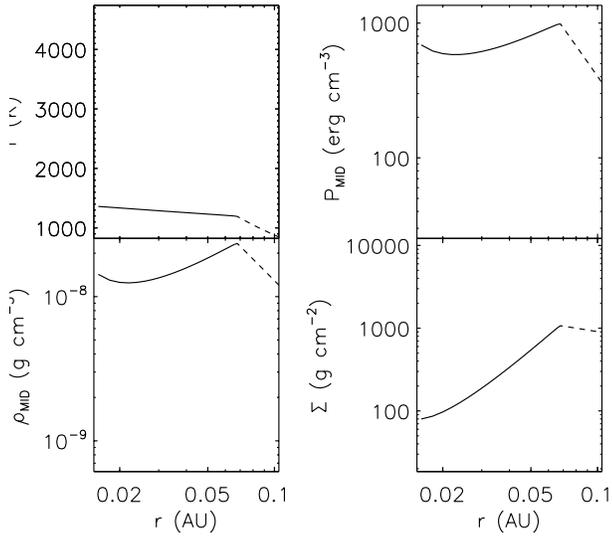} 
\caption{Various disk paramaters as a function of radius: same as Fig.\,9, but now for an accretion rate \mdot\, = 10$^{-8}$\,\msun\,yr$^{-1}$ ({\it top}) and 10$^{-10}$\,\msun\,yr$^{-1}$ ({\it bottom}).
}
\label{fig:fig24}
\centering
\end{figure}

\begin{figure} 
\centering
\includegraphics[width = \columnwidth, trim={2cm 1cm 2cm 2cm}, clip]{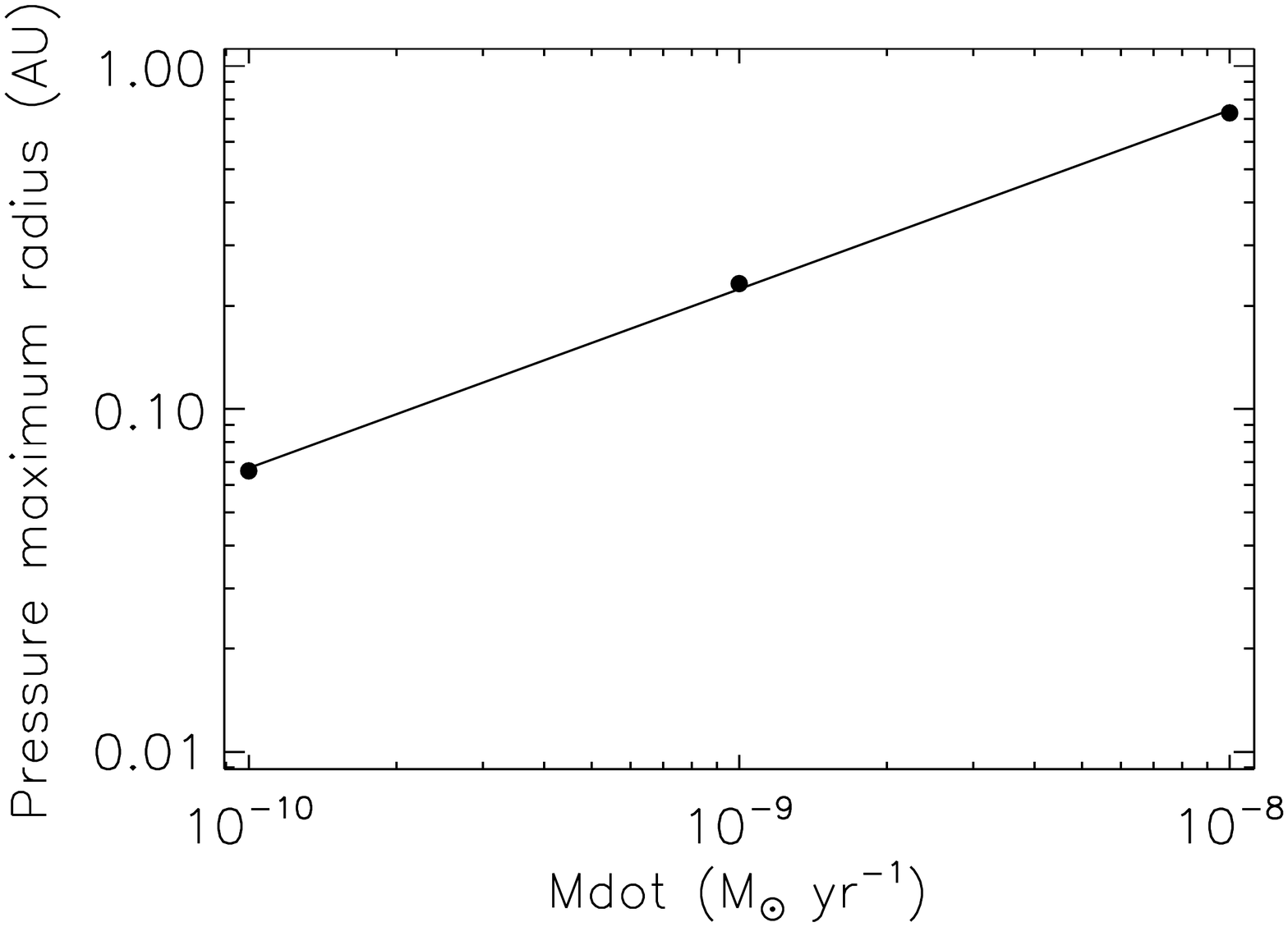}
\caption{Radial location of the pressure maximum as a function of the accretion rate \mdot, showing the approximate power law dependence $r_{P_{\rm max}} \propto {\dot{M}}^{1/2}$.
}
\label{fig:fig25}
\centering
\end{figure}

First (Figs.\,21 and 24), the pressure maximum is pushed out to $\sim$0.7\,AU when \mdot\,= $10^{-8}$\,\msun\,yr$^{-1}$, and in to $\sim$0.07\,AU when \mdot\,= $10^{-10}$\,\msun\,yr$^{-1}$, compared to $\sim$0.25 AU for the fiducial accretion rate. An increase (decrease) in ionisation fraction yields a higher (lower) $\bar\alpha$ at a fixed radius, so the pressure maximum (achieved where $\bar\alpha$ falls to $\adz$) occurs at a larger (smaller) radius for a given $\adz$. As Fig.\,25 shows, the radial location of the pressure maximum as a function of the accretion rate is approximately a power law: $r_{P_{\rm max}} \propto$ \mdot$^{1/2}$.

Second (Fig.\,21), for the higher \mdot\,= $10^{-8}$\,\msun\,yr$^{-1}$, the inner edge of the dead zone recedes to a larger radius ($\sim$0.4\,AU, versus 0.09\,AU for the fiducial case). The inner edge of the Hall zone is pushed out as well, but not as much, resulting in this zone now intruding on the MRI-active zone. For the lower \mdot\,= $10^{-10}$\,\msun\,yr$^{-1}$, on the other hand, the dead zone extends all the way to the disk inner edge; the active zone only occurs sandwiched between the dead and zombie zones, and never extends to the midplane. Interestingly, the very low ionisation fractions in this solution also allow the appearance of a region where Ohmic resistivity $\eta_O$ dominates over both $\eta_H$ and $\eta_A$ (red sliver at the disk outer edge in Fig.\,22; the only time such a region appears in our solutions).      

Third (Fig.\,23, right panel), $\bar\alpha$ saturates at $\sim$0.1 at small radii as the accretion rate increases to $\gtrsim$ $10^{-9}$\,\msun\,yr$^{-1}$ (left panel of Fig.\,22). The saturation occurs because, at these \mdot, potassium is almost completely ionized at small radii over the entire vertical extent of the disk (e.g., see top left panel of Fig.\,4, which shows that, near the disk inner edge in the fiducial case, $x_e \approx$ 1--2$\times$10$^{-7}$ from the midplane to the disk surface: very close to the maximum possible value of $x_e$ in our disks, equal to the abundance of K, of $\sim$2$\times$10$^{-7}$; for $10^{-8}$\,\msun\,yr$^{-1}$, $x_e$ (not shown) is even closer to this upper limit at small radii). This explains why we found, in \S8.1.6, that the innermost regions of our fiducial disk are viscously {\it stable}: this instability is mainly controlled by the change in $\bar\alpha$ with \mdot\,(see equation (32) and Fig.\,14), and this change is by definition very small when $\bar\alpha$ is close to saturation. Note as well that $\bar\alpha$ is saturated out to a much larger radius for $10^{-8}$\,\msun\,yr$^{-1}$ compared to the fiducial case (because the ionisation fraction grows with \mdot), implying that the inner disk becomes viscously stable over an increasing radial extent as the accretion rate climbs.   

\subsection{Variations in ${\boldsymbol M}_{\boldsymbol \ast}$}
Figs.\,26--28 show our disk solutions for the same \mdot\,and $\adz$ as the fiducial case, but with \mstar\, = 0.1\,\msun\,(instead of 1\,\msun). Note that the inner edge of the disk, assumed to lie at the stellar surface in our calculations, is also smaller in this case ($R_{in} = R_{\ast} \approx 1$\,\rsun, compared to $\sim$2.3\,\rsun\,for the fiducial mass).   

We see that the solutions for the two different stellar masses are nearly identical, except that the solutions for the lower mass are compressed radially inwards by a $\sim$constant multiplicative factor (i.e., shifted inwards by a constant additive factor, on the logarithmic radial scale in the plots). This is explained by the functional form of the fundamental parameters $\rho$, $P_{\rm gas}$ and $T$ in a steady-state $\alpha$-disk (equations (4-6)). Specifically, the dependence of each of these parameters on the stellar mass \mstar\,and orbital radius $r$ can be expressed as a dependence on the combined parameter \mstar/$r^3$ (the additional dependence on $r$ via $f_r$ is negligible for $r \gg R_{in}$). Dependencies on $\bar\alpha$ and the opacity $\kappa$ do not change this fact, since the latter quantities are themselves functions of $\rho$, $P_{\rm gas}$ and $T$. As such, for a fixed accretion rate, the solution at any radius $r_a$, for a stellar mass $M_{{\ast}a}$, is identical to that at radius $r_b \equiv r_a (M_{{\ast}b} / M_{{\ast}a})^{1/3}$ for a stellar mass $M_{{\ast}b}$.   

\begin{figure}
\centering
\includegraphics[width = \columnwidth, trim={2cm 2cm 2cm 2cm}, clip]{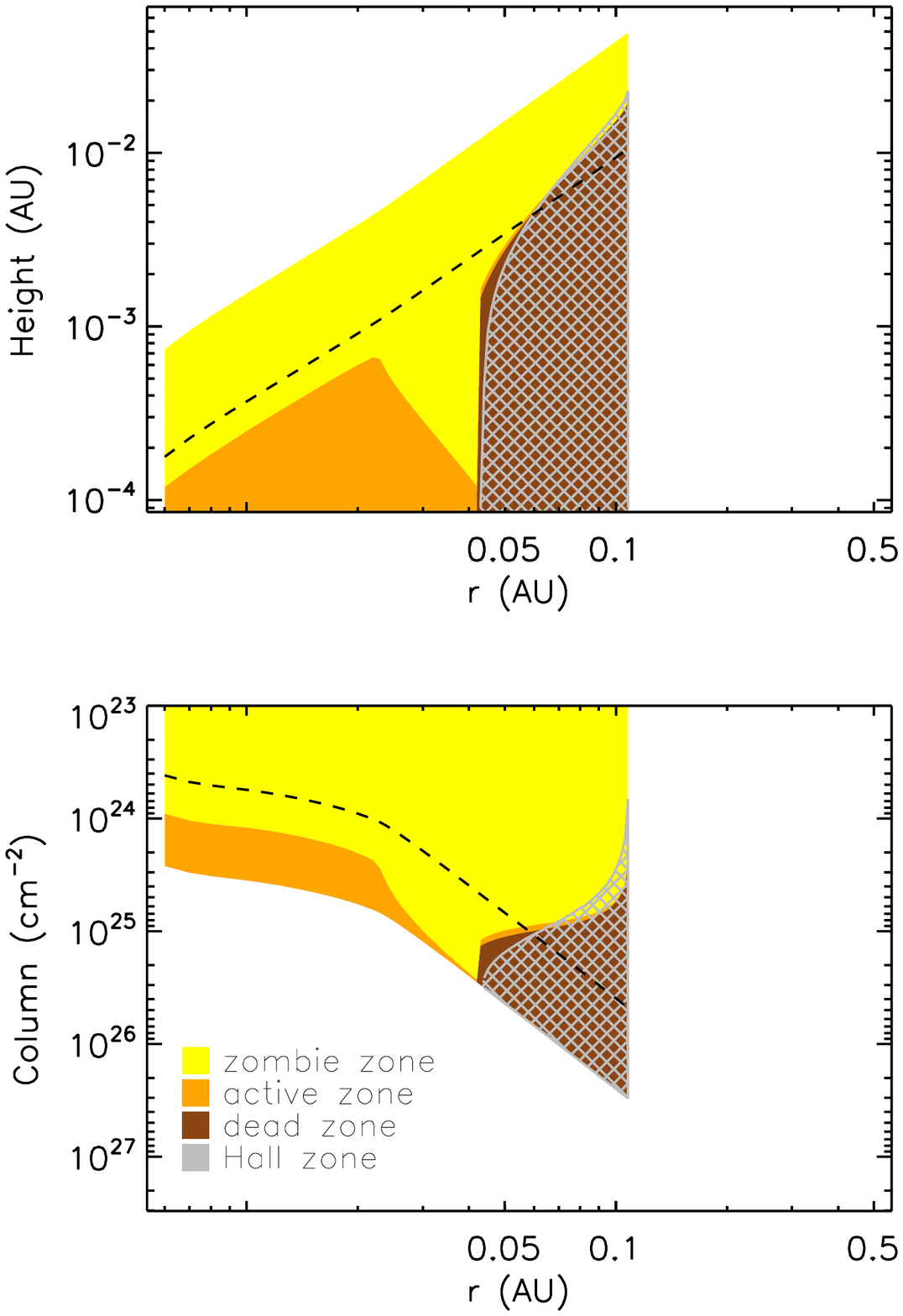} 
\caption{Same as Fig.\,5, but for now for a stellar mass \mstar\,= 0.1\,\msun.
}
\label{fig:fig26}
\centering
\end{figure}

\begin{figure*}
\centering
\includegraphics[width = \columnwidth, trim={2cm 1cm 2cm 2cm}, clip]{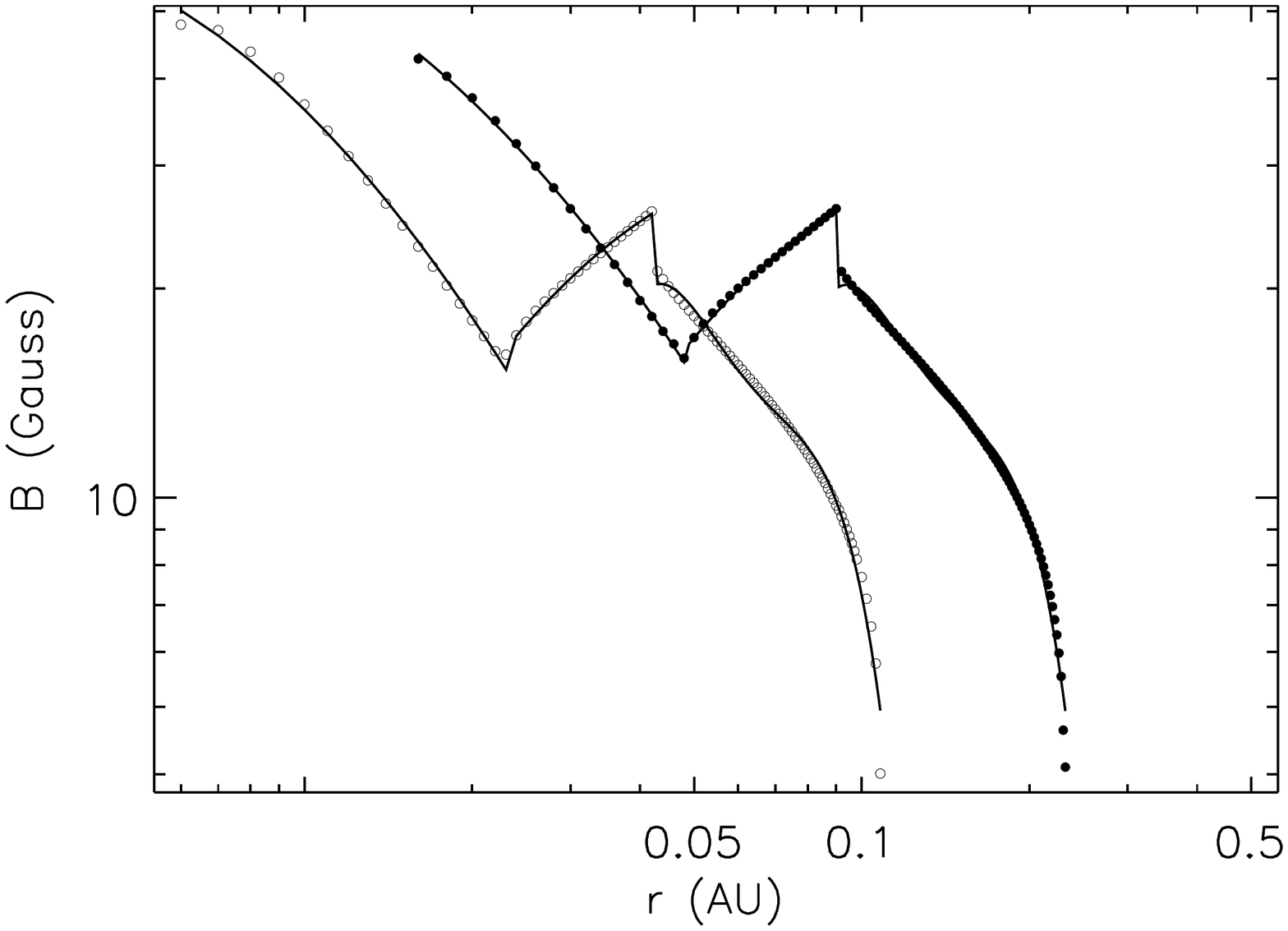}
\includegraphics[width = \columnwidth, trim={2cm 1cm 2cm 2cm}, clip]{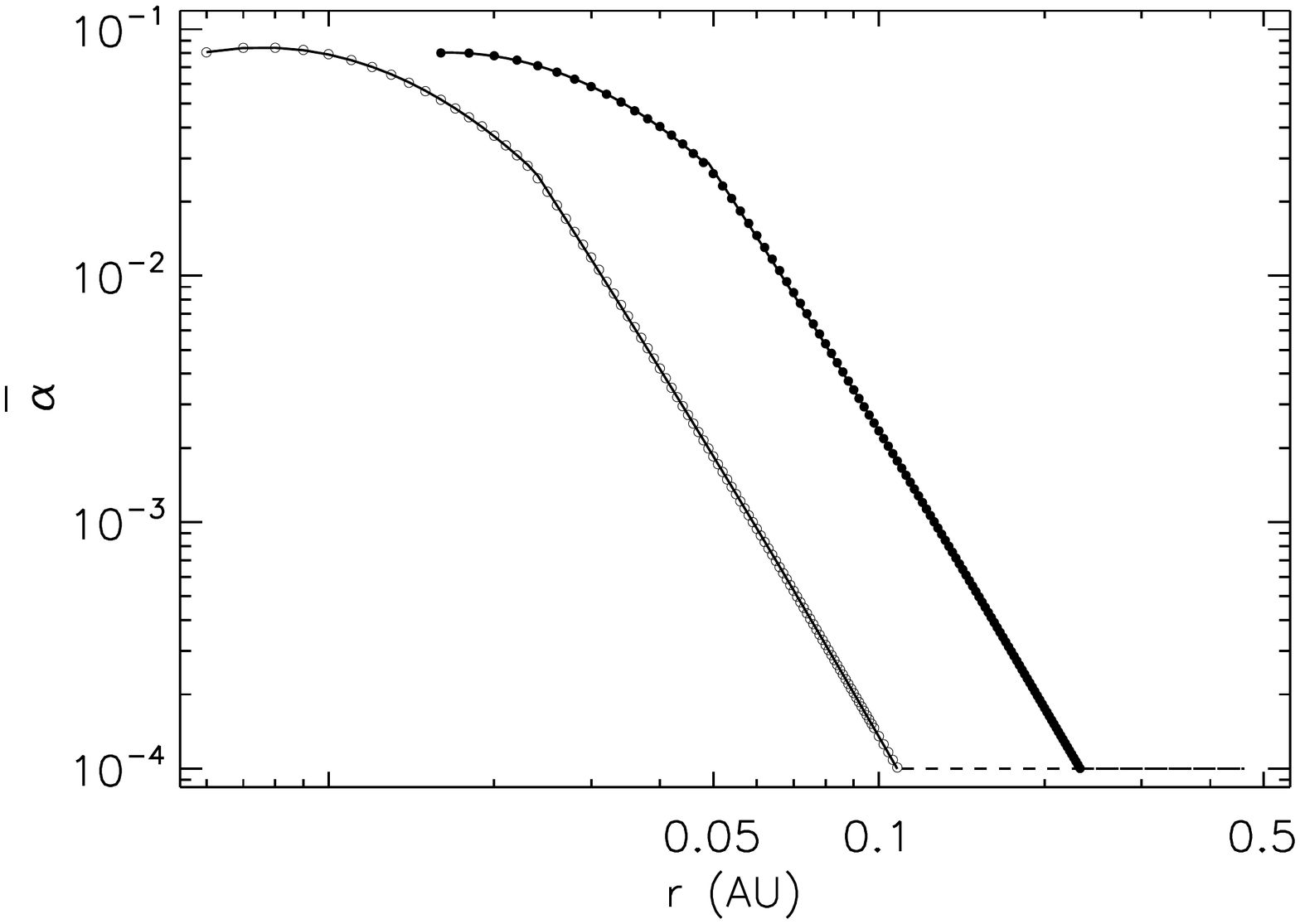} 
\caption{{\it Left}: Field strength $B$ as a function of radius. {\it Right}: $\bar\alpha$ as a function of radius. In both plots, results for \mstar\,= 0.1\,\msun ({\it empty circles}) are overplotted on the results for our fiducial model with \mstar\,= 1\,\msun ({\it filled black circles}; the fiducial results are the same ones shown in Figs.\,6 and 8 respectively).
}
\label{fig:fig27}
\centering
\end{figure*}

\begin{figure}
\centering
\includegraphics[width = \columnwidth, trim={4cm 0cm 1cm 1cm}, clip]{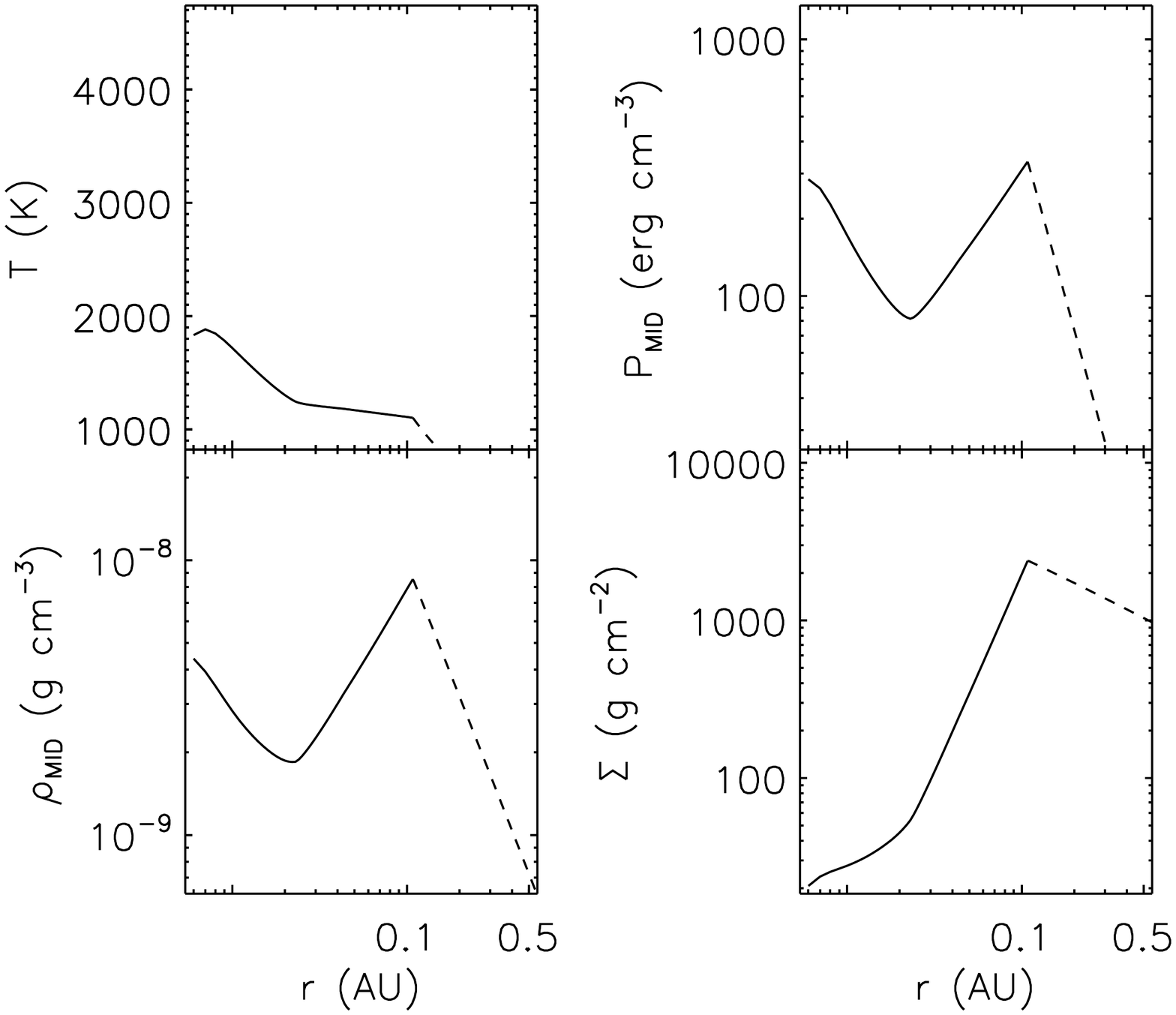}
\caption{Various disk parameters as a function of radius: same as Fig.\,9, but now for a stellar mass  \mstar\,= 0.1\,\msun.
}
\label{fig:fig28}
\centering
\end{figure}

\section{Discussion and Conclusions}
The Inside-Out Planet Formation (IOPF) mechanism depends upon the presence of a midplane pressure maximum, arising initially from the change in viscosity between the MRI-active innermost disk and the adjacent dead zone. We have investigated the formation and location of this first pressure maximum, by solving the coupled equations for MRI-driven viscosity with thermal ionization and an $\alpha$-disk structure in steady-state. We examine a range of disk accretion rates (\,$10^{-10}$--$10^{-8}$\,\msun\,yr$^{-1}$) and stellar masses (0.1--1\,\msun). Within the dead and zombie zones, where the viscosity comes from non-MRI hydrodynamic and/or gravitational stresses, we assume a constant viscosity parameter $\adz$ (which also sets a ``floor'' on the MRI-driven $\alpha$), set to a fiducial value in the range 10$^{-3}$--10$^{-5}$. We find that:

\noindent {\it (1)} A midplane pressure maximum does form, but it is located {\it within} the dead zone, rather than at the dead zone inner boundary (DZIB) as usually assumed. This is a general consequence of two factors: first, the midplane pressure does not depend on the local value of $\alpha$, but rather on its vertically averaged effective value $\bar\alpha$; second, the MRI-active zone does not end abruptly at the DZIB, but instead continues outwards above the dead zone, so that $\bar\alpha$ falls to its minimum value $\adz$ (causing a pressure maximum) {\it beyond} the DZIB. 

\noindent {\it (2)} The radial location of the pressure maximum has approximately power law dependencies on $\alpha_{\rm DZ}$, stellar mass and accretion rate: $r_{P_{\rm max}} \propto \alpha_{\rm DZ}^{-1/4}$, \mstar$^{1/3}$ and \mdot$^{1/2}$.
 
\noindent {\it (3)}  Inward of the pressure maximum, the surface density $\Sigma$ in our steady-state solutions {\it decreases} radially inwards, instead of increasing monotonically as usually assumed (e.g., in the Minimum Mass Solar Nebula (MMSN)). This is a general feature of all solutions with a constant accretion rate and an $\alpha$ that increases inwards (since a lower $\Sigma$ is required to produce the same \mdot\,with a larger $\alpha$). The very low $\Sigma$ that results in these inner disk regions has two consequences, as follow (points 4 and 5 below): 

\noindent {\it (4)} At these low $\Sigma$, Hall diffusion rather than Ohmic resitivity dominates near the midplane. Specifically, for the range of \mstar, \mdot\,and $\alpha_{\rm DZ}$ considered here, the Hall Elsasser number $\chi < 1$ within the Ohmic dead zone in our solutions. As such, in the presence of a net vertical background field aligned with the disk spin axis, the Hall effect can reactivate the dead zone, thereby removing the pressure maximum and suppressing the IOPF mechanism. This might explain why close-in small planets are {\it not} found in roughly half of all systems: any background stellar or interstellar field threading the disk will be randomly aligned / anti-aligned with the disk spin axis, yielding alignment in statistically half the cases.       

\noindent {\it (5)} At these low $\Sigma$, X-ray ionization can become competitive with thermal ionization, contrary to the standard assumption the X-rays may be ignored here. In our analysis, where X-rays are not included (their effects are only investigated a posteriori), the MRI-active zone eventually ends where thermal ionization peters out, and $\bar\alpha$ falls to the floor value $\alpha_{\rm DZ}$ (forming a pressure maximum there). In real disks, we expect the MRI-active layer above the dead zone will eventually become X-ray-supported, and thus continue outwards to join up with the active layer in the midplane beyond the outer boundary of the dead zone (e.g., see disk solutions with X-ray-driven MRI by Mohanty et al.\,2013). In this case, the minimum value of $\bar\alpha$ will be somewhat higher than $\alpha_{\rm DZ}$ (since the disk never becomes completely dead vertically); what this precise value is, and where it is achieved (and thus a pressure barrier is formed), will be X-ray dependent.   

\noindent {\it (6)} A linear stability analysis of our equilibrium disk solutions indicates that most of the inner disk is viscously unstable ($\partial\Sigma / \partial\dot{M} < 0$), with the inner edge of this unstable region moving outwards with increasing \mdot. The instability is driven primarily by the change in $\bar\alpha$ (due to variations in the ambipolar and Ohmic diffusivities) as a function of \mdot. To zeroeth order, this instability will cause the inner disk to break up into rings. A more detailed non-linear analysis, together with the inclusion of more realistic disk physics (i.e., inclusion of grain effects on the MRI, and a more rigorous treatment of the disk thermal structure, ionization incuding X-rays, and opacities) is required to verify the presence of the viscous instability.

\acknowledgements 
We thank Sourav Chatterjee, Xiao Hu, Zhaohuan Zhu, Mordecai-Mark Mac Low and Patrick Hennebelle for very helpful discussions. S.M. and J.T. acknowledge support from a Royal Society International Exchange grant IE131607 and from NASA ATP grant NNX15AK20G (JCT). M.R.J. acknowledges support from the Imperial College PhD Scholarship and the Dositeja stipend from the Fund for Young Talents of the Serbian Ministry for Youth and Sport.

\clearpage
\appendix
\section{A. Discussion of Conditions for Active MRI}
Our treatment of the conditions for active MRI generally follows that of \citet{mohanty13}; we summarize the salient points here. Magnetic torques are important for mass and angular momentum transport in the disk only if the gas is sufficiently coupled to the field, i.e., if gas motions can generate magnetic stresses faster than they can diffuse away due to a finite resitivity $\eta$. In a Keplerian disk, these stresses arise due to orbital shear, so the relevant timescale for field generation is the orbital period, i.e., the dynamical timescale, $t_{\rm dyn} \sim 1/\Omega$. 

For local tangled fields generated by MRI-induced turbulence, the height of the thin disk sets an upper limit on the wavelength of MRI modes, and hence on the dissipation timescale, so it is the vertical direction that is relevant. For a vertical mode with wavenember $k$, the Ohmic dissipation rate is $\sim k^2\eta_O$, while the growth rate is $k v_{{\mathcal A}z}$, where \vaz\,is the vertical component of the local Alfv\'{e}n velocity ($v_{{\mathcal A}z} \equiv B_z/\sqrt{4\pi \rho}$, for a local vertical field strength $B_z$ and gas density $\rho$ ; MRI simulations by \citet{sano04} indicate that $B_z^2 \sim B^2/25$, where $B$ is the r.m.s.\,field strength). Since the maximum growth rate is $\Omega$, the wavenumber of the fastest growing mode is $k =  \Omega/v_{{\mathcal A}z}$. Stipulating that the growth rate of this mode exceed its dissipation rate then yields the Ohmic Elsasser number criterion for active MRI:
$$ \Lambda \equiv \frac{v_{{\mathcal A}z}^2}{\eta_O \Omega} > 1 \eqno({\rm A}1) $$
whether the net background field is vertical, toroidal or zero \citep{sano02}.  Moreover, we will see below that efficient MRI additionally requires the gas pressure in the disk (\pgas) to substantially exceed the magnetic pressure (\pmag). For gas with sound speed \cs, \pgas\, ($\propto$\cssq) $\gg$ \pmag\, ($\propto$\vasq) implies \cs/\va\, $\gg$ 1, guaranteeing that the wavelength $\sim$\vaz/$\Omega$ ($<$\va/$\Omega$) of the fastest growing mode is indeed much smaller than the disk scale height $z_H \sim$ \cs/$\Omega$.    

The rationale for the Ohmic Elsasser criterion may be understood more clearly by considering the general induction equation for magnetic fields \citep[e.g.,][]{wardle07}:
$$ \frac{\partial {\mathbf B}}{\partial t}  = \nabla\times\left({\mathbf v}\times{\mathbf B}\right) - \nabla\times\left[\eta_O\left(\nabla\times{\mathbf B}\right) + \eta_H\left(\nabla\times{\mathbf B}\right)\times{\mathbf {\hat{B}}} + \eta_A\left(\nabla\times{\mathbf B}\right)_{\perp}\right] \eqno({\rm A}2) $$
where $\mathbf v$ and $\mathbf B$ are the neutral velocity and magnetic field vectors respectively, ``\^{}'' denotes a unit vector, and ``$\perp$'' indicates the component of a vector perpendicular to $\mathbf B$. The first expression on the right is the inductive term ($I$), while the second, third and fourth represent Ohmic ($O$), Hall ($H$) and ambipolar ($A$) diffusion respectively. Clearly, in magnitude, $I \sim \Omega B$ while $O \sim \eta_O B/L^2$; moreover, as noted above, $L \sim v_{{\mathcal A}z}/\Omega$ is the characteristic length scale of the fastest growing MRI mode (and thus the relevant scale for field diffusion too). Thus $\Lambda \sim I/O$, and equation (A1) simply expresses the intuitive notion that, for robust MRI-driven field amplification when Ohmic resistivity is the prime diffusive channel, the ratio of the inductive to the Ohmic term must exceed unity. A region where Ohmic diffusion kills the MRI, and thus equation (A1) is {\it not} satisfied (i.e., $\Lambda < 1$), is called a {\it Dead Zone}.

Analogously, when the Hall term dominates the diffusivities on the RHS of the induction equation, we expect it to strongly affect the MRI when the Hall Elsasser number $\chi$ satisfies 
$$ \chi \equiv \frac{v_{\mathcal A}^2}{|\eta_H| \Omega} < 1 \,\, . \eqno({\rm A}3) $$
We call a region satisfying equation (A3) the {\it Hall Zone}. The {\it nature} of the Hall term's effect on the MRI, however, is very different from that of Ohmic resistivity: in the presence of a net background vertical field threading the disk, the non-dissipative character of Hall diffusion implies that it may amplify or suppress the MRI depending on whether the field is aligned or anti-aligned with the rotation axis of the disk \citep[i.e., whether $\mathbf {B\cdot\Omega} > 0$ or $< 0$; this behaviour can be understood by noticing that flipping the direction of $\mathbf B$ changes the sign of every term in the induction equation, equation (A2), {\it except} the Hall one; e.g.,][]{wardle99, balbus01}. These issues have been explored in a linear analysis by \citet{wardle12}, and in various more recent non-linear simulations (see discussion of simulation results in \S2); \citet{xu16} have also explored similar Hall diffusion effects when, in the presence of a net field, grains cause a flip in the sign of \etah\,that mimics a reversal in field polarity. 

Including the quantitative effect of Hall diffusion on the MRI is thus non-trivial, and beyond the scope of this exploratory paper; as such, {\it we ignore it} here. We do calculate the magnitude of all three resistivities (\etao, \etah\, and \etaa), and show their relative strengths over our region of interest; however, in Hall-dominated areas, we use either the Ohmic Elsasser criterion (equation (A1)) or the ambipolar condition (discussed below; equations (A4,A5)) to evaluate the MRI efficiency, depending on whether \etao\, or \etaa\, is the next strongest resistivity. 

For ambipolar diffusion, the Elsasser number $Am$ is again defined analogously to the Elsasser number $\Lambda$ for Ohmic diffusion, but with \etaa\, replacing \etao\footnote{We use \va\,here instead of the \vaz\,employed in the Ohmic Elsasser number definition in equation (A1), since we will adopt (in equations (A5a,b)) the results of the numerical simulations by \cite{bai11a}, who use the total Alfv\'{e}n velocity to define $Am$.} :
$$ Am \equiv \frac{v_{\mathcal A}^2}{\eta_A \Omega} \,\, . \eqno({\rm A}4) $$ 
Note that $Am$ is independent of the field strength $B$, since \vasq\, $\propto B^2$ and so is \etaa\, (see \S4.3)\footnote{A typo in the text of \citet{mohanty13} suggests that $Am$ depends on $B$ through both \etaa\, and \va$^2$; while this is formally true, the two dependencies in fact cancel out. This does not vitiate any results in \citet{mohanty13}, since their actual calculations of $Am$ are correctly implemented.}.  \citet{wardle99} argued that the appropriate criterion for efficient MRI, when ambipolar diffusion dominates instead of Ohmic, should in fact mirror equation (A1), i.e., $Am > 1$. When electrons and ions are the only charged species, the latter condition reduces to $\gamma_i \rho_i / \Omega > 1$ (see \S4.3), where $\gamma_i$ is the neutral-ion collisional drag coefficient and $\rho_i$ is the ion density. This implies that the MRI can flourish in the presence of ambipolar diffusion (i.e., the field, to which the ions and electrons are frozen, will be sufficiently coupled to the mainly neutral fluid) only if a neutral particle collides at least once per orbit with an ion. This condition has often been used to investigate ambipolar dominated disk regions \citep[e.g.,][]{turner10}. 

On the other hand, \citet{hawley98} suggested that the above criterion is too lenient. Their 3D local shearing-box simulations, using an idealized two-fluid approximation (ions + neutrals; ionization and recombination are not considered, so ion and neutral numbers are individually conserved), indicated that efficient MRI with ambipolar diffusion requires neutral-ion collisions to be at least a hundred times more frequent, i.e., $\gamma_i \rho_i / \Omega \gtrsim 100$. 

However, more recently, \citet{bai11a} have argued that it is not the two-fluid approximation but the strong-coupling limit that is most applicable to protoplanetary disks. This limit holds when two criteria are met: {\it (a)} the neutral density vastly exceeds the ion density: $\rho_n \gg \rho_i$, a condition invariably satisfied in these disks; and {\it (b)} the recombination timescale is much shorter than the orbital period (dynamical timescale): $t_{\rm rcb} \ll t_{\rm dyn}\, (\sim 1/\Omega)$, which \citet{bai11b} demonstrates is true over most of the disk as well (see discussion below). In this case, the ion inertia may be neglected, the ion density is controlled entirely by ionization-recombination equilibrium with the neutrals, and the problem reduces to a single-fluid (of neutrals) approximation. In this strongly-coupled limit, with the ratio of the inductive to ambipolar term further given by the general expression for $Am$ in equation (A3) (instead of just the reduced value $\gamma_i \rho_i / \Omega$), \cite{bai11a} find that the MRI can operate at {\it any} value of $Am$, {\it provided} the field is sufficiently weak. Specifically, the MRI can be sustained as long as the plasma $\beta$-parameter, $\beta \equiv$ \pgas/\pmag, satisfies  
$$ \beta > \beta_{\rm min} \eqno({\rm A}5{\rm a}) $$
where the minimum value of $\beta$ is a function of $Am$:
$$ \beta_{\rm min}(Am) = \left[ \left(\frac{50}{Am^{1.2}}\right)^2 + \left(\frac{8}{Am^{0.3}} + 1\right)^2\right]^{1/2} , \eqno({\rm A}5{\rm b})  $$
and $P_{\rm B} = B^2/8\pi$. Note from (A5b) that \bmin\, approaches (50/$Am^{1.2}$) for $Am \lesssim 1$, and asymptotes to 1 from above as $Am \rightarrow \infty$. Hence condition (A5a) for active MRI demands that the gas pressure dominate over the magnetic pressure in the disk, as stated earlier. Following \citet{mohanty13}, we denote locations where equations (A5a,b) are {\it not} satisfied (i.e., $\beta <\beta_{\rm min}$), and thus the MRI is quenched by ambipolar diffusion, as {\it Zombie Zones}. 

\citet{bai11b} shows that $t_{\rm rcb} \ll t_{\rm dyn}$, i.e., the strong-coupling limit applies, when complex chemical networks are invoked and grains are abundant; he cautions that this limit may not hold in simpler formulations, as recombination pathways become limited. This warning is potentially germane to us, since, in our simplified treatment of thermal ionization here, the chemical network comprises only one channel (M $\rightleftharpoons$ M$^+$ + $e^-$, where M is a single species of alkali metal), and grains are moreover omitted. We proceed by first {\it assuming} that the strong-coupling limit holds for us as well, and thus use the conditions (A5a,b) to determine whether the MRI can operate in ambipolar-dominated regions; we then {\it check} whether $t_{\rm rcb} \ll t_{\rm dyn}$ holds in these regions in the disk solutions derived, to verify consistency (see discussion and equation (15), in \S4.2, and detailed discussion in \S8.1.8.). 



\section{B. Connecting the MRI and $\alpha$-Disk Formulations}
For a general shear stress $T_{r\phi}$ in the disk, the viscosity parameter $\alpha_T$ is defined by the relation
$$ T_{r\phi} \left(\equiv \nu \rho r \frac{d\Omega}{dr}\right) = -\alpha_T P_{\rm gas} = -\alpha_T \rho c_s^2 \eqno({\rm B}1) $$ 
where the first equivalence in parantheses is the definition of $T_{r\phi}$, and $\nu$ is the viscosity. The $-$ve sign on the R.H.S. terms enforces the convention that the $\alpha$-parameter be positive (since $T_{r\phi} \propto d\Omega/dr < 0$ in a Keplerian disk). We have labelled the $\alpha$-parameter with the subscript `$T$' to explicitly denote that it is defined here in terms of $T_{r\phi}$, instead of in terms of the viscosity $\nu$ as done in the Shakura-Sunyaev model (equation (1) in the main text; we connect the two definitions further below). For any particular driver of shear stress (e.g., the MRI), we will find it mathematically convenient to define an {\it effective} viscosity parameter $\bar{\alpha}_T$, given by the pressure-weighted vertical average of $\alpha_T$:
$$ \bar{\alpha}_T \equiv  \frac{\int_{2h} \alpha_T P_{\rm gas}\, dz}{\int_{2h} P_{\rm gas}\, dz} = \frac{\int_{2h} \alpha_T \rho\, dz}{\int_{2h} \rho\, dz} \eqno({\rm B}2) $$
where the integrals are over the total thickness of the layer (summed over both sides of the midplane) where the specified shear stress operates. The second equality above, which defines $\bar{\alpha}_T$ as a density-weighted vertical average, holds when the disk is vertically isothermal (so that $c_s$ is constant with height). Using equation (B1), this yields the useful form
$$ \int_{2h} T_{r\phi}\, dz = -\bar{\alpha}_T \int_{2h} P_{\rm gas}\, dz = -\bar{\alpha}_T c_s^2 \int_{2h} \rho\, dz \eqno({\rm B}3) $$   
where the first equality is general and the second again true for a vertically isothermal disk. 
For the specific case of MRI-driven turbulence, the vertical integral of the turbulent stress is given by \citep[e.g.,][]{wardle07}   
$$ \int_{2h} T_{r\phi {\rm , MRI}}\,dz = -\frac{h}{2\pi}\langle -B_r B_{\phi} \rangle \approx -\frac{hB^2}{8\pi} \eqno({\rm B}4) $$
where $B_r$ and $B_{\phi}$ are the radial and azimuthal components of the field respectively, $\langle -B_r B_{\phi}\rangle \equiv -(2h)^{-1} \int_{2h} B_r B_{\phi} dz$, and $B$ is the r.m.s.\,field strength. The second equality flows from the result of MRI simulations by \citet{sano04} that $B^2 \sim 4 \langle -B_r B_{\phi} \rangle$. Replacing the vertical integral on the left by using the first equality in equation (B3), noting that the vertical average of $P_{\rm gas}$ over the MRI layer is $\langle P_{\rm gas} \rangle \equiv \left(\int_{2h} P_{\rm gas}\, dz \right)/2h$, recognizing that $P_{\rm B} = B^2/8\pi$ is the magnetic pressure, and using the definition of the plasma beta parameter $\beta \equiv P_{\rm gas}/P_{\rm B}$, we finally arrive at
$$ \bar{\alpha}_{T_{\rm MRI}} \approx \frac{1}{2\langle \beta \rangle} \eqno({\rm B}5) $$
for MRI-driven turbulent stresses (as denoted by the subscript on $\bar{\alpha}_T$ on the L.H.S.). Here $\langle \beta \rangle \equiv \langle P_{\rm gas} / P_{\rm B} \rangle = \langle P_{\rm gas} \rangle / P_{\rm B}$ is the vertical average of the plasma beta parameter over the active layer thickness $2h$ (the second equality comes from our assumption that the field strength is constant over this thickness). \citet{bai11a} also arrive at equation (B5); it is essentially a restatement of the assertion above that $B^2 \sim 4 \langle -B_r B_{\phi} \rangle$, as they discuss.   

Now, at any radial location in the disk, we expect the vertical structure to be multi-layered, with the most general structure being a dead zone (where Ohmic resistivity suppresses the MRI) straddling the midplane, a zombie zone (where ambipolar diffusion shuts off the MRI) near the disk top and bottom surfaces, and an MRI-active layer sandwiched in between. The shear stress within the MRI-active and inactive layers is driven by different physical mechanisms, and hence $\bar{\alpha}_T$ within these layers will be (very) different. It is therefore convenient, in analogy with equation (B2) for the individual disk layers, to define an average viscosity parameter $\bar{\alpha}_{T_{\rm avg}}$ over the {\it entire} thickness of the disk:
$$ \bar{\alpha}_{T_{\rm avg}} \equiv \frac{\int\limits_{-\infty}^{+\infty} \alpha_T P_{\rm gas}\, dz}{\int\limits_{-\infty}^{+\infty} P_{\rm gas}\, dz} = \frac{\int\limits_{-\infty}^{+\infty} \alpha_T \rho\, dz}{\int\limits_{-\infty}^{+\infty} \rho\, dz}\,\, . \eqno({\rm B}6) $$
For a vertically isothermal disk, which is assumed in this paper (and where the second equality above applies), $\bar{\alpha}_{T_{\rm avg}}$ can be put in a very simple form by noting that
$$ \int\limits_{-\infty}^{+\infty} T_{r\phi}\, dz =  -\bar{\alpha}_{T_{\rm avg}} c_s^2 \int\limits_{-\infty}^{+\infty} \rho\, dz = \sum_i \left( \int_{2h_i} T_{r\phi,i}\, dz \right)  \eqno({\rm B}7) $$
where the first equality comes from combining equations (B1) and (B6). The second equality simply breaks up the vertical integral over the total disk thickness into a sum of integrals over zones with different shear-stress mechanisms; $2h_i$ and $T_{r\phi,i}$ denote respectively the thickness of the $i$-th zone (summed over both sides of the midplane) and the form of the shear stress tensor there. Using the second equality in equation (B3) to replace the individual integrals under the summation above, and dividing throughout by the mean molecular mass $\mu$, we get
$$ \bar{\alpha}_{T_{\rm avg}} = \frac{\sum_i \left( N_i\,\bar{\alpha}_{T_i}\right)}{N_{\rm tot}} \eqno({\rm B}8) $$        
where $N_i \equiv \int_{h_i} \rho/\mu\,dz$ is the (one-sided) column density of each $i$-th zone, and $N_{\rm tot} \equiv \int_0^{+\infty}\rho/\mu\,dz = \sum_i N_i$ is the (one-sided) total column density from the surface to the midplane (we assume the disk is symmetric about the midplane). Thus, for a vertically isothermal disk, $\bar{\alpha}_{T_{\rm avg}}$ is the {\it column-weighted vertical average} of the effective viscosity parameters $\bar{\alpha}_{T_i}$ within each zone (MRI-active, dead and zombie; we will denote these zones respectively by $i$ = MRI, DZ and ZZ). 

Now, the parameter $\alpha$ that is used to derive the Shakura-Sunyaev disk equations is defined in terms of the viscosity $\nu$ (equation (1) in the main text), while $\alpha_T$ is defined in terms of the shear stress $T_{r\phi}$ (equation (B1)). Combining these two equations with the definition $ T_{r\phi} \equiv \nu \rho r\, d\Omega/dr$, we see that
$$ \alpha = \frac{2}{3}\, \alpha_T \eqno({\rm B}9) $$
where the factor of 2/3 comes from $d\Omega/dr$ in a Keplerian disk.  

Moreover, the Shakura-Sunyaev equations ((3)--(6) in the main text) are derived on the basis of vertically integrated quantities ($\Sigma$ and \mdot; see H16). As such, it is not $\alpha$ that enters directly into these equations, but more precisely the {\it effective} parameter $\bar\alpha$, which is a vertical average over the entire disk thickness  defined analogously to equation (B6): 
$$ {\bar{\alpha}} \equiv \frac{\int\limits_{-\infty}^{+\infty} \alpha P_{\rm gas}\, dz}{\int\limits_{-\infty}^{+\infty} P_{\rm gas}\, dz} = \frac{\int\limits_{-\infty}^{+\infty} \alpha \rho\, dz}{\int\limits_{-\infty}^{+\infty} \rho\, dz}\,\, . \eqno({\rm B}10) $$
Combining this with equations (B6), (B9) and (B8) yields
$$ {\bar{\alpha}} = \frac{2}{3}\, \bar\alpha_{T_{\rm avg}} =  \frac{{\frac{2}{3}}\sum_i \left( N_i\,\bar{\alpha}_{T_i}\right)}{N_{\rm tot}} \eqno({\rm B}11) $$  
where the last equality holds for the vertically isothermally case. Note that $\bar{\alpha}_{T_{\rm MRI}} \approx 1/(2\langle \beta \rangle)$ by equation (B5). In the dead and zombie zones, the effective parameters $\bar{\alpha}_{T_{\rm DZ}}$ and $\bar{\alpha}_{T_{\rm ZZ}}$ are set by hydrodynamic and/or gravitational instabilities, and we set their values guided by the results of numerical simulations (see below). Furthermore, without detailed simulations of how the viscosity in the dead and zombie zones might differ, we assume that the effective viscosity parameters are the same in both zones: $\bar{\alpha}_{T_{\rm DZ}} = \bar{\alpha}_{T_{\rm ZZ}}$. Then, for the vertically isothermal conditions that we adopt, we may write
$$ {\bar{\alpha}} =  \frac{N_{\rm MRI}\,\bar{\alpha}_{\rm MRI} + (N_{\rm DZ} + N_{\rm ZZ})\bar{\alpha}_{\rm DZ} }{N_{\rm tot}} \eqno({\rm B}12) $$ 
where $\bar{\alpha}_{\rm MRI} \equiv 2\bar{\alpha}_{T_{\rm MRI}}/3 \approx 1/(3\langle \beta \rangle)$ and $\bar{\alpha}_{\rm DZ}\, (= \bar{\alpha}_{\rm ZZ}) \equiv 2\bar{\alpha}_{T_{\rm DZ}}/3$. Based on simulations (e.g., Dzyurkevich et al.\,2010; Dzyurkevich et al.\,2013 and references therein; Malygin et al.\,2017 and references therein), we adopt a fiducial value of $\bar{\alpha}_{\rm DZ} = 10^{-3}$ or 10$^{-4}$ or 10$^{-5}$.         

Finally, the accretion rate (positive inwards) due to the shear stress within any $i$-th zone (MRI-active, dead, or zombie) is given by
$$ \dot{M}_i = -\frac{2}{r\Omega}\, \frac{\partial}{\partial r}\left[2\pi r^2\int_{2h_i} T_{r\phi,i} \,dz\right] \,\, . \eqno({\rm B}13) $$
For the vertically isothermal case, we can replace the stress-integral using the last equality in equation (B3), which yields
$$ \dot{M}_i = \frac{12\pi\,\mu\,m_{\rm H}}{r\Omega}\, \frac{\partial}{\partial r} \left(r^2 c_s^2\, N_i {\bar{\alpha}}_i\right) \eqno({\rm B}14) $$
where $i$ = MRI, DZ or ZZ, and $\bar{\alpha}_{\rm MRI}$ and $\bar{\alpha}_{\rm DZ}$ ($= \bar{\alpha}_{\rm ZZ}$ by assumption here) are defined above.  We use this formula to calculate the accretion rates within the individual disk zones. Note that, for the specific case of accretion within the MRI zone, we can combine equations (B13) and (B4) to alternatively write
$$ \dot{M}_{\rm MRI} \approx \frac{1}{2r\Omega}\, \frac{\partial}{\partial r} \left(r^2 h B^2\right)\,\, . \eqno({\rm B}15) $$
This shows explicitly how, given a disk structure and chemistry (ionization), the field strength $B$ controls the accretion rate through the MRI-active layers: directly via its appearance in the above formula, and indirectly by influencing the magnitudes of the Ohmic Elsasser number $\Lambda$ and the plasma $\beta$ parameter, which in tandem set the active layer thickness $h$.   

For a general shear stress $T_{r\phi}$, the total accretion rate at any radius is (analogous to equation (B13) but now integrated over the entire disk thickness)
$$ \dot{M} = -\frac{2}{v_K} \frac{\partial}{\partial r}\left[2\pi r^2\int\limits_{-\infty}^{+\infty} T_{r\phi} \,dz\right] \,\,. \eqno({\rm B}16) $$
If the total accretion rate is {\it radially constant} (as we shall demand for our equilibrium solutions), then, multiplying throughout by $v_K$ and integrating both sides over radius, from the disk inner edge $R_{in}$ out to any desired radius $r$, we get
$$ \dot{M} = -\frac{2\pi}{f_r\,\Omega} \int\limits_{-\infty}^{+\infty} T_{r\phi} \,dz \,\,. \eqno({\rm B}17) $$
The factor $f_r \equiv (1 - \sqrt{R_{in}/r})$ is the same one that appears in the Shakura-Sunyaev equations in \S3; it arises from the radial integral of $v_K$. Note that there is no equivalent contribution from the disk inner edge when radially integrating the $\partial/\partial r$ term on the R.H.S. of equation (B16), since $T_{r\phi} \propto d\Omega/dr = 0$ at the inner edge: in the $\alpha$-disk model, $R_{in}$ is by definition the location where the angular velocity $\Omega$ plateaus and turns over.  

Finally, combining the first equality in equations (B7) and (B11) to replace the vertical integral of the shear-stress in (B17) above, and using the definition of surface density $\Sigma \equiv \int_{-\infty}^{+\infty} \rho\, dz$, we arrive at
$$ \dot{M} = \frac{3\pi\,\bar{\alpha}\,c_s^2\,\Sigma}{f_r\,\Omega} \,\,, \eqno({\rm B}18) $$
the standard expression for a constant accretion rate in a vertically isothermal $\alpha$-disk model.

\section{C. Polynomial Fits to Solutions for $\bar{\alpha}(r)$ and $B(r)$}

We fit our $\log \bar{\alpha}(\log r)$ and $\log B(\log r)$ solutions with piece-wise polynomials over 1, 2 or 3 distinct intervals in radius $r$. The fits are of the form $y = c_0 + c_1x + c_2x^2 + ...$, where $y = \log \bar{\alpha}$ or $\log B$, $x = \log r$, and $c_n$ is the $n$-th polynomial coefficient. We list the radius intervals and polynomial coefficients for our various disk models in Tables \ref{tab:poly1} to \ref{tab:poly6}. The starting radius for the innermost interval for all models is $R_{in}$=$R_{\ast}$ (= 1\,\rsun\, for \mstar\, = 0.1\,\msun\, and 2.3\,\rsun\, for \mstar\, = 1\,\msun).  

\begin{table}[h!]
\caption{disk model: \mstar\,= 1\,\msun, \mdot\, = 10$^{-9}$\,\msun\,yr$^{-1}$, $\alpha_{\rm DZ} = 10^{-4}$.
\label{tab:poly1}}
\centering
 \begin{tabular}{|c c c c c c c|} 
 \hline
 Fitted function & Interval end radius (AU) & $c_0$ & $c_1$ & $c_2$ & $c_3$ & $c_4$ \\ [0.5ex] 
 \hline\hline
 $\log \bar{\alpha}(\log r)$ &	0.048 &      -8.5526396 &     -9.0739395 &     -3.1434803 &    -0.22451420 & \\
						   &	0.232 &      -6.6333076 &     -4.3837696 &    -0.37942259 & & \\
 \hline
 $\log B(\log r)$ &	0.048 &       -1.3996038 &     -2.7297839 &    -0.57806547 & & \\
				  &	0.090 &       0.90110720 &     -1.4531042 &    -0.92303502 & & \\
				  &	0.232 &       -65.562920 &     -307.15327 &     -531.13562 &     -407.62824 &     -116.78615 \\ [1ex] 
 \hline
 \end{tabular}
\end{table}

\begin{table}[h!]
\caption{disk model: \mstar\,= 0.1\,\msun, \mdot\, = 10$^{-9}$\,\msun\,yr$^{-1}$, $\alpha_{\rm DZ} = 10^{-4}$.
\label{tab:poly2}}
\centering
 \begin{tabular}{|c c c c c c c|} 
 \hline
 Fitted function & Interval end radius (AU) & $c_0$ & $c_1$ & $c_2$ & $c_3$ & $c_4$ \\ [0.5ex] 
 \hline\hline
 $\log \bar{\alpha}(\log r)$	&	0.023	&	-12.472096	&	-12.102624	&	-3.8119263	&	-0.3015077	&		\\
								&	0.108	&	-8.103168	&	-4.5772352	&	-0.34497049	&		&		\\
 \hline
 $\log B(\log r)$	&	0.023	&	-2.6963071	&	-3.4619552	&	-0.66750054	&		&		\\
					&	0.042	&	0.40642769	&	-1.9598652	&	-0.89520128	&		&		\\
					&	0.108	&	-233.05517	&	-779.1073	&	-972.0254	&	-538.41475	&	-111.58188	\\ [1ex] 
 \hline
 \end{tabular}
\end{table}

\begin{table}[h!]
\caption{disk model: \mstar\,= 1\,\msun, \mdot\, = 10$^{-10}$\,\msun\,yr$^{-1}$, $\alpha_{\rm DZ} = 10^{-4}$.
\label{tab:poly3}}
\centering
 \begin{tabular}{|c c c c c c c|} 
 \hline
 Fitted function & Interval end radius (AU) & $c_0$ & $c_1$ & $c_2$ & $c_3$ & $c_4$ \\ [0.5ex] 
 \hline\hline
 $\log \bar{\alpha}(\log r)$ &	0.066 & -11.604954 & -9.0560947 & -2.2069353 & & \\
 \hline
 $\log B(\log r)$ & 0.066 & -169.31517 & -454.52007 & -454.24907 & -201.59358 & -33.457131\\ [1ex] 
 \hline
 \end{tabular}
\end{table}

\begin{table}[h!]
\caption{disk model: \mstar\,= 1\,\msun, \mdot\, = 10$^{-8}$\,\msun\,yr$^{-1}$, $\alpha_{\rm DZ} = 10^{-4}$.
\label{tab:poly4}}
\centering
 \begin{tabular}{|c c c c c c c|} 
 \hline
 Fitted function & Interval end radius (AU) & $c_0$ & $c_1$ & $c_2$ & $c_3$ & $c_4$ \\ [0.5ex] 
 \hline\hline
 $\log \bar{\alpha}(\log r)$	&	0.190	&	-6.7811175	&	-12.02901	&	-8.2440562	&	-1.8594328	&		\\
								&	0.730	&	-4.5536009	&	-4.1222938	&	-0.17861849	&		&		\\
 \hline
 $\log B(\log r)$	&	0.190	&	-0.18901295	&	-1.6613468	&	-0.22499278	&		&		\\
					&	0.362	&	1.2930281	&	-0.098087489	&	-0.85474294	&		&		\\
					&	0.730	&	-1.2770216	&	-25.759443	&	-116.39124	&	-238.42566	&	-180.2177	\\ [1ex] 
 \hline
 \end{tabular}
\end{table}

\begin{table}[h!]
\caption{disk model: \mstar\,= 1\,\msun, \mdot\, = 10$^{-9}$\,\msun\,yr$^{-1}$, $\alpha_{\rm DZ} = 10^{-5}$.
\label{tab:poly5}}
\centering
 \begin{tabular}{|c c c c c c c|} 
 \hline
 Fitted function & Interval end radius (AU) & $c_0$ & $c_1$ & $c_2$ & $c_3$ & $c_4$ \\ [0.5ex] 
 \hline\hline
 $\log \bar{\alpha}(\log r)$	&	0.048	&	-8.9655034	&	-9.8830202	&	-3.6673762	&	-0.33667805	&		\\
								&	0.415	&	-6.5991431	&	-4.2982932	&	-0.33100659	&		&		\\
 \hline
 $\log B(\log r)$	&	0.048	&	-1.4088427	&	-2.7388476	&	-0.58018993	&		&		\\
					&	0.092	&	0.99828119	&	-1.2919053	&	-0.85648004	&		&		\\
					&	0.415	&	-3.0815134	&	-20.240474	&	-39.070548	&	-34.357312	&	-11.153956	\\ [1ex] 
 \hline
 \end{tabular}
\end{table}

\begin{table}[h!]
\caption{disk model: \mstar\,= 1\,\msun, \mdot\, = 10$^{-9}$\,\msun\,yr$^{-1}$, $\alpha_{\rm DZ} = 10^{-3}$.
\label{tab:poly6}}
\centering
 \begin{tabular}{|c c c c c c c|} 
 \hline
 Fitted function & Interval end radius (AU) & $c_0$ & $c_1$ & $c_2$ & $c_3$ & $c_4$ \\ [0.5ex] 
 \hline\hline
 $\log \bar{\alpha}(\log r)$	&	0.048 &	-8.9910621	&	-9.9615069	&	-3.7341083	&	-0.35427178	&		\\
								&	0.127 &	-6.7619972	&	-4.6539156	&	-0.51395667	&		&		\\
 \hline
 $\log B(\log r)$	&	0.048	&	-1.3612921	&	-2.6728148	&	-0.55758751	&		&		\\
					&	0.086	&	0.80147769	&	-1.6020543	&	-0.97716225	&		&		\\
					&	0.127	&	-3619.2649	&	-14562.646	&	-21960.398	&	-14712.958	&	-3694.7273	\\ [1ex] 
 \hline
 \end{tabular}
\end{table}

\end{document}